\documentclass[aps,pre,showpacs,floatfix,superscriptaddress]{revtex4-1}
\usepackage{graphicx}
\usepackage{epstopdf}
\usepackage{natbib}
\usepackage{amsmath}
\usepackage{epsfig,xcolor,tikz}
\usepackage{bm,ulem}

\newcommand\Rey{\mbox{{Re}}}  % Reynolds number
\newsavebox{\astrutbox}
\sbox{\astrutbox}{\rule[-5pt]{0pt}{20pt}}

\newcommand{\ADD}[1]{{#1}}

\newcommand\etal{\mbox{\textit{et al.}}}

\newcommand\ie{i.e.\ }
\def\red{\textcolor{black}}

\begin{document}

\title{Dynamics of circular arrangements of vorticity in two dimensions}

\author{Rohith V. Swaminathan}
\affiliation{MIT/WHOI Joint Program in Oceanography, Massachusetts Institute of Technology, Cambridge, MA 02139, USA}
\author{S. Ravichandran}
\affiliation{TIFR Centre for Interdisciplinary Sciences, Tata Institute of Fundamental Research, Hyderabad 500075, India}
\author{Prasad Perlekar}
\affiliation{TIFR Centre for Interdisciplinary Sciences, Tata Institute of Fundamental Research, Hyderabad 500075, India}
\author{Rama Govindarajan}
\affiliation{TIFR Centre for Interdisciplinary Sciences, Tata Institute of Fundamental Research, Hyderabad 500075, India}
\email{rama@tifrh.res.in}

\begin{abstract}
The merger of two like-signed vortices is a well-studied problem, but in a turbulent flow, we may often have more than two like-signed vortices interacting. We study the merger of three or more identical co-rotating vortices initially arranged on the vertices of a regular polygon. At low to moderate Reynolds numbers, we find an additional stage in the merger process, absent in the merger of two vortices, where an annular vortical structure is formed and is long-lived. Vortex merger is slowed down significantly due to this. Such annular vortices are known at far higher Reynolds numbers in studies of tropical cyclones, which have been noticed to a break down into individual vortices. In the pre-annular stage, vortical structures in a viscous flow \red{are found here to} tilt and realign in a manner similar to the inviscid case, but the pronounced filaments visible in the latter are practically absent in the former. Five or fewer vortices initially elongate radially, and then reorient their long axis closer to the azimuthal direction so as to form an annulus. With six or more vortices, the initial alignment is already azimuthal. \red{Interestingly} at higher Reynolds numbers, the merger of an odd number of vortices \red{is found to} proceed very differently from that of an even number. The former process is rapid and chaotic whereas the latter proceeds more slowly via pairing events.

The annular vortex takes the form of a generalised Lamb-Oseen vortex (GLO), and diffuses inwards until it forms a standard Lamb-Oseen vortex. For lower Reynolds number, the numerical (fully nonlinear) evolution of the GLO vortex follows exactly the analytical evolution until merger. At higher Reynolds numbers, the annulus goes through instabilities whose nonlinear stages show a pronounced difference between even and odd mode disturbances. Here again, the odd mode causes an early collapse of the annulus via decaying turbulence into a single central vortex, whereas the even mode disturbance causes a more orderly progression into a single vortex. Results from linear stability analysis agree with the nonlinear simulations, and predict the frequencies of the most unstable modes better than they predict the growth rates.

It is hoped that the present findings, \red{that multiple vortex merger is qualitatively different from the merger of two vortices,} will motivate studies on how multiple vortex interactions affect the inverse cascade in two-dimensional turbulence.
\end{abstract}

\maketitle
%%\begin{keywords}
%%Authors should not enter keywords on the manuscript, as these must be chosen by the author during the online submission process and will then be added during the typesetting process (see http://journals.cambridge.org/data/\linebreak[3]relatedlink/jfm-\linebreak[3]keywords.pdf for the full list)
%%\end{keywords}

\section{Introduction}
\begin{figure}
\includegraphics[scale=0.45]{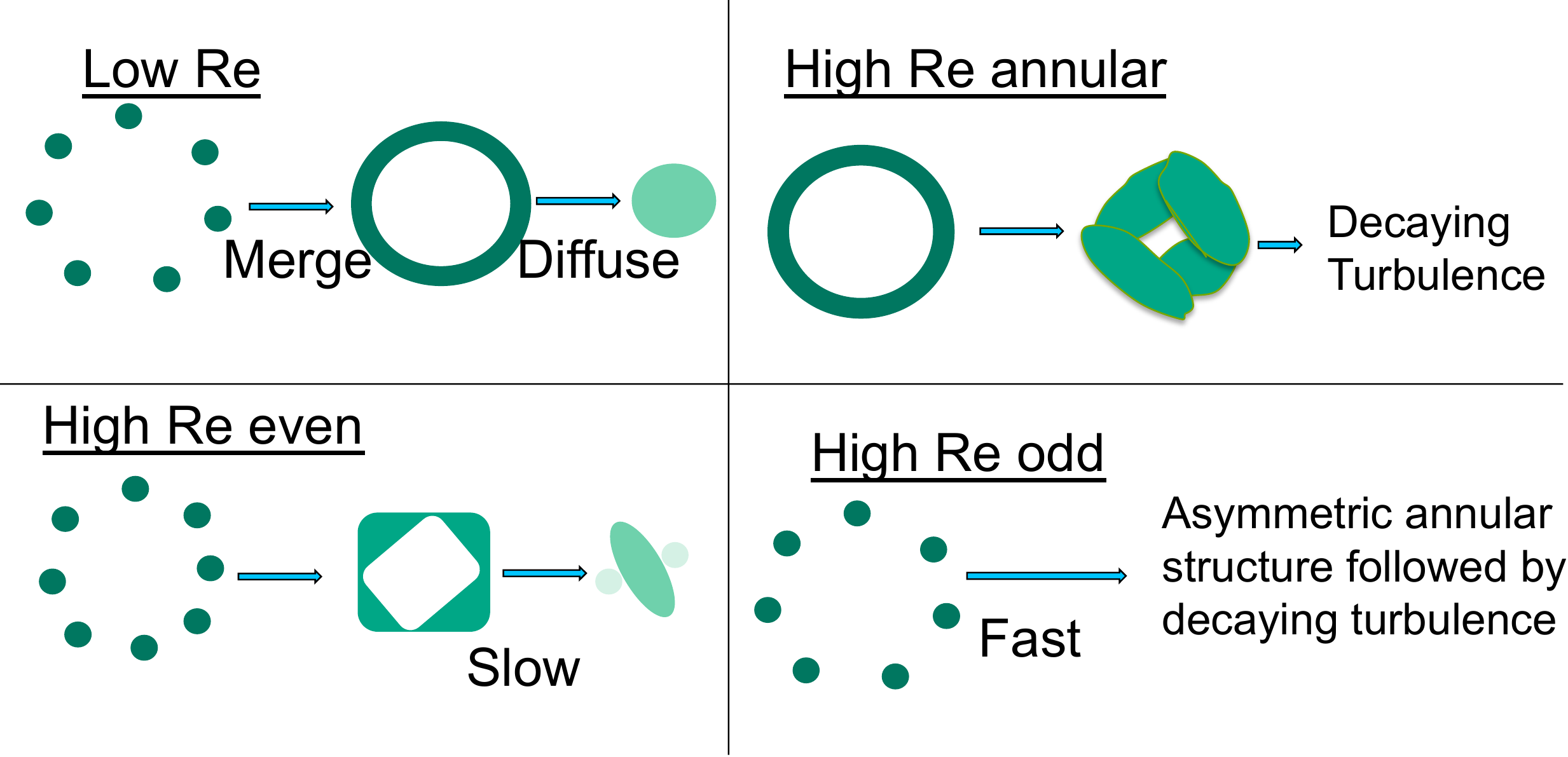}
%\vskip-1truein
\caption{ A schematic summarizing the dynamics. At low Reynolds number, $n$ co-rotating vortices will merge to form an annular structure, which then diffuses inwards to form a single Lamb-Oseen vortex. At high Reynolds number, an annular vortex is not stable. It breaks into several vortices, and creates decaying turbulence. $n$ vortices at high Reynolds number will form an annulus-like structure which is symmetric when $n$ is even and asymmetric when $n$ is odd. The former will decay into a tripolar structure while the latter will break down quickly into decaying turbulence.}
\label{schematic}
\end{figure}

Vortices may be considered to be fundamental building blocks of turbulent flows. \ADD{The study of collections of point vortices has been found to be useful in understanding two-dimensional turbulence \cite{Aref2007}.} Merger events are believed to be of central importance in the inverse cascade of turbulent kinetic energy in two-dimensional turbulence (\cite{Boffetta_Ecke2012,pan09,Danilov_Gurarie2000}). In a turbulent flow dense with vorticity, the most common form of interaction is between two vortices, but it is not unusual for three or more vortices to interact. While the merger process of two identical vortices is extremely well-studied (see e.g. \citet{Melander_etal1988,Cerretelli_Williamson2003a}), the simultaneous interaction in viscous flow of three or more vortices has not been studied in the context of merger, to the best of our knowledge. In this paper, we investigate the merger and breakup of several vortices and the breakup of an annulus of vortex.

We study a model flow consisting of several (up to $9$) equal like-signed vortices placed initially on the vertices of regular polygons, and find that the merger of $N>2$ vortices shows several features distinct from the two-vortex case. The summary of our findings is depicted schematically in figure \ref{schematic}. Vortex merger depends on both whether the number of vortices is odd or even, and on the Reynolds number of the flow. At relatively low Reynolds number, we find an annular vortex stage in the merger process. This annular structure remains axisymmetric but reduces in radius to eventually produce a single merged Lamb-Oseen vortex. On the other hand the evolution at high Reynolds numbers is more complicated. An annular vortex loses symmetry and the flow becomes turbulent as it decays. The vortices in a system of an even number of vortices, e.g. $6$ or $8$, pair up to form a $4$ vortex configuration which then undergoes further mergers. More fascinating is the case of merger of an odd number of vortices. Here, all vortices cannot pair and the merger process is more turbulent. The system rapidly degenerates into decaying turbulence. Our system possesses of course many more degrees of symmetry than are likely to occur too often in turbulent flow. A more general initial condition would not have identical vortices exactly on the vertices of regular polygons. However annular structures that break up into vortices are seen in real atmospheric flows (e.g. \cite{Menelaou2013a}). A regular structure such as ours reduces the number of variables to be specified as a first step.

We also numerically investigate the stability of an annular vortex. This problem was first studied by \cite{Schubert_etal1999}, who performed inviscid stability analysis and numerical simulations with broadband initial perturbations and found a ring of elevated vorticity perturbed with azimuthally broad-banded initial conditions to go unstable and form multiple vortices, which undergo further rearrangement to a near monopolar circular vortex. We study how the wavenumber of the initial perturbation affects the stability of the annular vortex, and find similar growth rates for the $m=4$ mode as Schubert \etal.

At low Reynolds number the annular vortex is stable and peak vorticity decays via diffusion. However at higher Reynolds number the annular vortex becomes unstable. The linear growth rate is positive for a range of azimuthal wavenumber, and the variation with wavenumber is smooth. In the nonlinear stages of disturbance growth, however, we see some interesting behaviour. Similar to vortex merger at high Reynolds number, the nonlinear destabilisation of the annular vortex strongly depends on whether the initial perturbation is even or odd. This final stage of the merger process is the similar to what happens in inviscid flow, where a thin annulus often goes unstable and bunches up into several patches of vorticity. 

We first examine what is known about the inviscid case because it provides a comparison and contrast with the viscous study to follow.

\section{Dynamics of multiple vortices in inviscid flow}

\begin{figure}
\centering
\includegraphics[trim = 250 0 250 0, clip, scale=0.235]{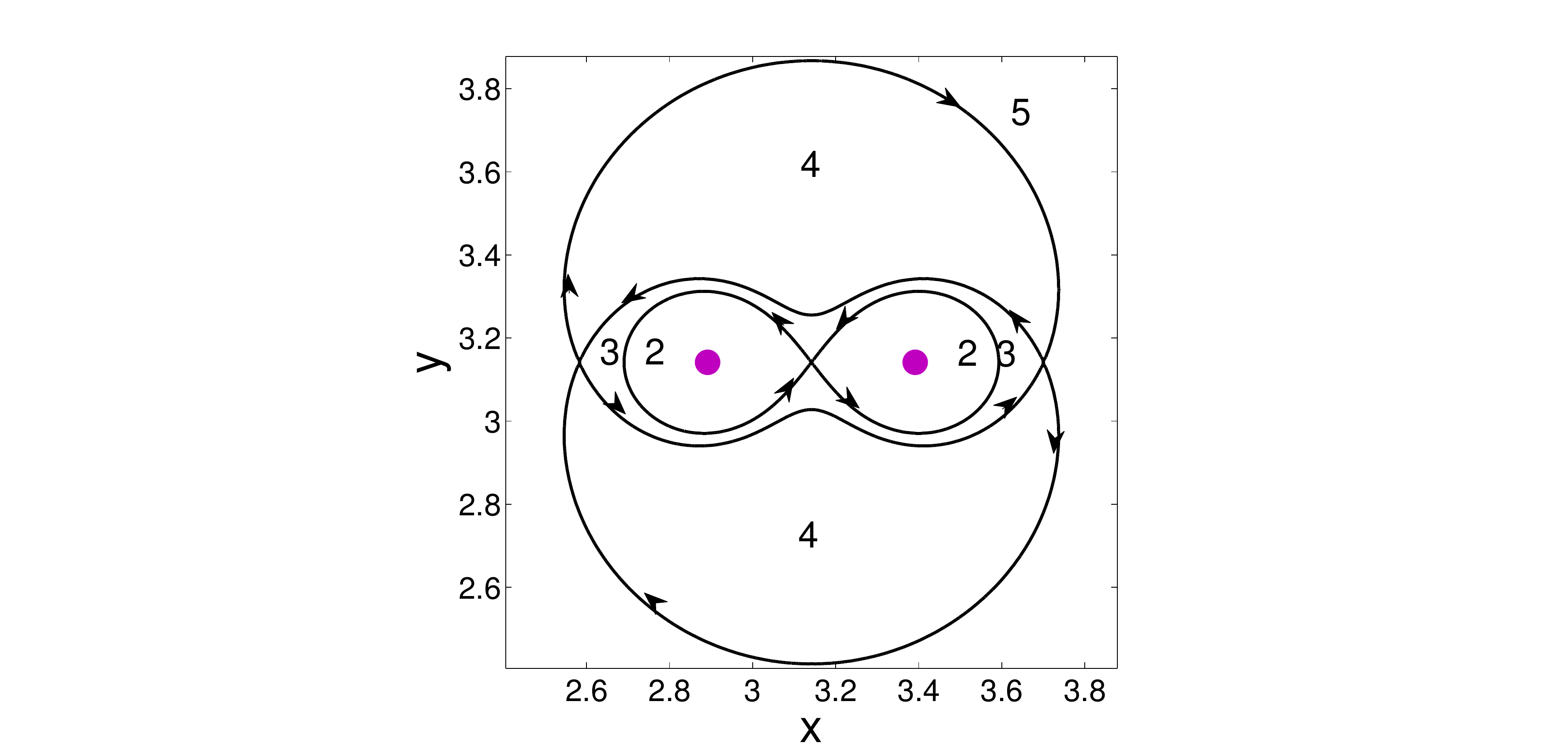}
{\includegraphics[trim = 250 0 250 0, clip, scale=0.235]{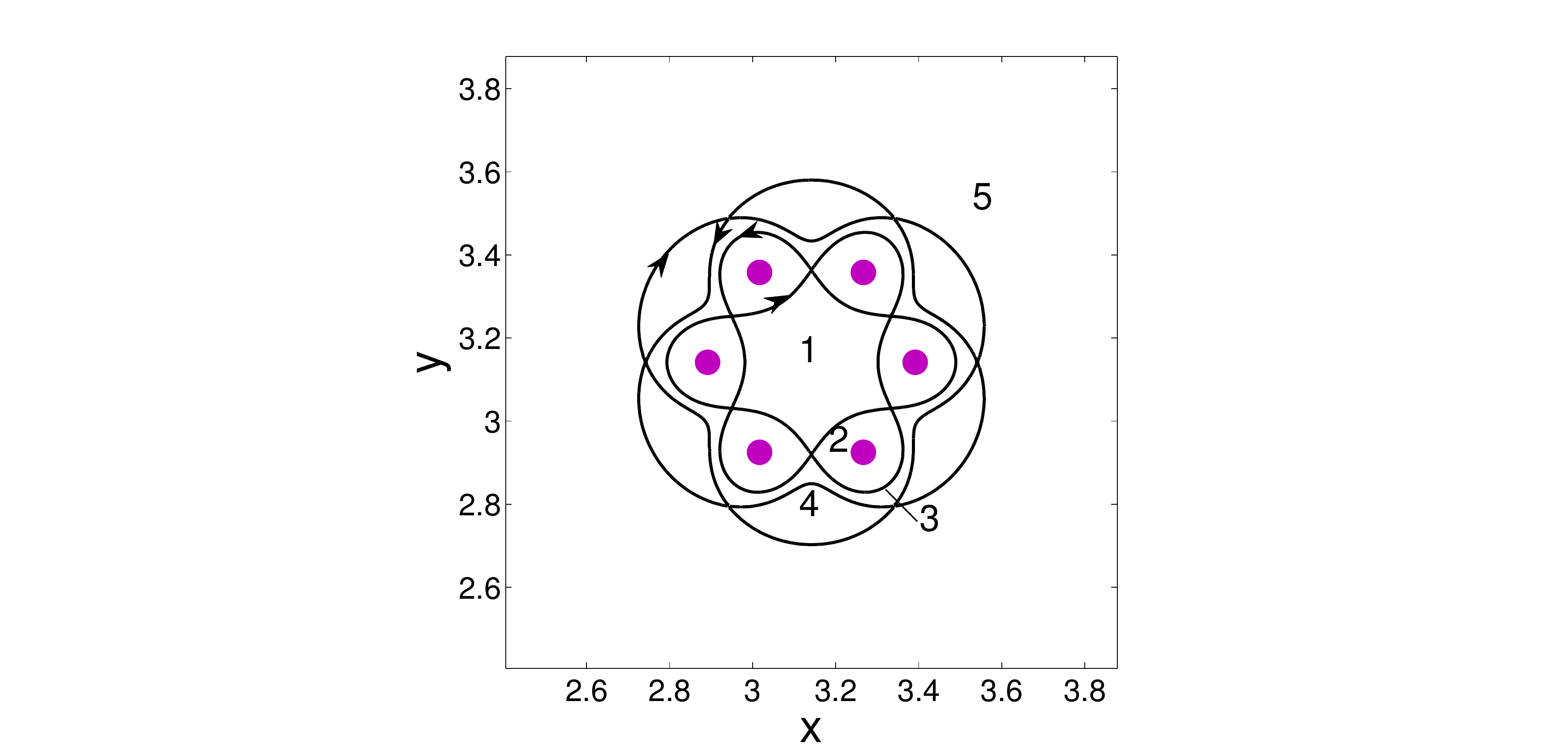}}
{\includegraphics[trim = 245 0 250 0, clip, scale=0.235]{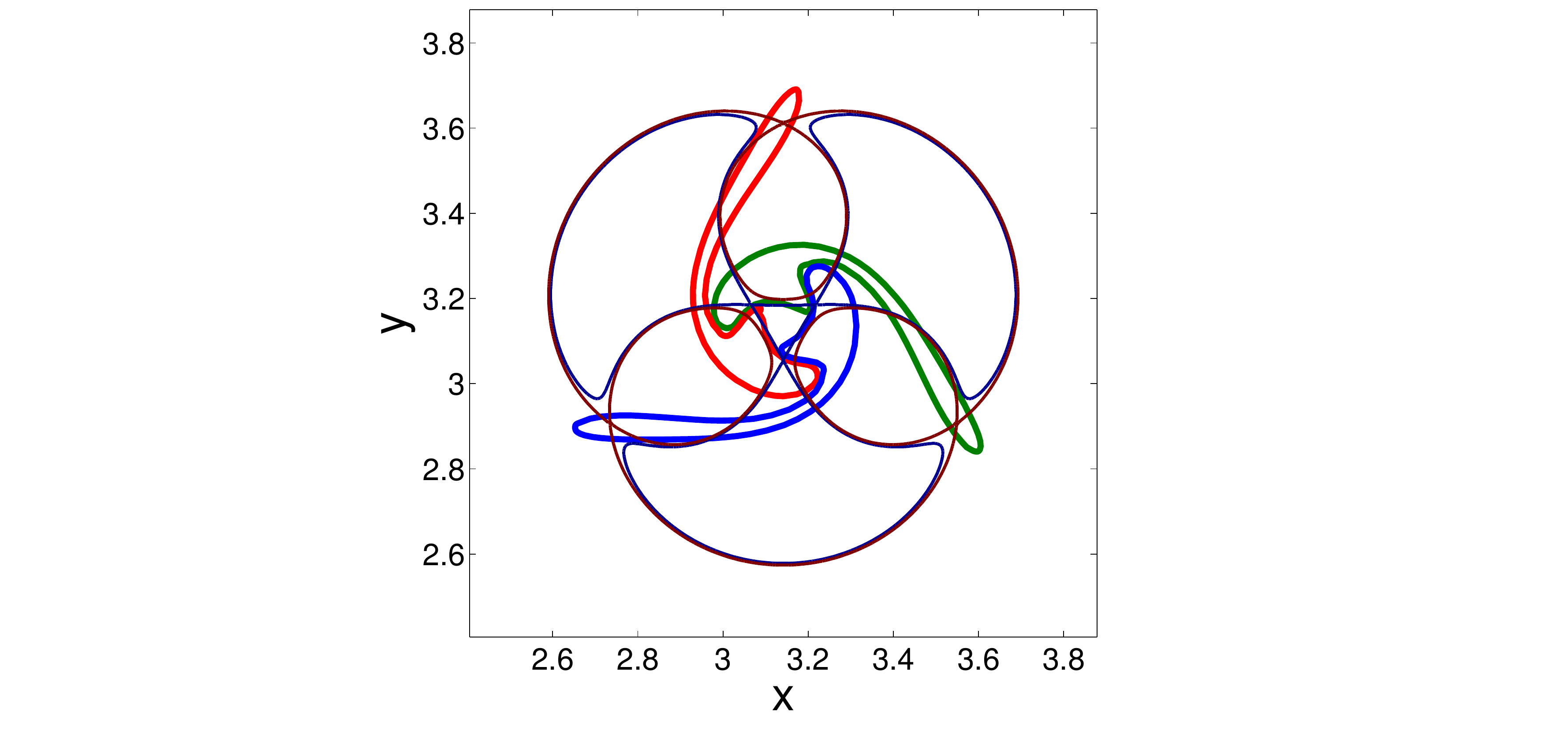}}
\caption{Separating streamlines in the co-rotating frame for two (left) and six (centre) point vortices. The vortices are indicated by large (pink online) dots. The regions are distinguished as follows: (1) central (not present for 2 vortices), (2) inner core, (3) exchange band, (4) outer recirculating and (5) external flow. (Right) Separatrices for three point vortices (thin lines) overlaid on three patches of uniform vorticity which evolved starting from circular patches at the same positions as the \ADD{point vortices} (solid lines). }
\label{fig:pointvortsf2}
\end{figure}

Consider an inviscid system with $n$ co-rotating point vortices, each of circulation $\Gamma/n$, and separated by a distance $d/2$ from the centroid of the system, as shown in figure \ref{fig:pointvortsf2}. These vortices are confined by the Biot-Savart law to motion on a circle of diameter $d$, with an azimuthal velocity of $\displaystyle u_\theta=(n-1)\Gamma/(n \pi d)$. In a coordinate frame rotating with the same angular velocity as the vortices, the system is in steady state, with the positions of the vortices, as well as the streamlines, frozen. There are several saddle points in this frame of reference, and the vortices themselves act as centres. Streamlines connecting the saddle points, either homoclinically or heteroclinically, form separatrices in the flow, as shown in figures \ref{fig:pointvortsf2}(a) and (b) for two and six vortices respectively. The separatrices divide the fluid flow into distinct regions: (1) central, (2) inner core, (3) exchange band, (4) outer recirculating and (5) external flow \cite{Dritschel1985}. \ADD{The regions 2 comprise the immediate proximity of the vortices, so the vortex dominates the motion and closed streamlines are seen. In regions 4 too,} flow follows closed streamlines in the rotating coordinate system, and the fixed points \ADD{(not shown)} within these are centres. The outer recirculating region has fluid rotating in a sense opposite to the the inner core and exchange band regions. Only region 2 contains vorticity. The central region exists only when we have three or more vortices, this is evident in figures \ref{fig:pointvortsf2} (b) and (c). The exchange band \ADD{is seen to be a thin band of fluid, which will be seen to become important in filament formation. It} is well-known in the two-vortex viscous case to be of importance during the merger process, we will see how it affects the dynamics in the multiple vortex case.

\begin{figure}
\centering
\includegraphics[width=\linewidth]{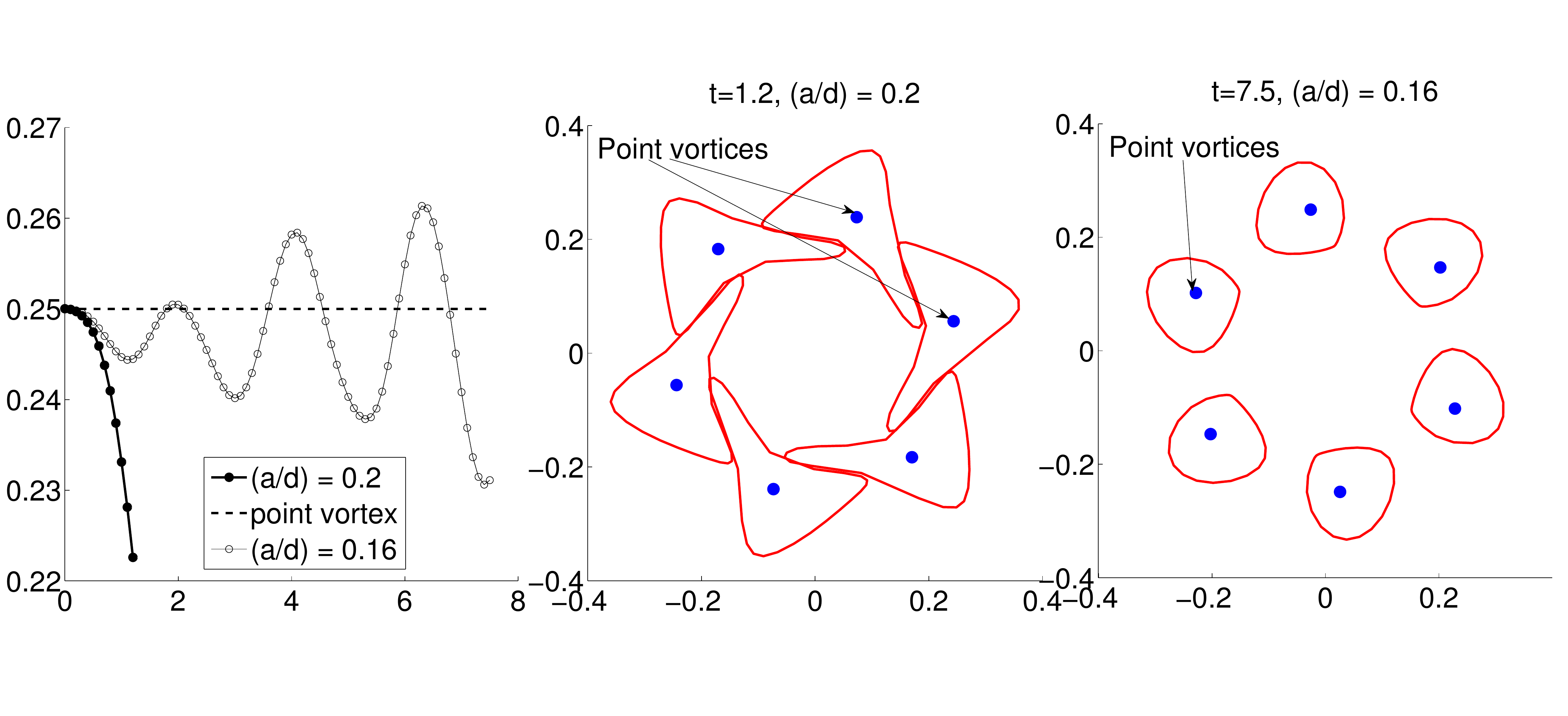}
\includegraphics[width=\linewidth]{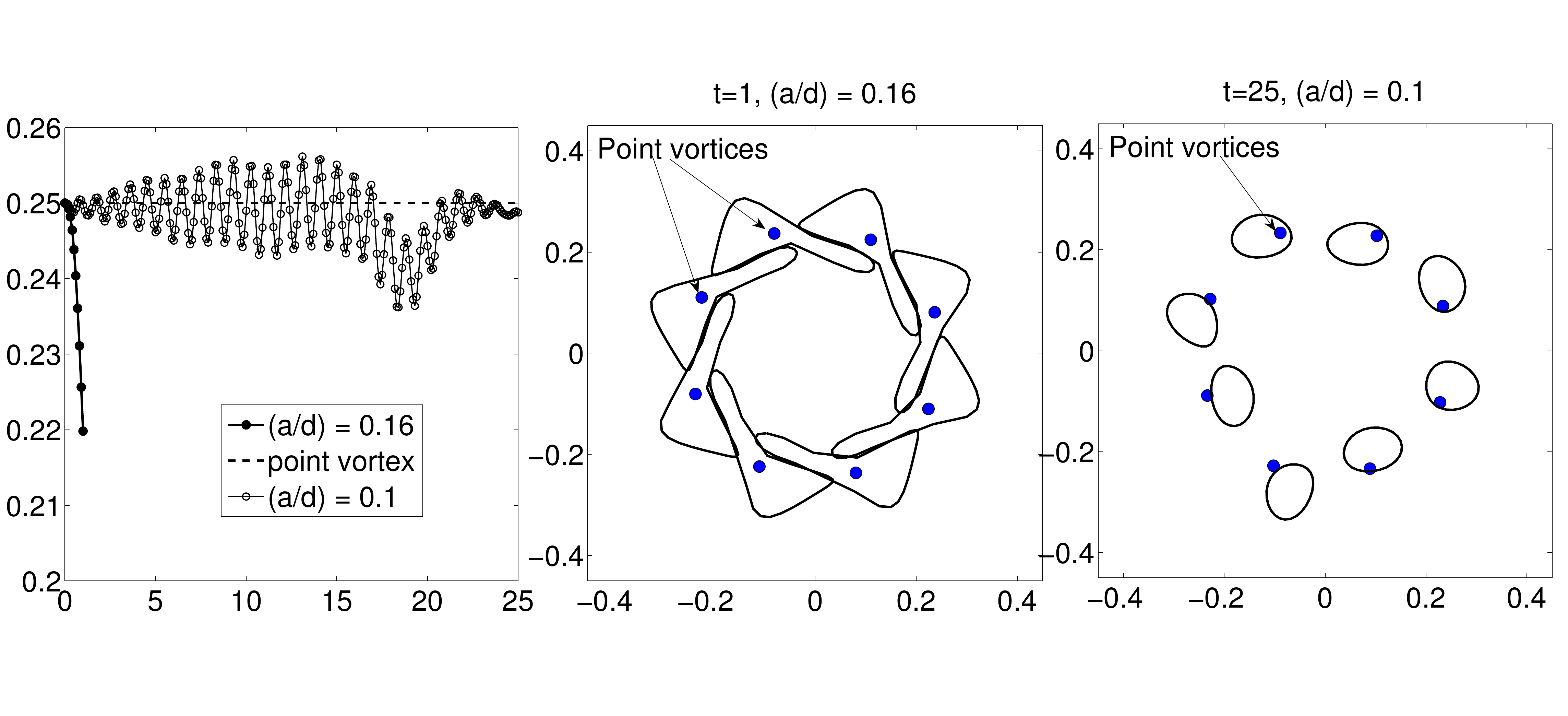}
\caption{Inviscid simulations of vortex patches with $(a/d)_{i} > (a/d)_{crit}$ and $(a/d)_{i} < (a/d)_{crit}$ for $N=6$ (top) and $N=8$ (bottom). The average radial distance of the patches from the origin is shown on the left panels. For $N=6$, the $(a/d)_{crit}$ is higher than the value for $N=8$. It is seen that that the dynamics up to the time shown, in terms of the radial location, is similar to that of the point vortices when the patches are far apart from each other at the initial time. Also the shape deviates from the circular, but remains convex. Note however that eight patches which start $(a/d)_{i} < (a/d)_{crit}$ go unstable and therefore do not remain axisymmetric, whereas the point vortices occupy (by the analytical expressions) the vertices of a rotating octagon at all times. When the initial separation is smaller, the patches for $N=6$ as well as for $N=8$ undergo rapid changes in shape, and the average distance from the centre decreases rapidly.}
\label{fig:patchsims}
\end{figure}

The point vortex system was shown long ago \citep{Thomson1883,Havelock1932} to be unstable for $n > 7$. Among the host of studies that followed, we pick our way through only the most relevant for the present work. Primary among these is \citet{Dritschel1985} who extended the work of Thomson to the case of $n$ finite patches of uniform vorticity rather than point vortices, placed initially at the vertices of a regular polygon, and allowed to undergo inviscid dynamics. If the vortices have finite but small radius $a$, such that $\left(a/d\right)$ is much less than a critical ratio $\left(a/d\right)_{crit}$, the dynamics is similar to that of point vortices. Larger inviscid patches do not remain circular but their shapes may be obtained exactly. If $\left(a/d\right)\ge\left(a/d\right)_{crit}$, the vortices, while rotating around each other, deform and also move towards each other, but Kelvin's circulation theorem prevents any reconnection of their boundaries. In the case of vortices with finite size, non-linear inviscid studies \citep{Dritschel1986} so far have indicated that these vortices collapse into an annulus-like structure.

\ADD{To differentiate finite size effects from viscous diffusion effects, we start with circular patches of uniform vorticity, rather than point vortices, placed at the vertices of regular polygons, and allow their shapes to evolve inviscidly. The inviscid patch vorticity simulations have been carried out using a contour dynamics code from \citet{Pozrikidis2011}. Figure \ref{fig:pointvortsf2}(c) shows how vortex patches have evolved in a three-vortex configuration. An annular vortical structure is formed surrounding a small central region without vorticity. The patches have developed long tails as well. Separatrices of the point vortex configuration are shown in the same figure for comparison. It is seen that the central region is the same for both patch and point vortices. }

\ADD{Figure \ref{fig:patchsims} shows how an annular structure forms for $N=6$ and $N=8$ patches. For small initial patch radii $a_i$, the vortex patches are strained from their initially circular shapes, but remain separate. The left panels of the figures show that the vortex centers oscillate about their initial radial distance with a systematic increase in amplitude in the six vortex case and chaotically in the eight vortex case. On the other hand, for $(a/d)_i > (a/d)_{crit}$, the patches reorient themselves azimuthally and undergo a distinct change in shape. The long tails of the patches are seen to be folded inwards for $N=6$ and $N=8$, unlike at $N=3$ where they were pointed outwards. It may be inferred from figure \ref{fig:patchsims} that $(a/d)_{crit}$ is between $0.16$ and $0.2$ for $N=8$, and between $0.1$ and $0.16$ for $N=8$; \ie $(a/d)_{crit}$ decreases with increasing $N$. When the patches are too close or the filaments too thin, inviscid simulations become prohibitively expensive.}

An inviscid annular vortex on the other hand goes unstable to azimuthal disturbances and splits into smaller multiple vortices \cite{Dritschel1986}. Symmetric neutral disturbances of large azimuthal wavenumber are the only ones that resist instability. 

\section{Simulations of vortex merger}

\subsection{Numerical scheme and initial conditions}

Since the flow is incompressible and two-dimensional, we may work with the vorticity and stream function equations:
\begin{align}
\label{eq:vortstream}
\frac{\partial \omega}{\partial t}+\frac{\partial\psi}{\partial y}\frac{\partial \omega}{\partial x}-\frac{\partial\psi}{\partial x}\frac{\partial \omega}{\partial y}&=\nu \nabla^{2} \omega,\\\nonumber\\
\nabla^2\psi&=-\omega,
\end{align}
where $\omega$ and $\psi$ are the vorticity and stream function  respectively, $x$ and $y$ are spatial coordinates and $t$ is time. We solve this system by a Fourier pseudo-spectral technique \citep{Canuto2006, per13} for spatial discretisation, with a standard de-aliasing by zero padding using the $3/2$ rule. Simulations at high Reynolds numbers and those for the stability of an annular vortex structure are time evolved with an exponential Adams-Bashforth temporal discretisation. Unless stated otherwise, we work with a square domain of 
area $4\pi^2$ and discretise it with $N^2$ collocation points with $N=2048$. 

%To run for long times our code has been ported on to GPU's.
%For our moderate Reynolds number vortex merger simulations we time evolve using an explicit Runge-Kutta fourth-order temporal. 

At the initial time, $n$ identical co-rotating vortices ($n$ varying from 2 to 8) are arranged at equal intervals on the circumference  of a circle of radius $R$ i.e. on the vertices of a regular polygon. Each of these vortices has a Gaussian initial vorticity profile given by $\omega=\omega_{0}e^{-r^2/a^2}$, with a total circulation in the system given by $\Gamma$. Here $r$ is the radial location measured from the centre of each vortex. In our simulations, unless explicitly mentioned otherwise, we prescribe $\left(a/d\right)_i=0.1$ at $t=0$, where the subscript $i$ refers to initial conditions. To make fair comparisons between cases of different $n$, we follow two approaches. In the first case, we keep the total initial circulation $\Gamma$ constant across the set of simulations. The Reynolds number is then given by
\begin{align}
\Rey_{\Gamma}=\frac{\Gamma}{\nu},
\end{align} 
and the time scale is the period of rotation of point vortices of the same strength, given by
\begin{align}
T_{\Gamma}=\frac{2n \pi^2 d^2}{(n-1)\Gamma}.
\end{align}
In the following, wherever we refer to $\Rey$ without a subscript, we mean $\Rey_\Gamma$.

In the second case, we keep the total initial energy, within a domain of diameter $d_0$, constant across the set of simulations. The Reynolds number is based on the root mean square velocity $u_{rms}$ within this domain, and its size:
\begin{align}
\Rey_E=\frac{u_{rms}d_{0}}{\nu},
\end{align}
with the time scale $T_E=d_0/u_{rms}$.

\begin{figure}[!h]
\includegraphics[width=0.45\linewidth]{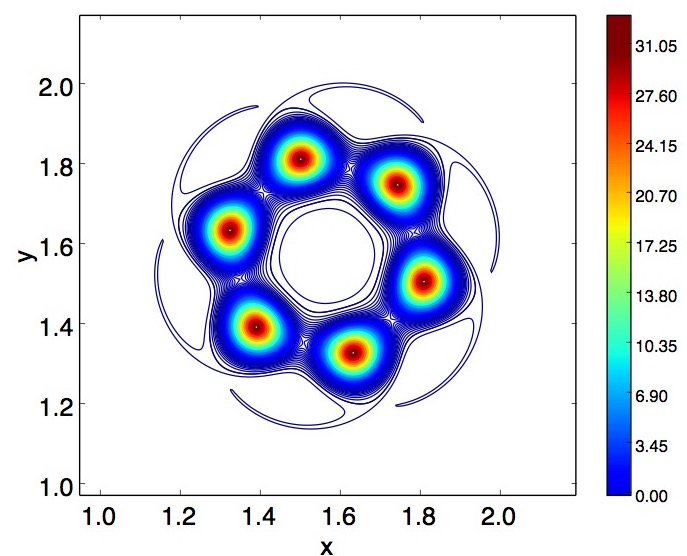}
\includegraphics[width=0.45\linewidth]{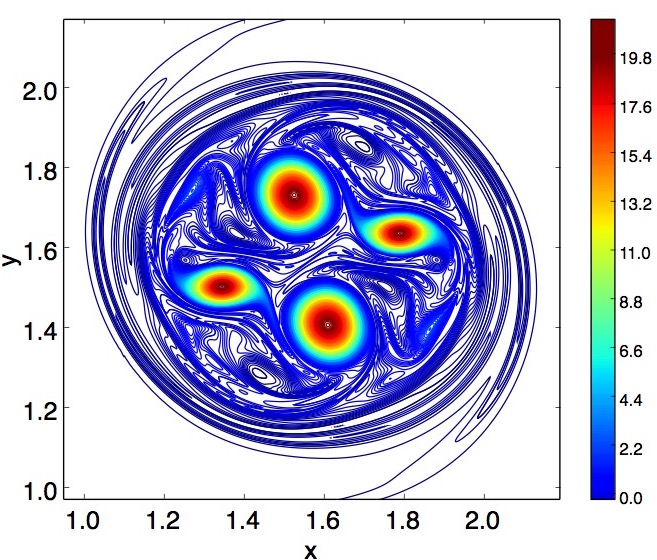}
\includegraphics[width=0.45\linewidth]{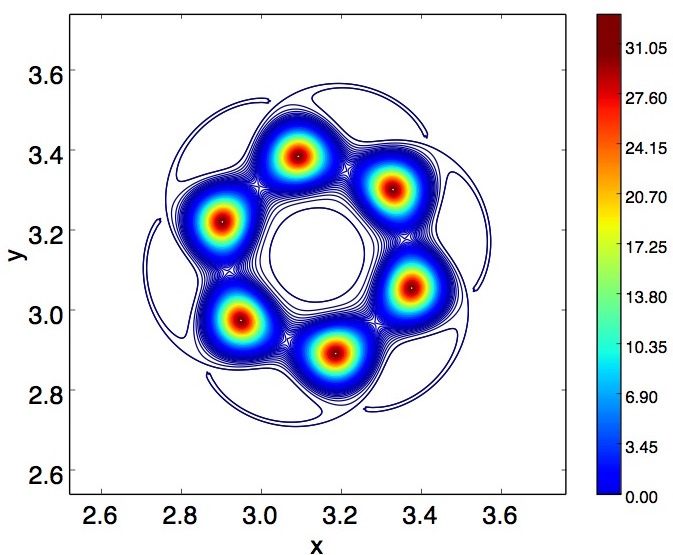}
\includegraphics[width=0.45\linewidth]{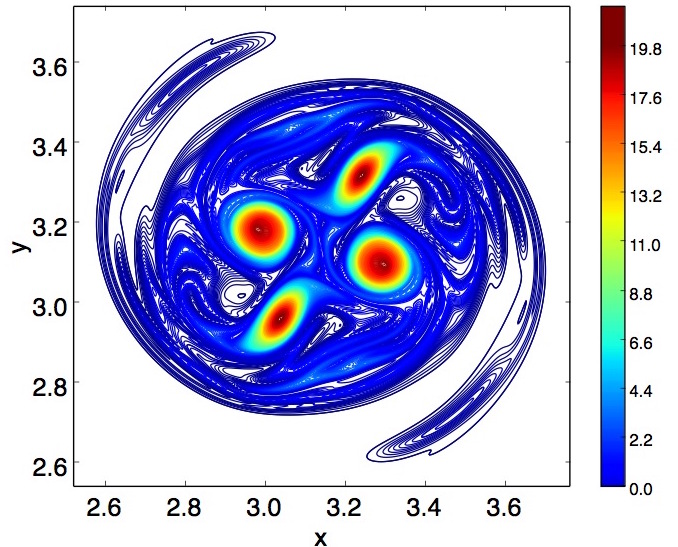}
\caption{\label{fig:box_size_test} Contour plots (in colour online) of the vorticity field for a box of size $\pi^2$ (left) and $4 \pi^2$ (right) at times $t=4.6$ (top row) 
and $t=14.86$  (bottom row). We see that main vortex structures are very similar however because of the large-scale counter vortex 
being different in the two cases, the details of the vortex wings are different.}
\end{figure}

In viscous flow, two vortices which come close enough to touch each other merge into one vortex. This process has been described in great detail, see e.g., \citet{Melander_etal1988,Cerretelli_Williamson2003a,Meunier_etal2002,Meunier_Leweke2005,Fuentes2005,Huang2005}. If the initial radius of each vortex is much smaller than the distance separating their centres, the vortices rotate at a constant frequency as described above. Simultaneously they grow in size by viscous diffusion. Once their radius reaches a critical fraction ($=0.22$) of their separation distance, enough vorticity has entrained into the exchange band, which creates filamentary structures that induce a radially inward velocity on the vortices, bringing them close to each other. This is followed by a second diffusive stage where the separation decreases slowly to $0$, whence an axisymmetrization of the merged vortex follows. The filamentary debris is erased by viscosity during this time. A typical history of the separation distance is shown in figure \ref{fig:bncomp} which also serves as a validation of our numerical approach.

All the results presented here use an initial separation $d_i=0.5$ for the vortices. This initial separation is chosen so that the vortices are all well within the periodic box, and image-effects are minimised. \ADD{That this is so was verified by checking that the results are similar in boxes of twice and half the present size. Representative results are shown in figure \ref{fig:box_size_test}. In the 6-vortex case, symmetry breaks by two adjacent vortices merging, and the corresponding pair diametrically opposite doing the same. Numerical noise decides which among the adjacent vortices will merge. Changing the domain size changes this choice in this case, so an apparent change in orientation is seen in the merged structures. Apart from this there are minor differences in detail. We note that the flow system we study has a net vorticity, whereas by applying periodic boundary conditions we force the induced velocity to zero at the boundaries. This does produce small numerical errors.}

\begin{figure}[H]
\includegraphics[width=0.5\columnwidth]{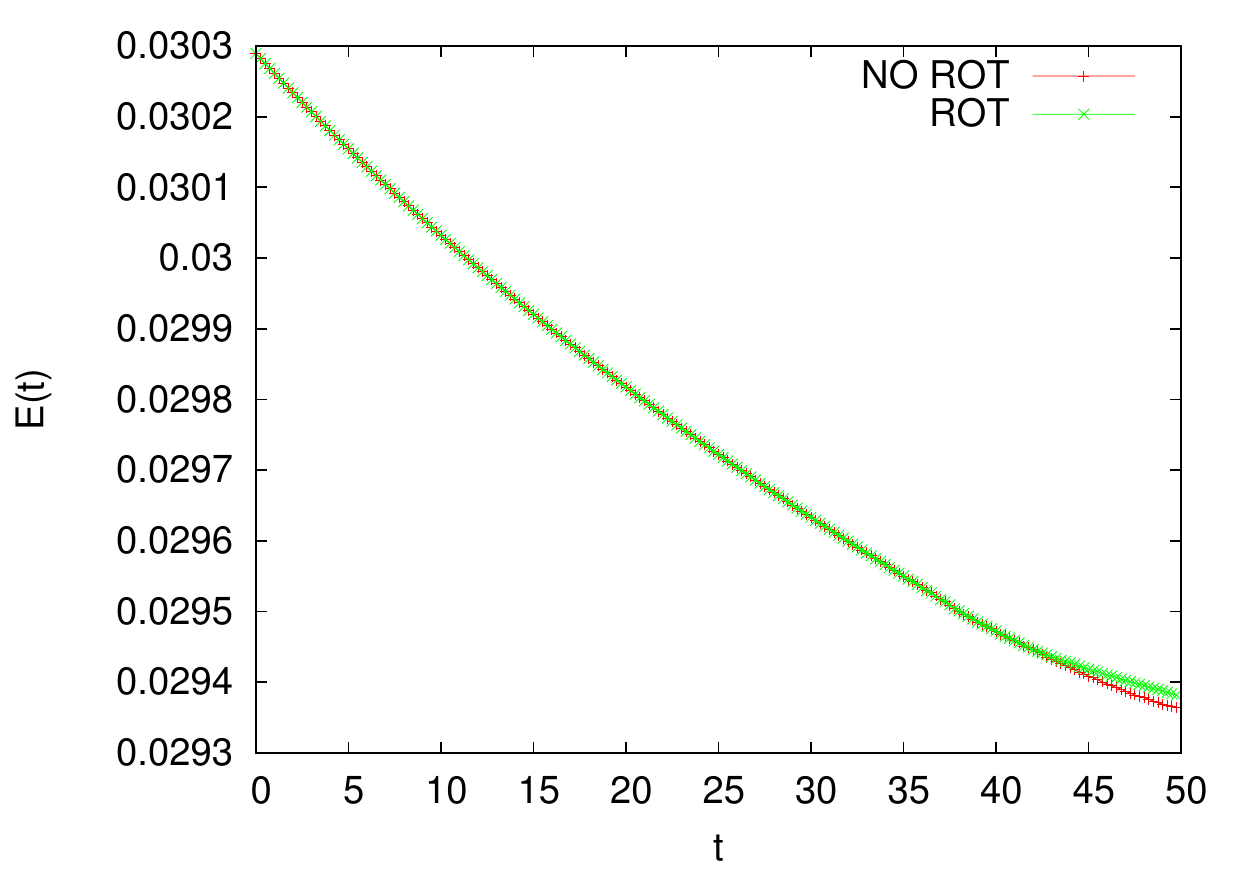}
\caption{\label{fig:energy_rotated} The energy in two simulations rotated by $\pi/5$ relative to each other. The energies are identical until the very end of the simulations where numerical noise becomes important and makes the two curves diverge.}
\end{figure}

\begin{figure}[H]
\includegraphics[width=0.5\columnwidth]{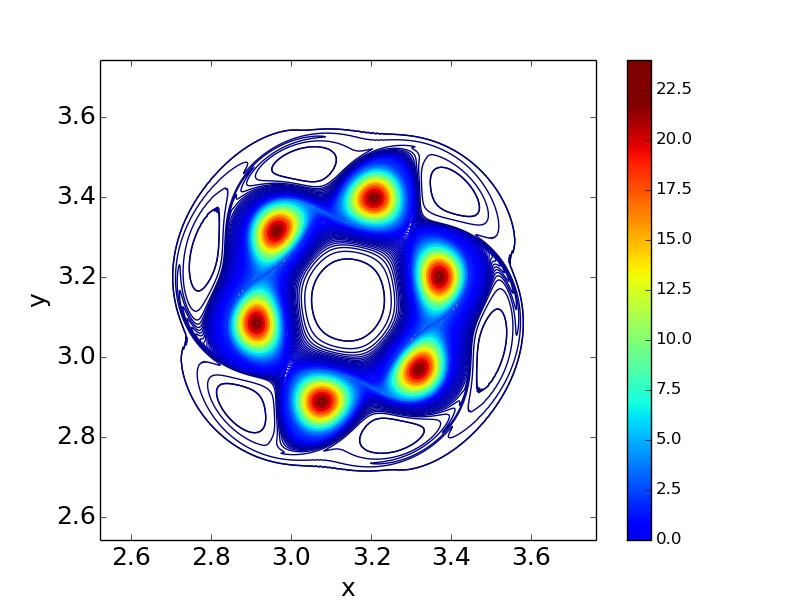}\includegraphics[width=0.5\columnwidth]{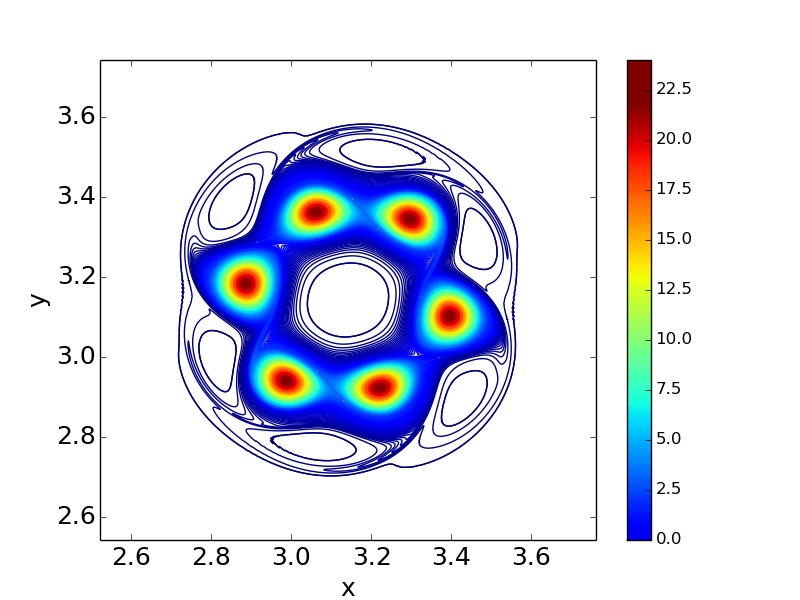}
\caption{\label{fig:Rotated-initial-conditions} Vorticity contours (in colour online) for the simulations from figure \ref{fig:energy_rotated} at time = 35. The figure on the right was started with the vortices rotated by $\pi/5$ relative to the vortices in the left figure. Note that the angle between the figures is now closer to $\pi/2$. See text for an explanation. }
\end{figure}

\begin{figure}
\centering
\includegraphics[trim = 0 0 0 0, clip, scale=0.3]{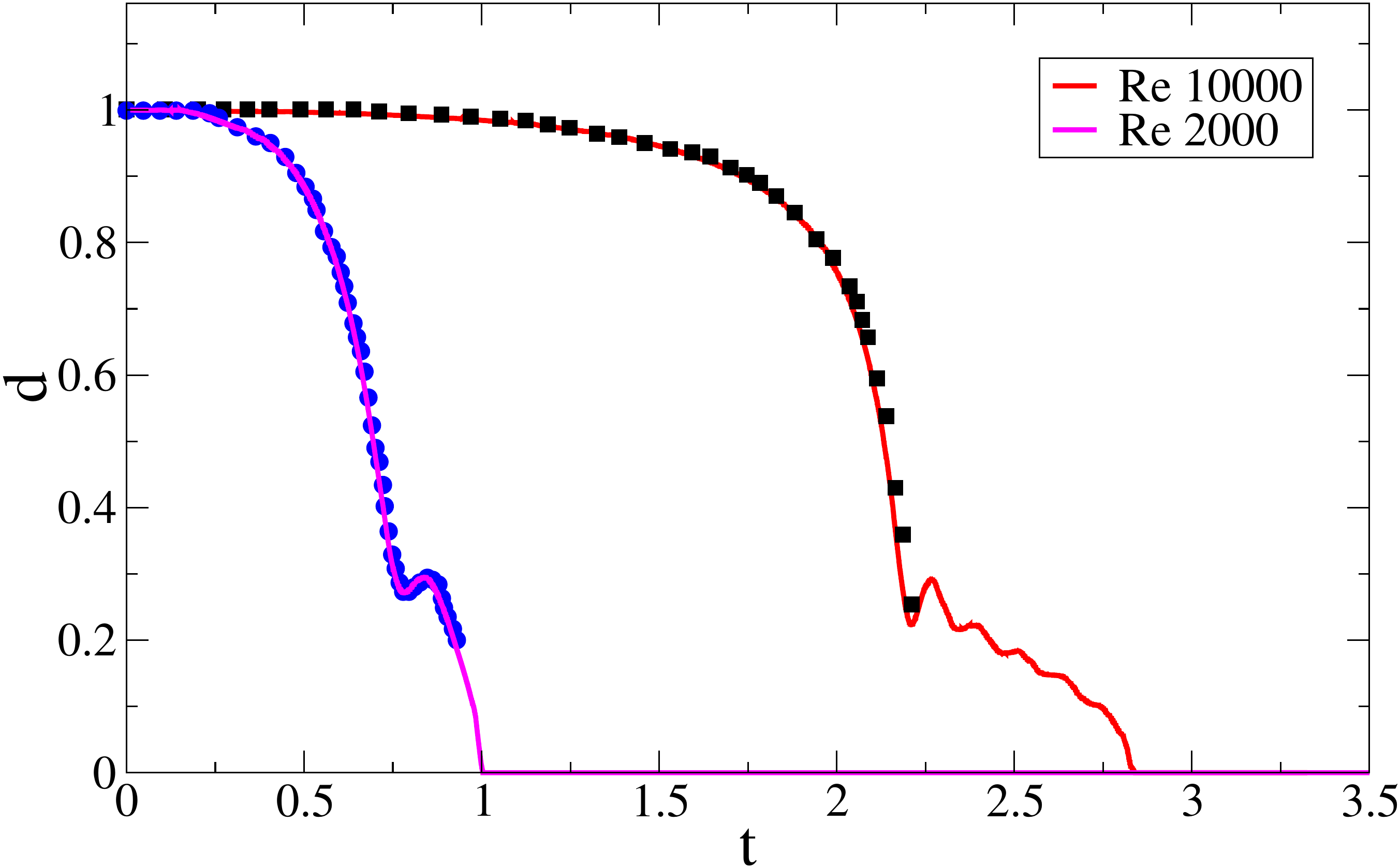}
\caption{Evolution of the separation distance in the merger of two identical vortices. The results of \citet{Brandt_Nomura2006} ($\Rey=2000$, blue filled circles) and \citet{Brandt_Nomura2007} ($\Rey=10000$, black filled squares) compared to present simulations.}
\label{fig:bncomp}
\end{figure}

\ADD{We also verify that the effects of using a Cartesian grid on an axisymmetric problem are negligible, as follows. Two initial conditions rotated relative to each other by an angle of $\pi/5$ are simulated. The results are compared in figure \ref{fig:energy_rotated}, which shows that the energies in the two simulations are identical until the very end.  Again the symmetry-breaking occurs due to numerical noise, and decides which pair of vortices merge.  In figure \ref{fig:Rotated-initial-conditions}, therefore, (counting from the top) vortices 2 \& 3 and 5 \& 6 merge first in the simulation on the left; whereas on the right, vortices 1 \& 2 and 4 \& 5 merge first.  This leads to merged structures that appear to be rotated by more than $\pi/5$.}

\begin{figure}
{\includegraphics[trim = 245 0 250 0, clip, scale=0.235]{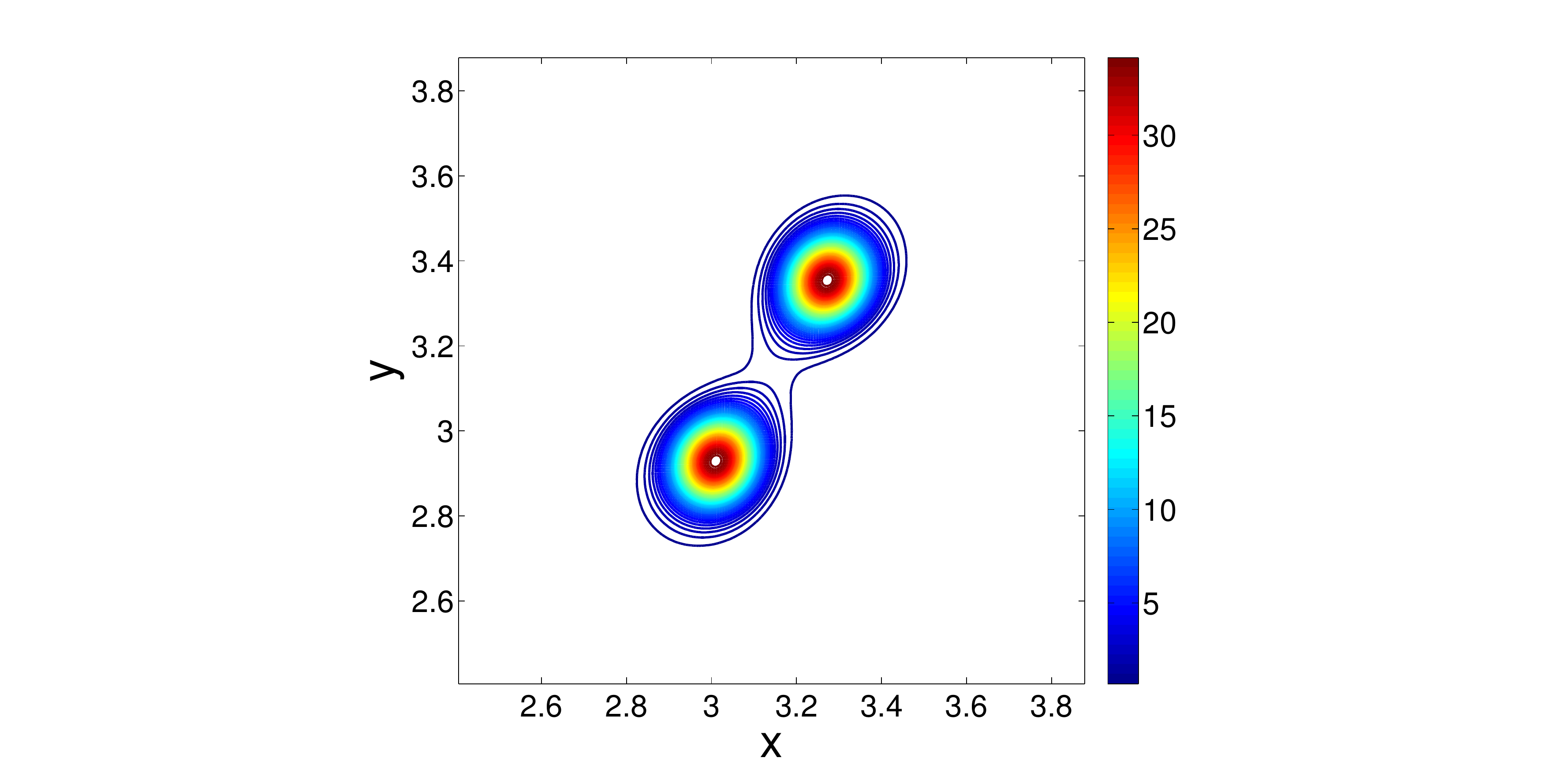}}
{\includegraphics[trim = 245 0 250 0, clip, scale=0.235]{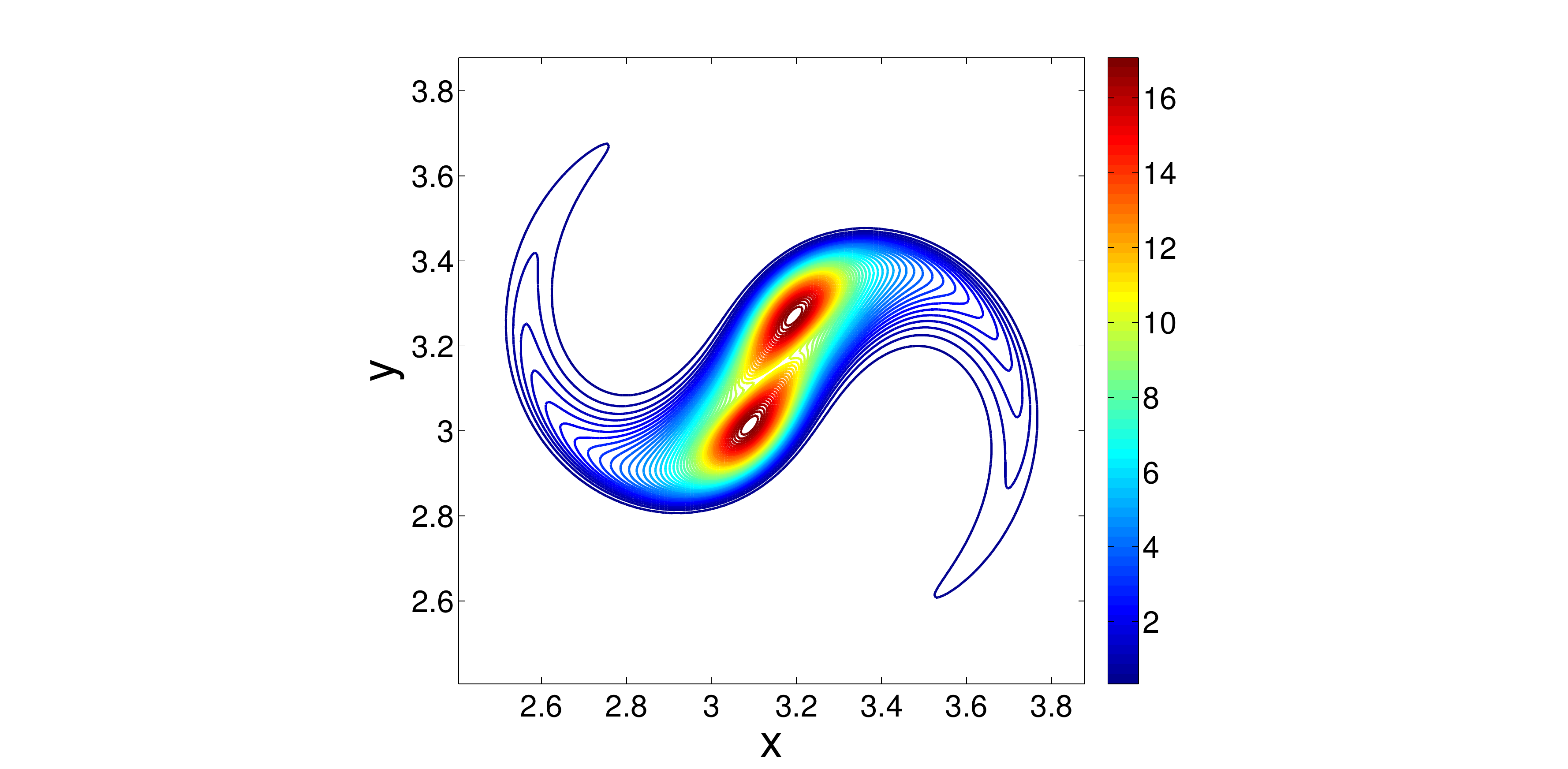}}
{\includegraphics[trim = 245 0 250 0, clip, scale=0.235]{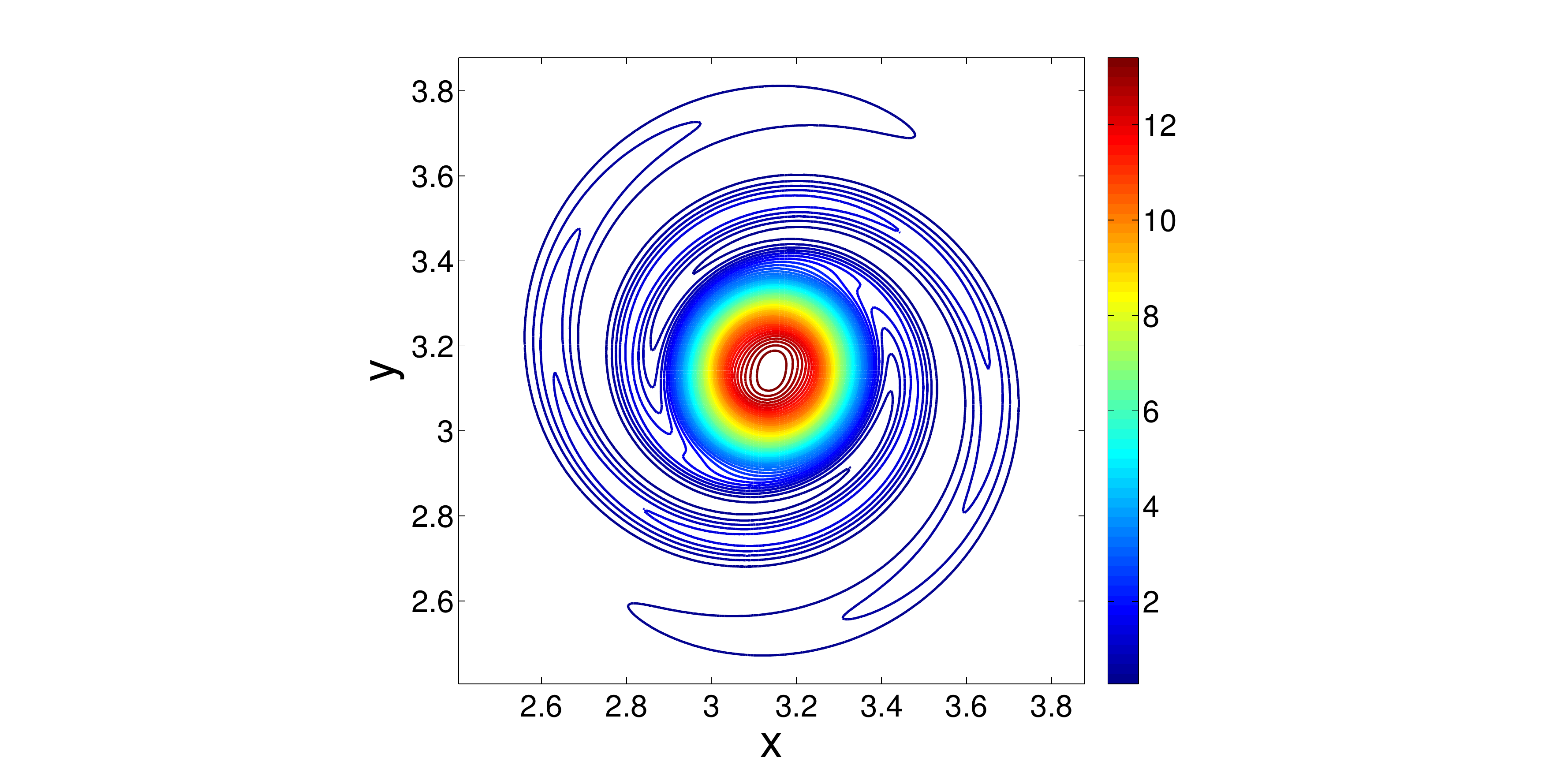}}
\caption{Vorticity contours (in colour online) showing the viscous evolution of two Gaussian vortices with $\left(a/d\right)_{i}=0.1$ and $\Rey_{\Gamma}=4000$. Left: t=0.669, centre: t=1.52, right: t=2.}
\label{fig:2vort}
\end{figure}

To contrast later with multiple-vortex merger, we present in figure~\ref{fig:2vort} the viscous evolution of two Gaussian vortices. Notice the tilt of the vortices with respect to the radial direction at intermediate times, and the filaments in the outer region. The tilt of the vortices plays an important role in merger, and leads, as we will see below, to vortices merging azimuthally, instead of radially.

\subsection{Merger of multiple vortices at moderate Reynolds number}

To show why three or more vortices are different from two, we return to figure~\ref{fig:pointvortsf2}. The annular nature of the exchange band region and the consequent existence of a central passive region means that for $n \ge 3$ vortices tend to align azimuthally rather than radially. The formation of an azimuthally restrained vortical structure is evident in figure \ref{fig:pointvortsf2}(c). Such a structure will be contrasted with an axisymmetric annulus formed in viscous merger. In addition, we point out that although the merger of multiple vortices is fundamentally no different from the merger of two vortices, the well-known sequence of events in two-vortex merger is not followed and is replaced, instead, by the formation of the annulus and a much slower relaxation to a single vortex at the origin.
\begin{figure}
{\includegraphics[trim = 245 0 250 0, clip, scale=0.235]{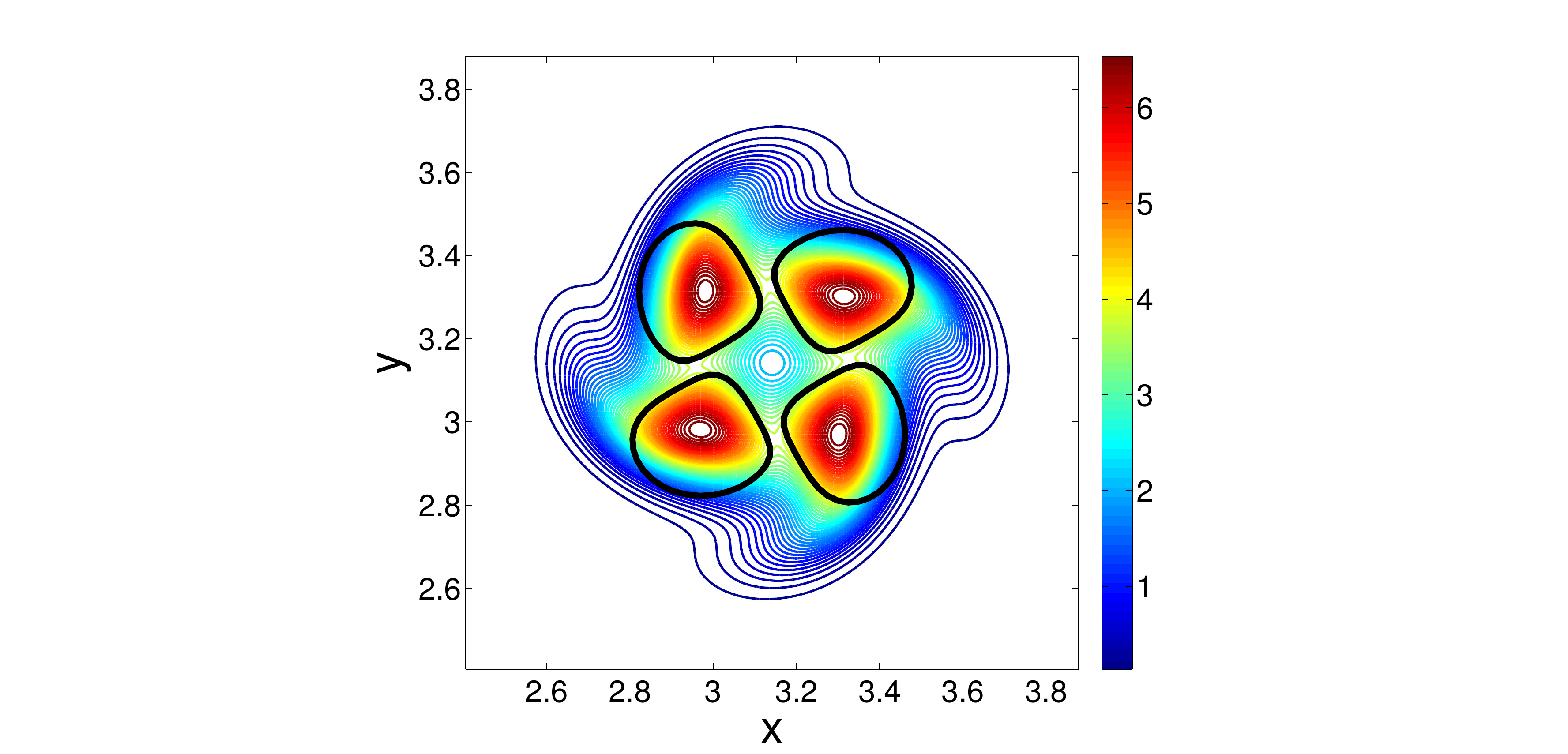}}
{\includegraphics[trim = 245 0 250 0, clip, scale=0.235]{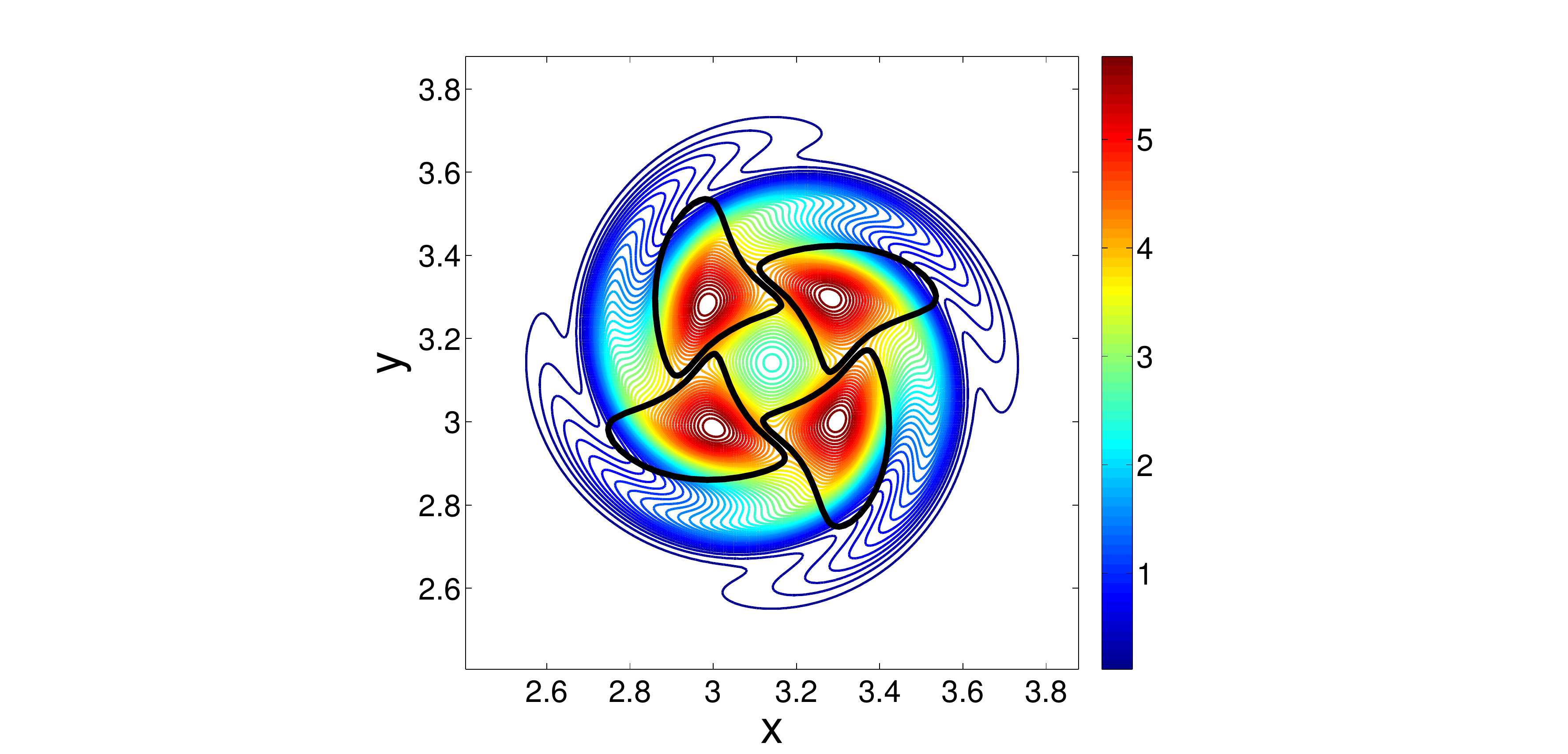}}
{\includegraphics[trim = 245 0 250 0, clip, scale=0.235]{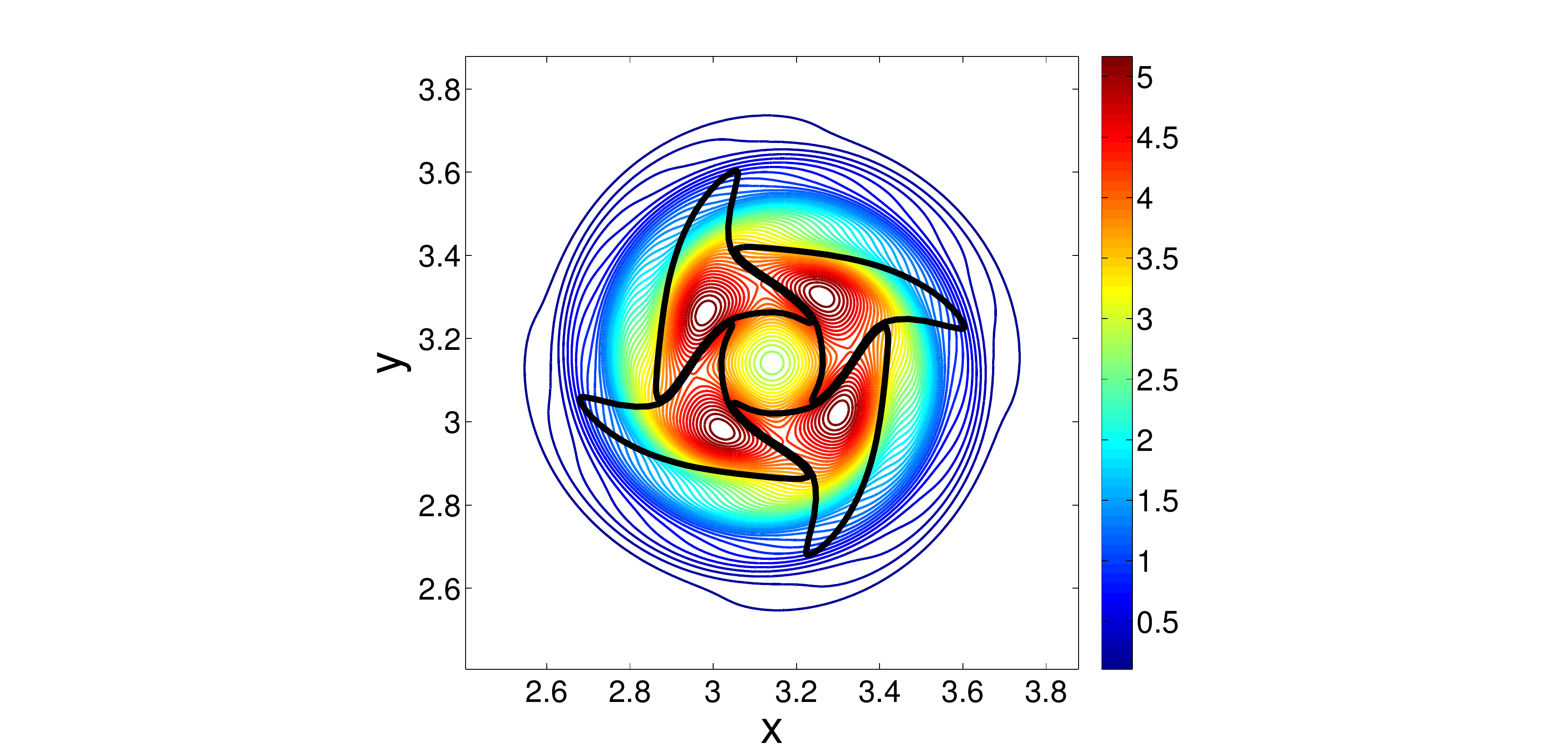}}
\caption{Vorticity contours (in colour online) showing the viscous evolution of four Gaussian vortices with $\left(a/d\right)_{i}=0.3$ and $\Rey_{\Gamma}=4000$. A gentle initial radial alignment is seen. Left: t=0.81, centre: t=2.43 and right: t=4.86. The solid black lines show the inviscid evolution of patches of vortices of uniform vorticity (and the same total strength) at the corresponding times.}
\label{fig:4vort}
\end{figure}

\begin{figure}
{\includegraphics[trim = 245 0 250 0, clip, scale=0.235]{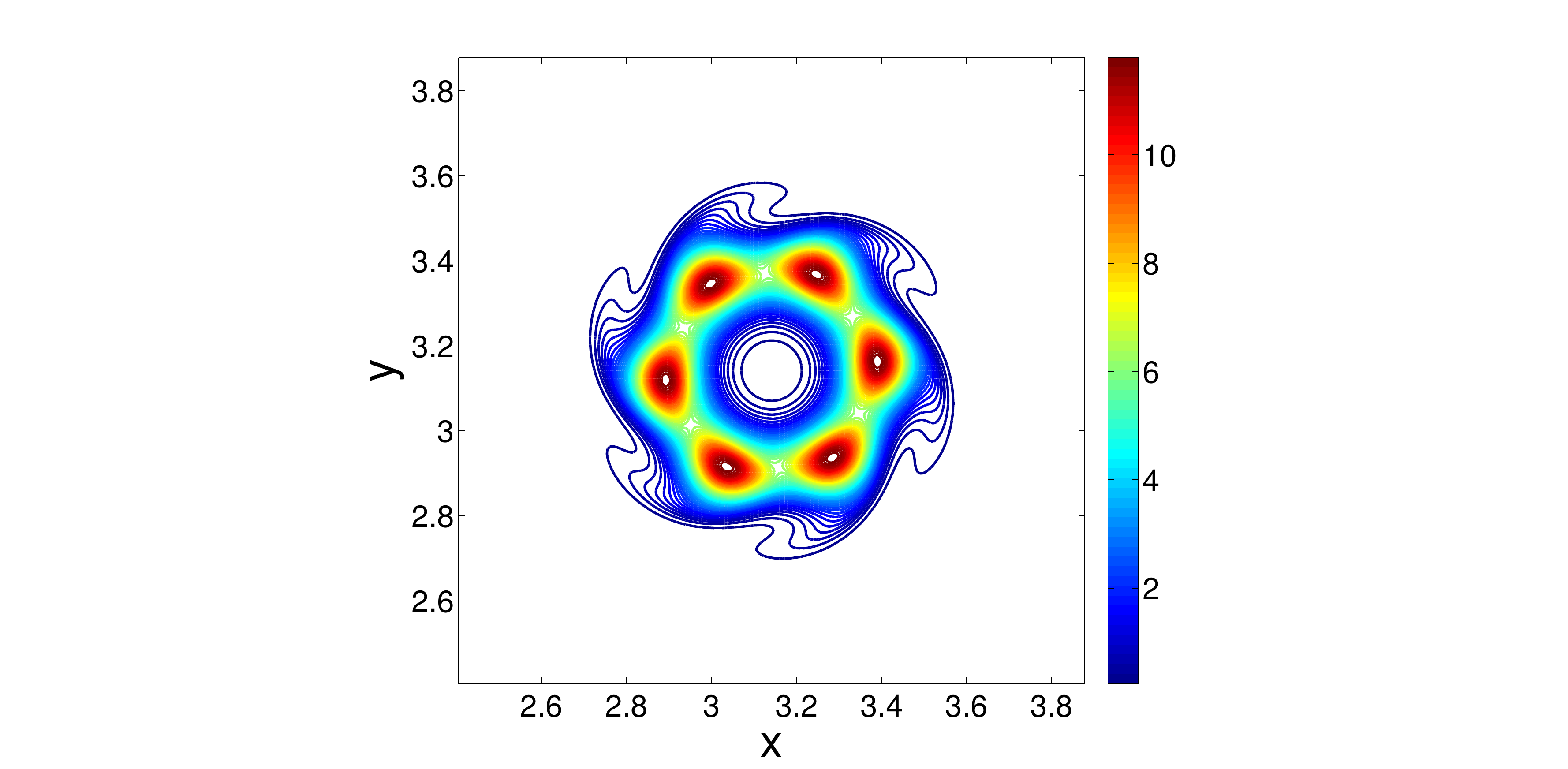}}
{\includegraphics[trim = 245 0 250 0, clip, scale=0.235]{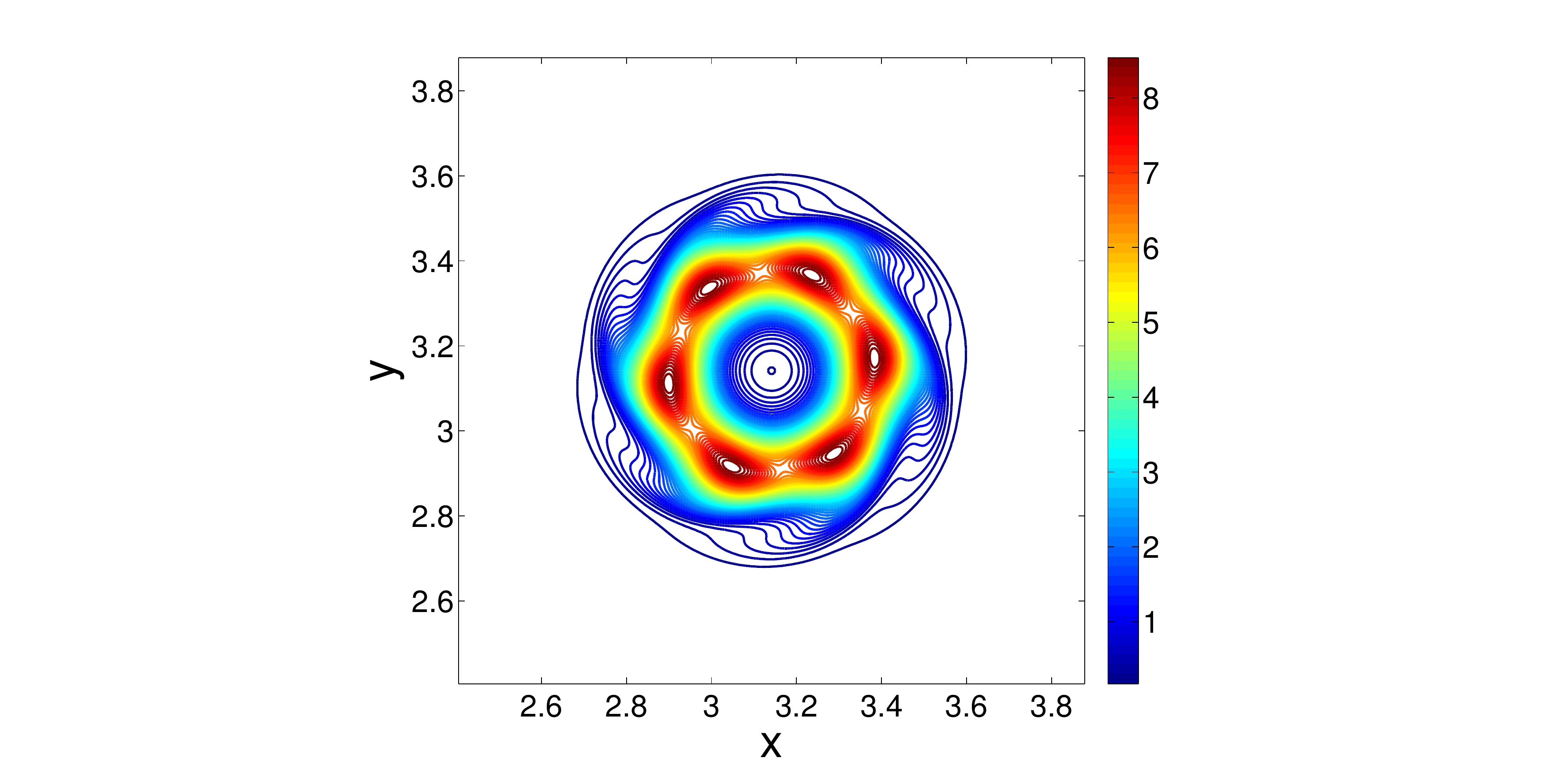}}
{\includegraphics[trim = 245 0 250 0, clip, scale=0.235]{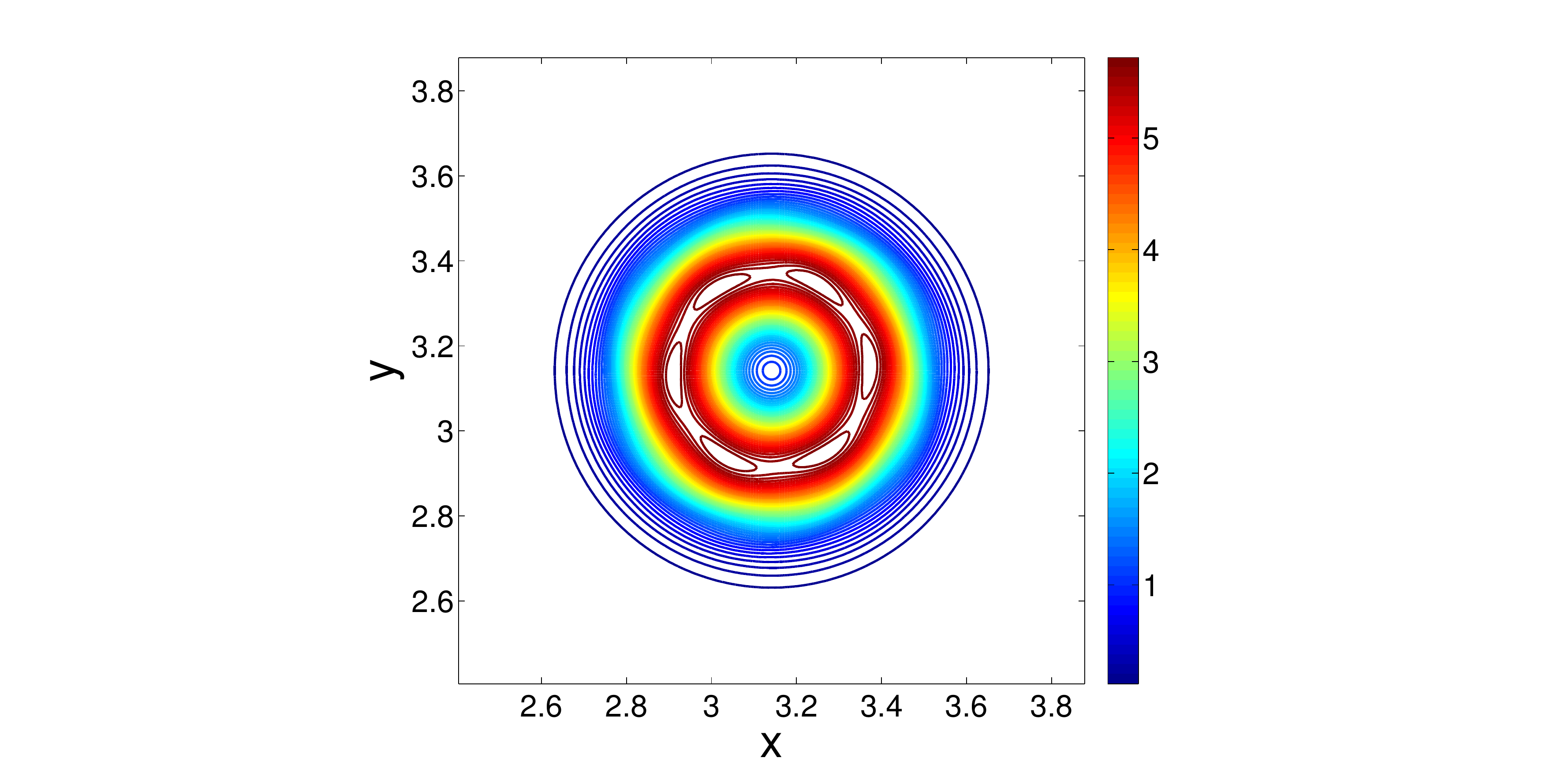}}

{\includegraphics[trim = 245 0 250 0, clip, scale=0.235]{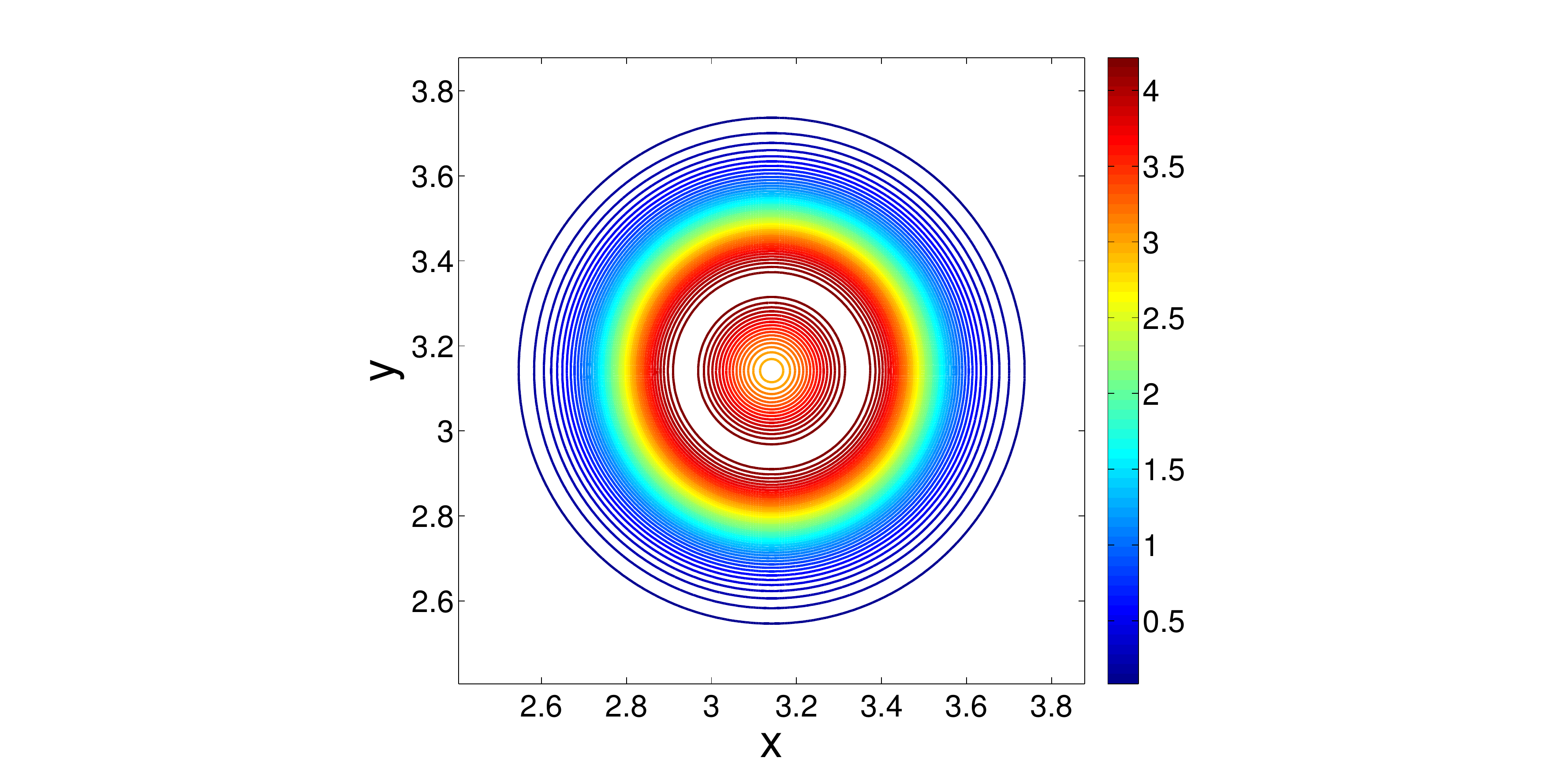}}
{\includegraphics[trim = 245 0 250 0, clip, scale=0.235]{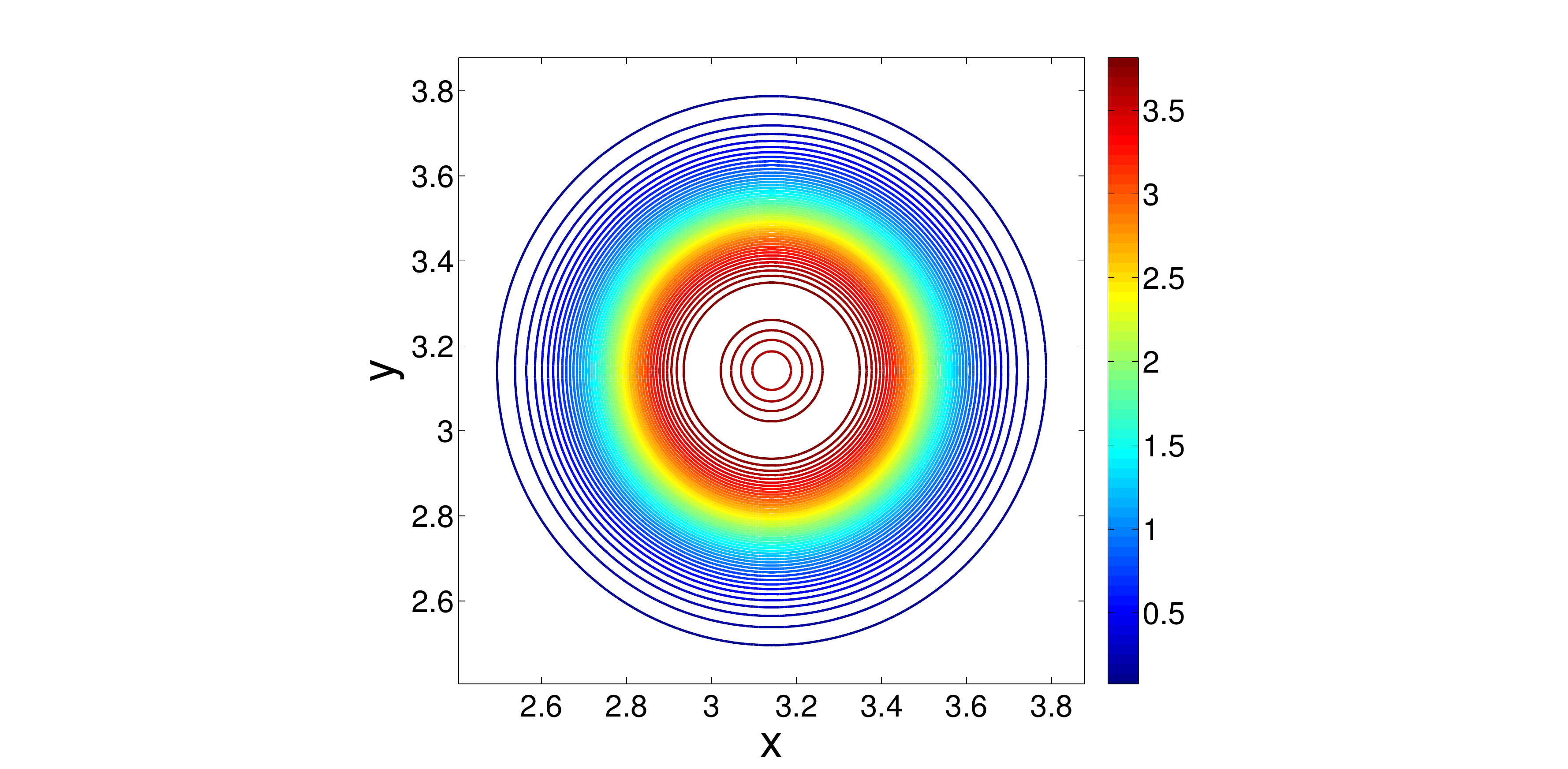}}
{\includegraphics[trim = 245 0 250 0, clip, scale=0.235]{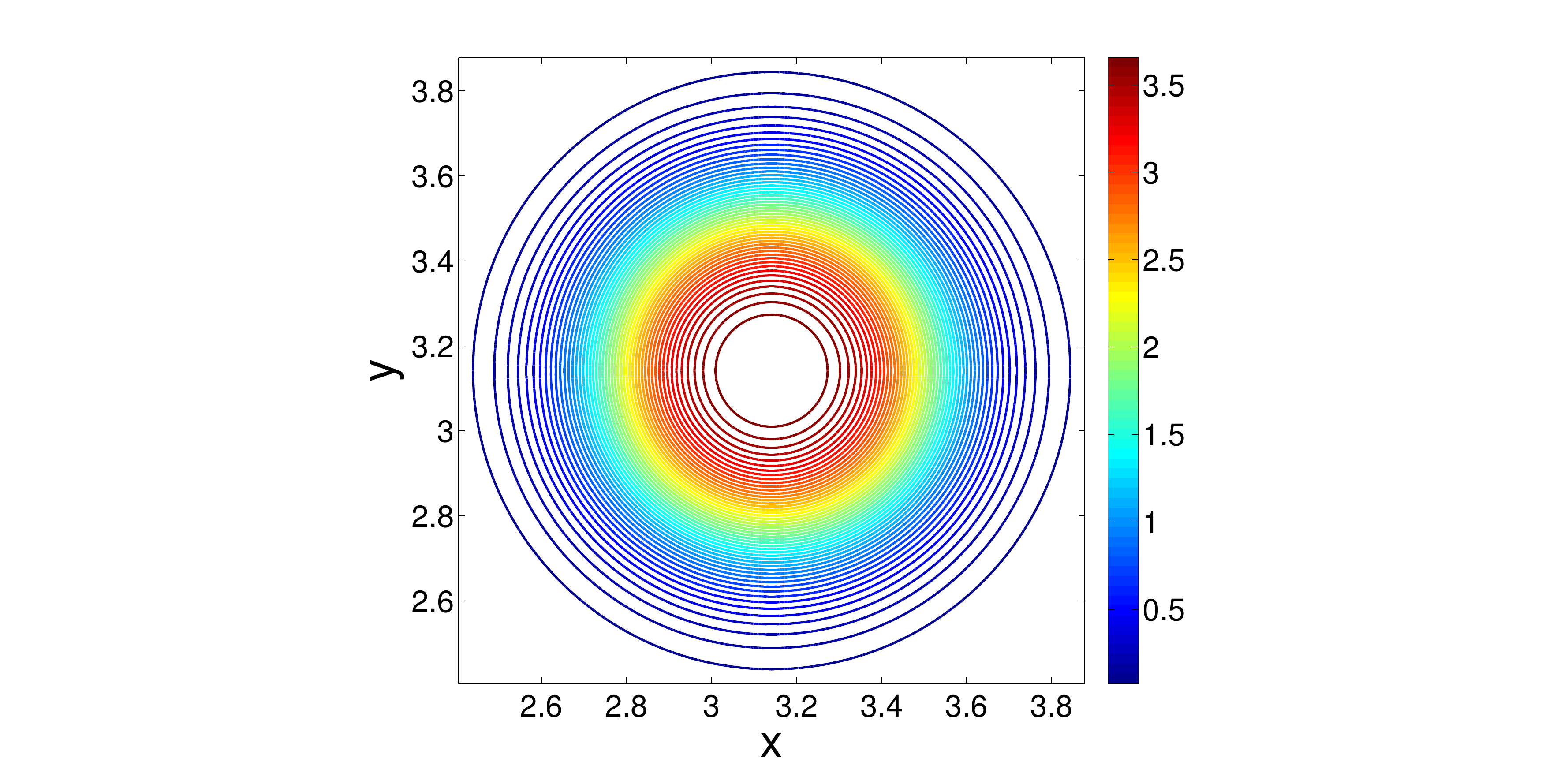}}
\caption{Vorticity contours (in colour online) showing the viscous evolution of 6 Gaussian vortices with $\left(a/d\right)_{i}=0.1$ and $\Rey_{\Gamma}=4000$. Observe initial azimuthal alignment, very clean annulus, and the subsequent inward motion of the annulus to form a single maximum. Times shown (left to right, top row) : t=1, t=1.52, t=2.7; bottom row: t=5.4, t=7.5, t=10.47.}
\label{fig:6vort}
\end{figure}
Four-vortex merger is shown in figure~\ref{fig:4vort}, \ADD{where the initial $(a/d)$ is chosen such that merger will occur. Patch vortices are shown superimposed on the separatrices. The initial convective stage of the dynamics is seen to be similar in the inviscid and viscous cases. However, the realignment and shape are different at later stages, and sharper filaments of exchange-band vorticity are seen. In particular, the tails of the vortex patches in the inviscid case are shrunk and blunted in the viscous case.} The central \ADD{vorticity-free} region is seen to be affecting the dynamics \ADD{in a passive manner}, effecting an azimuthal realignment, and in the viscous case, giving rise at later times to an axisymmetric annular structure (not shown). The case of three vortices is not shown for brevity, but it proceeds in a manner very familiar to the four vortex case.
Figure~\ref{fig:6vort} shows the viscous evolution of six Gaussian vortices. The strain field now is such that even at short times, the vortices align themselves azimuthally, forming an axisymmetric annulus. Note that such axisymmetry is impossible in inviscid evolution. There is then a slow reduction in radius of the annulus on a diffusive time scale, to finally form a single Gaussian vortex. The formation of filaments is minimal, unlike in the merger of two vortices.
To summarize, merger is now an azimuthal phenomenon, like people on a dance floor joining hands to form a circle, and does not result immediately in the formation of a vorticity maximum at the centre. The latter is attained on a diffusive time scale.

\begin{figure}
\centering
\includegraphics[trim = 0 0 0 0, clip, scale=0.25]{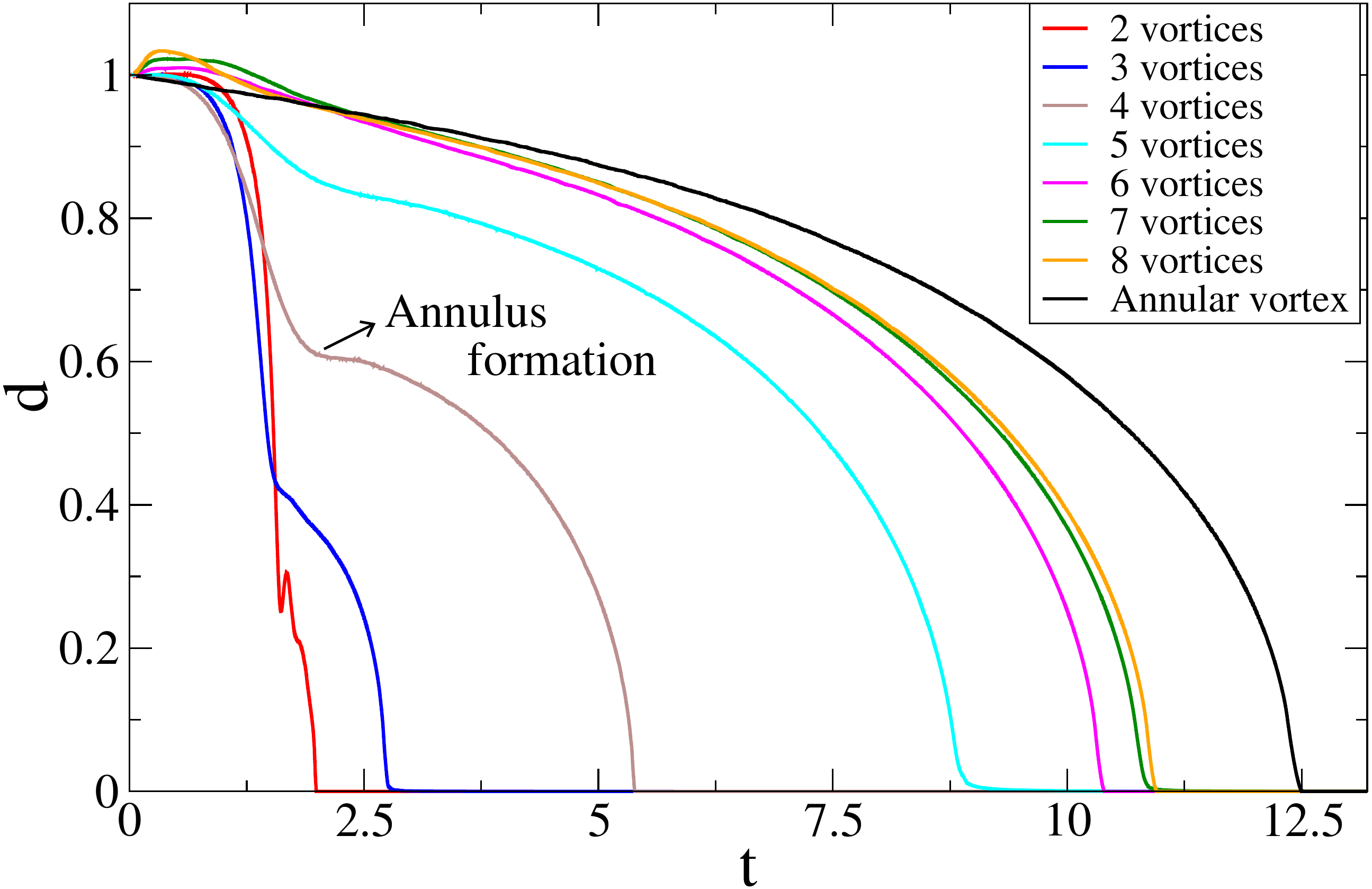}
\includegraphics[trim = 0 0 0 0, clip, scale=0.25]{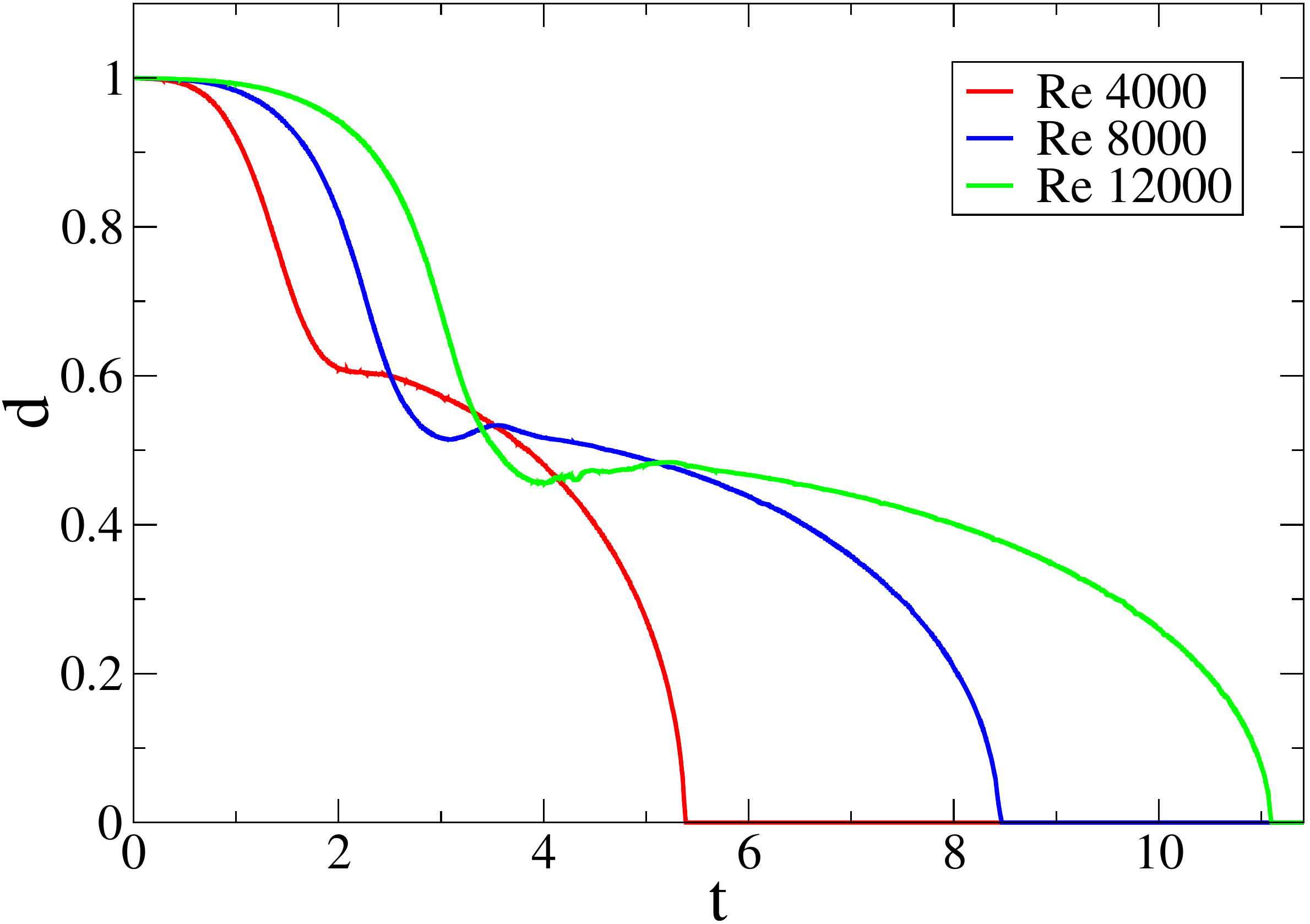}
\caption{(Left) Non-dimensional separation distance $d$ (scaled by the initial separation) as a function of the non-dimensional time $t$ at $\Rey_{\Gamma}=4000$. Starting with six or more vortices is qualitatively the same as starting with a pure annular vortex. Here $\left(a/d\right)_{i}=0.1$. (Right) Separation distance $d$ versus time $t$ for four vortices, for different $\Rey_{\Gamma}$.}
\label{fig:34allre}
\end{figure}
Figure~\ref{fig:34allre}(a) shows the distance $d$ of the maximum vorticity point of each vortex from the centre, as a function of time $t$, for different number of vortices, at $\Rey_{\Gamma}=4000$. The length and time are non-dimensionalised respectively by the initial separation $d_i$ and the time scale $T_{\Gamma}$. A bend in the curve, where the coming together of the vortices slows down, is evident at $n=3$ and $n=4$. This bend corresponds to the formation of the annulus. Consequently, the convective stage is shorter in these cases. The convective stage completely disappears for six or more vortices. We have instead a new annular stage which is diffusive. Thus, the annular stage dominates the merger process as we increase the number of vortices, and spans the entire merger process for six and more vortices. This is a slow stage and effectively delays the merger. It can also be seen that as we increase the number of vortices, the process ceases to depend on the number of vortices. As we increase the number of vortices, there is a reduction in filament formation, which helps to slow down the approach of the vortices towards each other. In the simulations above, we have kept the initial $\Gamma$ the same as we increase the number of vortices, and one may argue that in order to make a fair comparison between two different systems, it is the initial energy which has to be kept the same and not the initial circulation. We repeated all these simulations keeping the initial energy the same as we changed $n$, and we found that nothing changes qualitatively. The annular stage once again increasingly dominates the merger process as $n$ increases, and effectively delays the merger.  We note that holding $\left(a/d\right)$ constant while increasing $n$ makes the $n$-vortex system look more and more like an annular vortex with perturbations. This will naturally bias the system towards forming an annulus, as is evident from figure~\ref{fig:34allre}(a).

Next we study the effect of changing $\Rey_{\Gamma}$ on four vortex merger, this is shown in figure~\ref{fig:34allre}(b). As expected, the duration of the first diffusive stage increases as the Reynolds increases. This is because the critical $\left(a/d\right)$ is reached at a later time for lower relative viscosity. As is also true of the the two-vortex case, the convective stage is almost independent of the Reynolds number. Most important, the final (annular) stage of merger is strongly dependent on the Reynolds number, with the duration scaling approximately linearly with the Reynolds number, suggesting that the annular stage is a diffusive stage.

\begin{figure}
{\includegraphics[trim = 245 0 250 0, clip, scale=0.235]{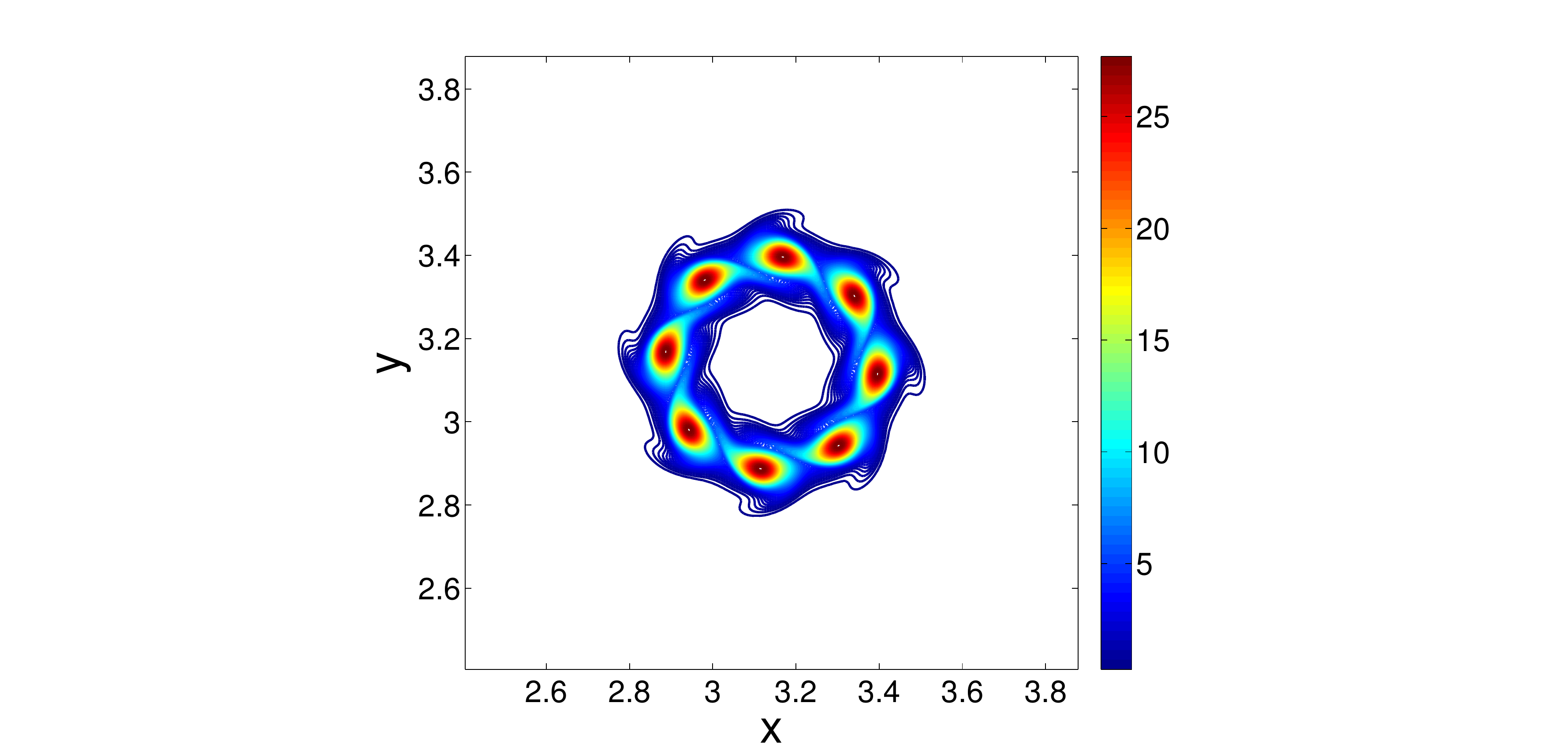}}
{\includegraphics[trim = 245 0 250 0, clip, scale=0.235]{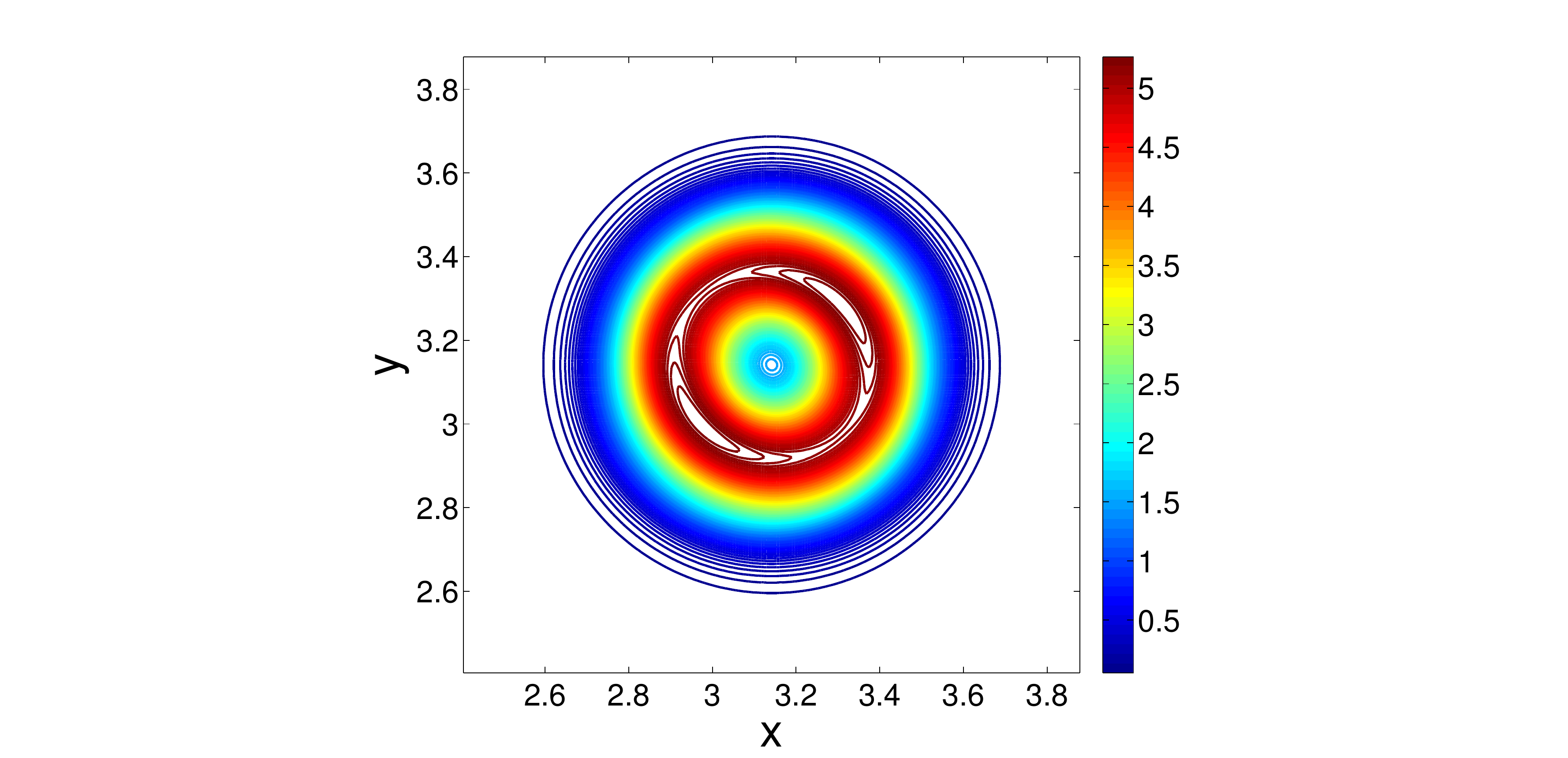}}
{\includegraphics[trim = 245 0 250 0, clip, scale=0.235]{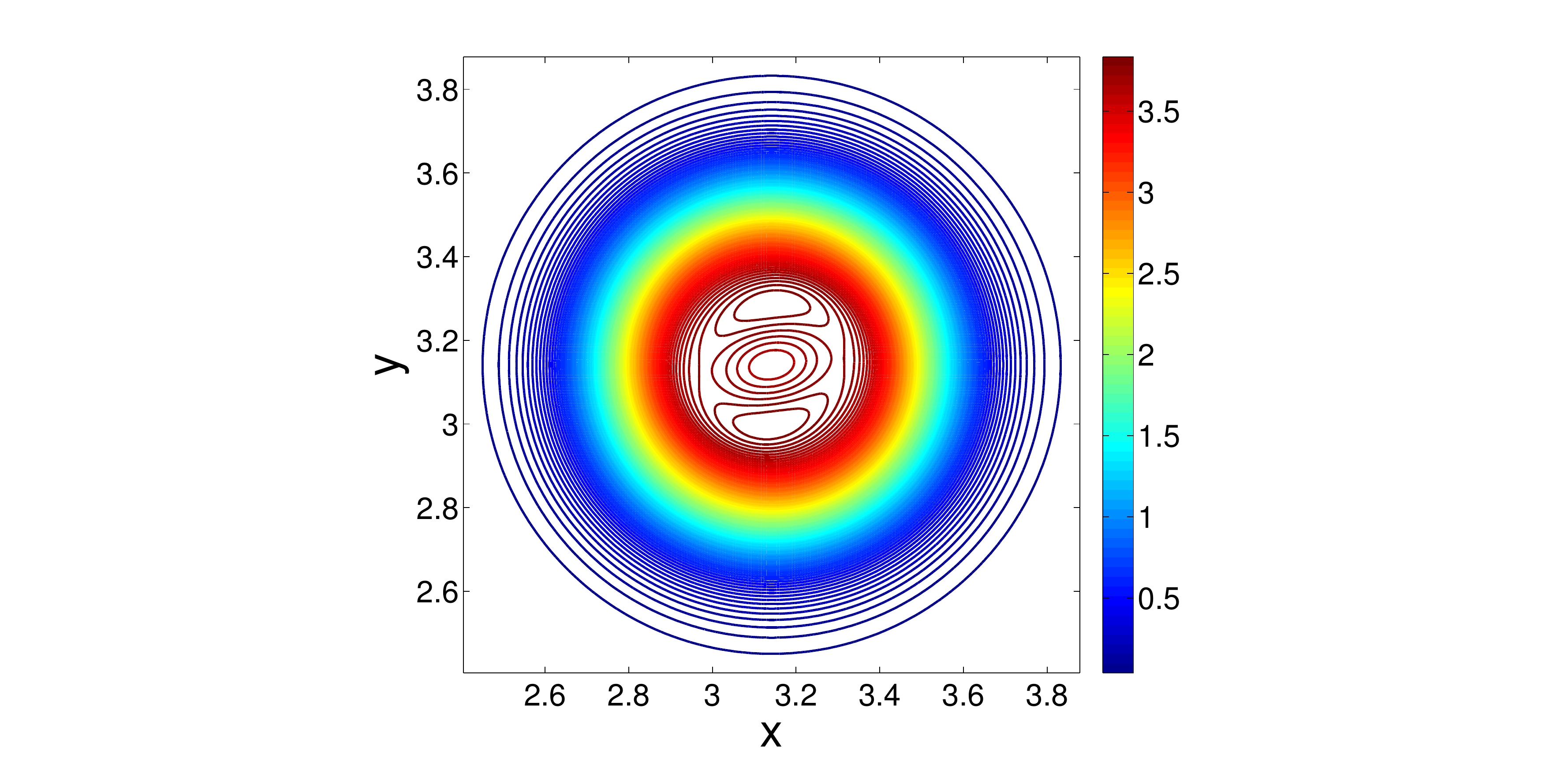}}
\caption{Vorticity contours (in colour online) showing the viscous evolution of 8 Gaussian vortices with $\left(a/d\right)_{i}=0.1$ and $\Rey_{\Gamma}=12000$. 
(Left) t=0.1685, (centre) t=10.0447, (right) t=25.0276. }
\label{fig:8vortre12k}
\end{figure}

\begin{figure}
\includegraphics[width=\linewidth]{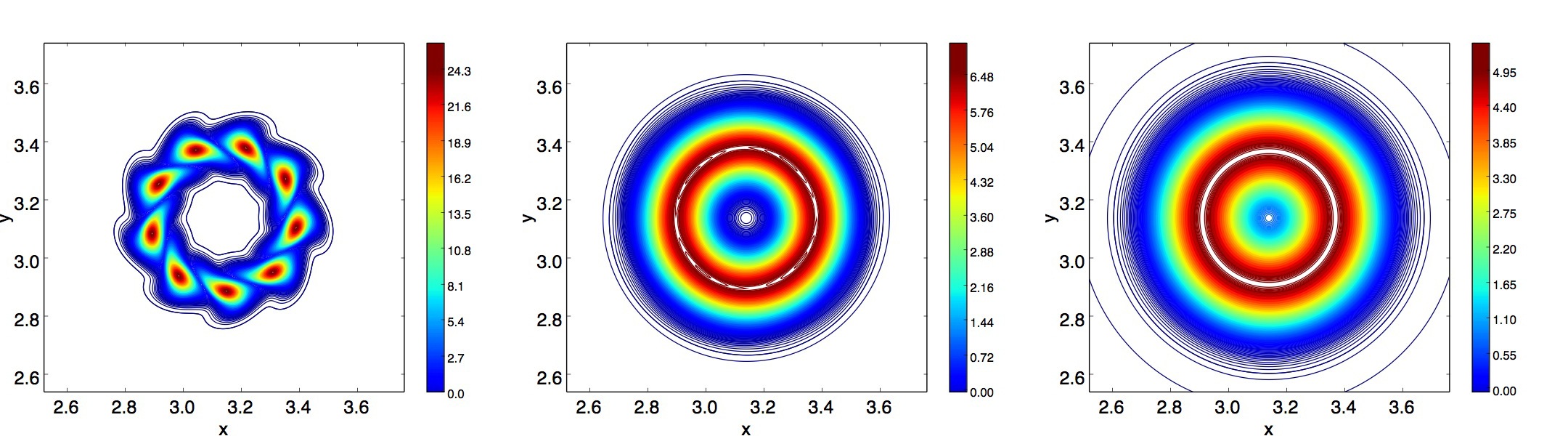}
\caption{\label{fig:9vortre12k}Vorticity contours (in colour online) showing the viscous evolution of nine Gaussian vortices with $\left(a/d\right)_{i}=0.1$ and $\Rey_{\Gamma}=12000$ at times (in units of $T_{\Gamma}$): $t=0.09$ (left panel) , $t=4.59$ (middle panel) , and $t=9.09$ (right panel).}
%\label{fig:9vortre12k}
\end{figure}

We now look at what happens with eight and nine vortices. We choose $\Rey_{\Gamma}=12000$ and $\left(a/d\right)_{i}=0.1$. The results are shown in figure~\ref{fig:8vortre12k} for $8$ vortices and figure~\ref{fig:9vortre12k} for $9$ vortices. In both cases, the merger proceeds as discussed before in that the annulus forms and migrates inwards, but in the eight vortex case, a small departure from axisymmetry, in the form of a tripolar vortex, is visible at late times. In the next section, we will contrast these results against eight and nine vortex dynamics at very high Reynolds number.

\subsection{High Reynolds number multiple vortex simulations}

At higher Reynolds number, nonlinearity dictates the dynamics and the viscosity is not effective in making the vortices diffuse and merge. To capture the small scale structures that appear at high Reynolds number, we discretised the computational domain with  $4096^2$ collocation grid points.  We find that the dynamics strongly depends on whether the number of vortices are even or odd (a feature not observed in linear stability).  

\subsubsection{Six-vortex merger,  $\Rey_{\Gamma}=1.5\times 10^5$}
\begin{figure*}
\includegraphics[width=\linewidth]{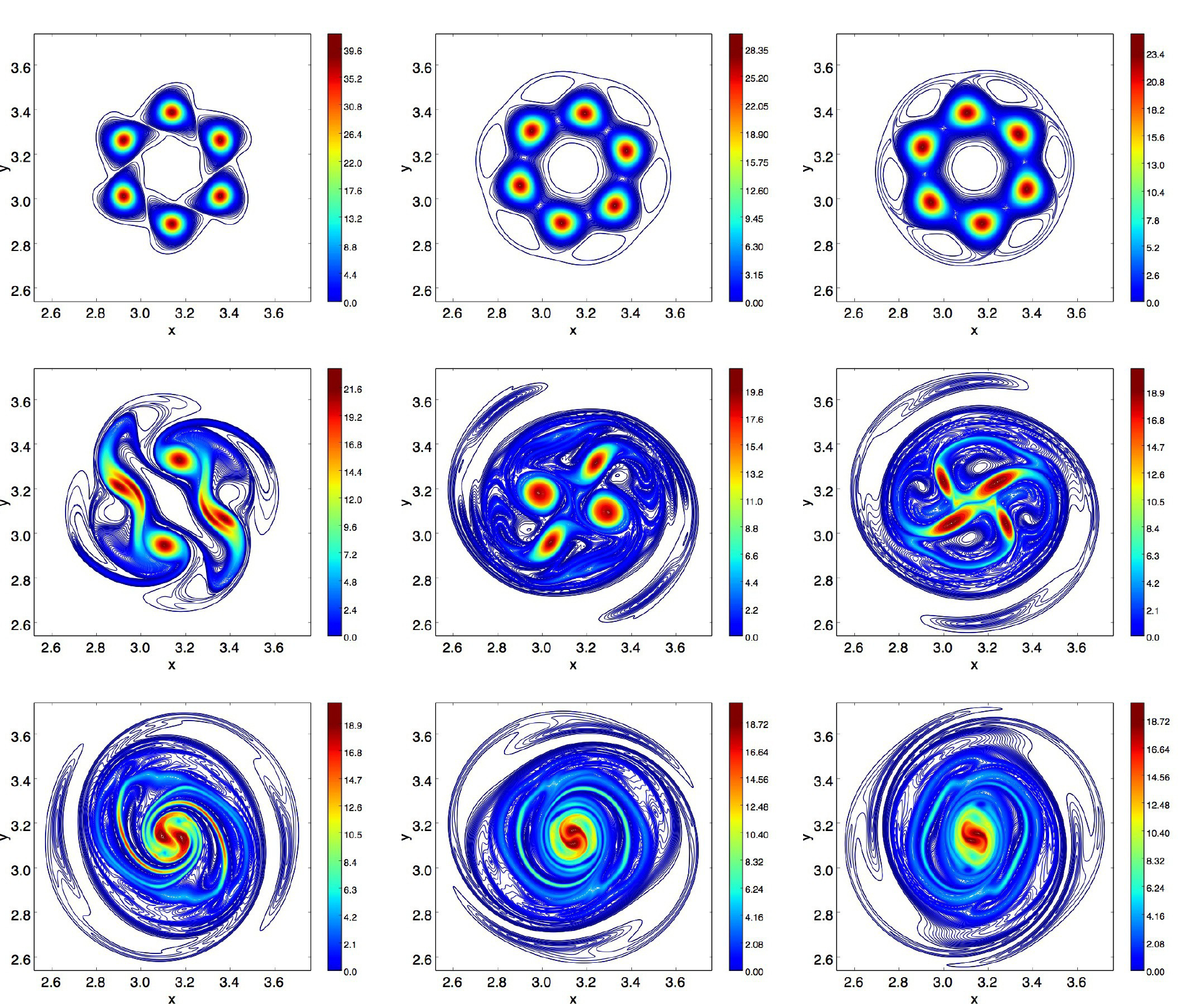}
\caption{\label{6vorm} Contours of vorticity (in colour online) showing the time evolution of a $6$ vortex configuration at $\Rey_{\Gamma}=1.5\cdot10^5$. The different snapshots are at times (in units of $T_\Gamma \approx 0.74$) $t=0.09$(top-left), $t=6.42$ (top-centre), $t=10.64$(top-right), $t=12.75$ (middle-left), $t=14.86$ (middle-centre), $t=15.71$ (middle-right), $t=16.13$ (bottom-left), and $t=16.55$ (bottom-centre), and $t=16.89$ (bottom-right).}
\end{figure*}

We begin with an even number (six) of vortices arranged, as before, on the vertices of a regular polygon. The plot in figure~\ref{6vorm} shows the time evolution of the vorticity contours during the merger process. Initially, the dynamics proceeds in a manner similar to that at moderate Reynolds number. The vortices start to rotate in the counter-clockwise direction and try to form an annular like structure. The annulus, however, is never formed. The six vortices undergo an asymmetric merger to first form four vortices, which are not identical. The two located diametrically opposite to each other, which came from the merger events are larger. This is followed by another pair of mergers to give two vortices, with a lot of fine filamentary structure in the neighbourhood. \ADD{Thus it is seen that reducing the viscosity reduces the propensity to form an annulus, and other, less regular dynamics intervenes.}

\subsubsection{Eight-vortex merger,  $\Rey_{\Gamma}=2\times 10^5$}
\begin{figure*}
\includegraphics[width=\linewidth]{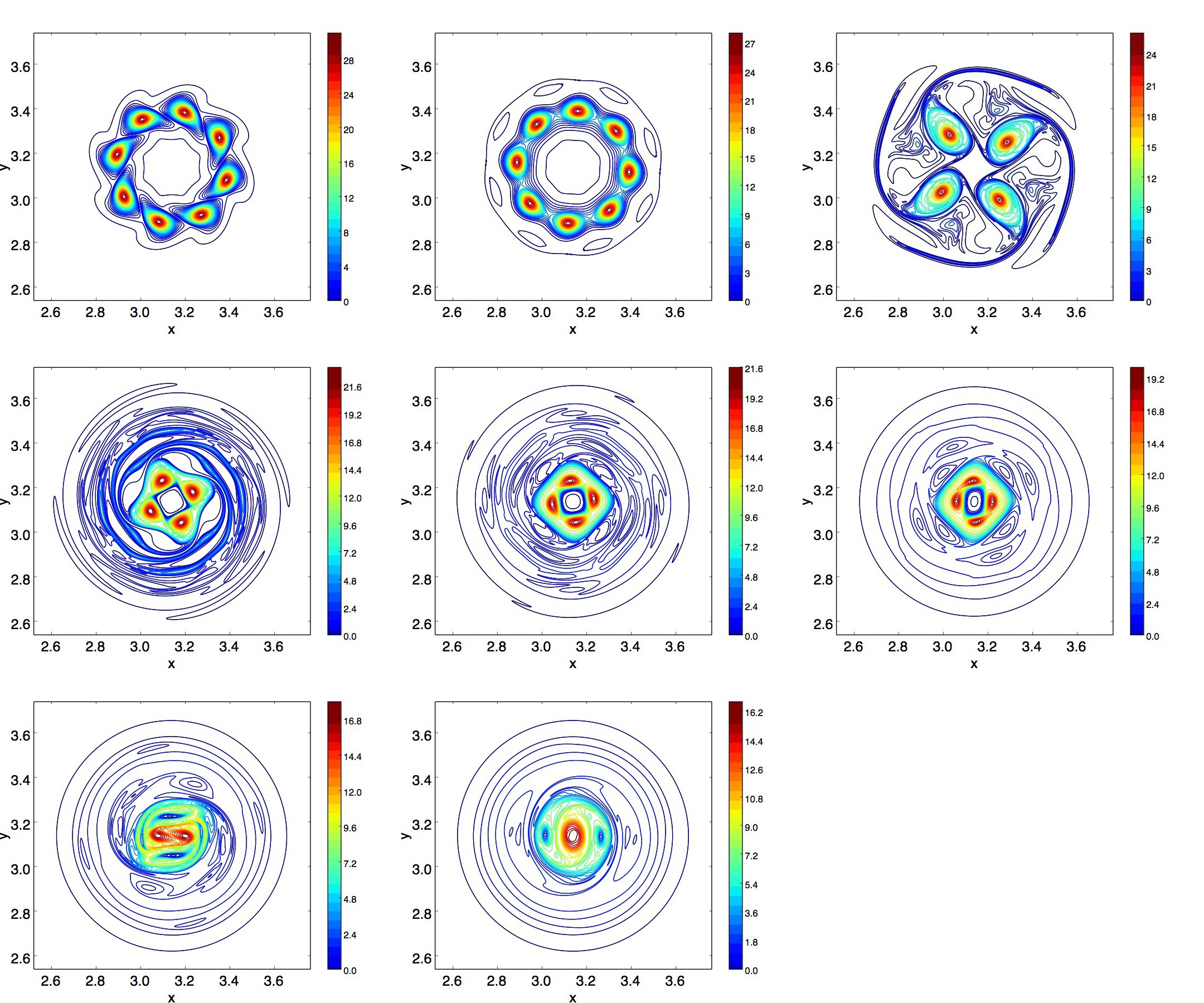}
\caption{\label{8vorm} Contours of vorticity (in colour online) showing the time evolution of an $8$ vortex configuration at $\Rey_{\Gamma}=2\cdot10^5$. The different snapshots are at times (in units of $T_\Gamma$) $t=0.09$(top-left), $t=2.31$ (top-centre), $t=4.52$(top-right), $t=6.74$ (middle-left), $t=8.96$ (middle-centre), $t=11.17$ (middle-right), $t=13.39$ (bottom-left), and $t=15.60$  (bottom-right).}
\end{figure*}

We now examine a multiple of four (i.e. eight) vortices on the vertices of a regular polygon, in figure~\ref{8vorm}. Now the first merger event is perfectly symmetric, with four pairwise mergers yielding four identical vortices. The eight-vortices pair-wise merge event to form four vortices that are surrounded by thin filamentary structures. The perfect symmetry of this case is preserved through most of the later dynamics, whereas this is not possible in the six and nine vortex cases. Here the four vortex structure continues to decay until, at a later stage, these vortices further merge to form a tripolar vortex.  

\subsubsection{Nine-vortex merger, $\Rey_{\Gamma}=2\times 10^5$}
\begin{figure*}
\includegraphics[width=\linewidth]{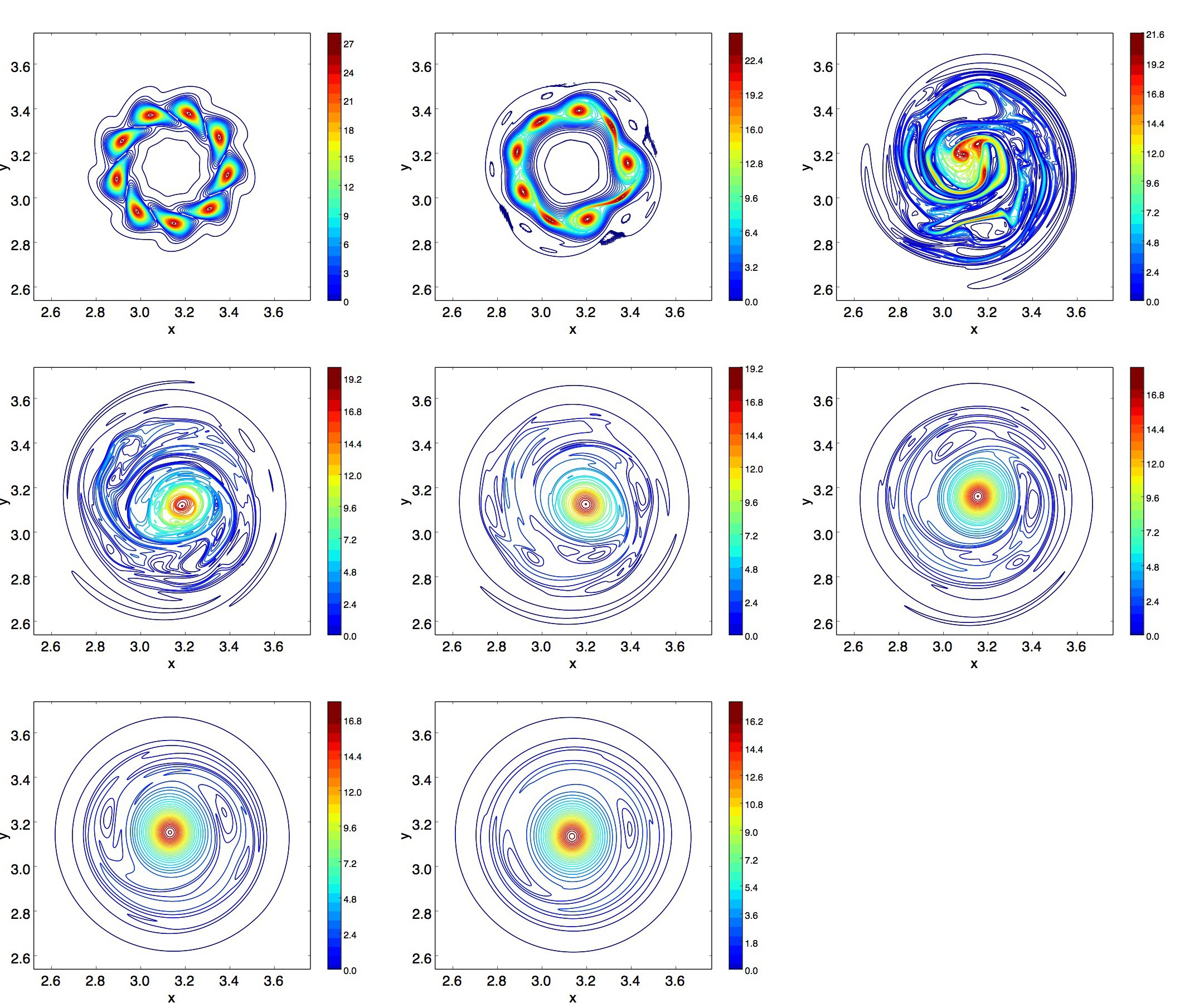}
\caption{\label{9vorm} Contours of vorticity (in colour online) showing the time evolution of a nine vortex configuration at $\Rey_\Gamma=2\cdot10^5$. The different snapshots are at times (in units of $T_\Gamma$) $t=0.09$(top-left), $t=2.34$ (top-centre), $t=4.59$(top-right), $t=6.85$ (middle-left), $t=9.09$ (middle-centre), $t=11.35$ (middle-right), $t=13.59$ (bottom-left), and $t=15.85$ (bottom-right).}
\end{figure*}

The dynamics of an odd number (nine) of vortices arranged initially on the vertices of a regular polygon is very different from its even-number counterpart, \ADD{since simple symmetric mergers can no longer take place}. The plot in figure~\ref{9vorm} shows the time evolution of the vorticity contours during the merger process. Initially the vortices start to rotate in the counter-clockwise direction and form a annular structure. However, because of an additional unpaired vortex this structure very quickly turns unstable and chaotic. We point out that, because of rapid mixing, the total time of merger is about half what is expected. At late times a single vortex at the centre of the domain is formed.

\begin{figure}
\begin{center}
\includegraphics[width=0.45\linewidth]{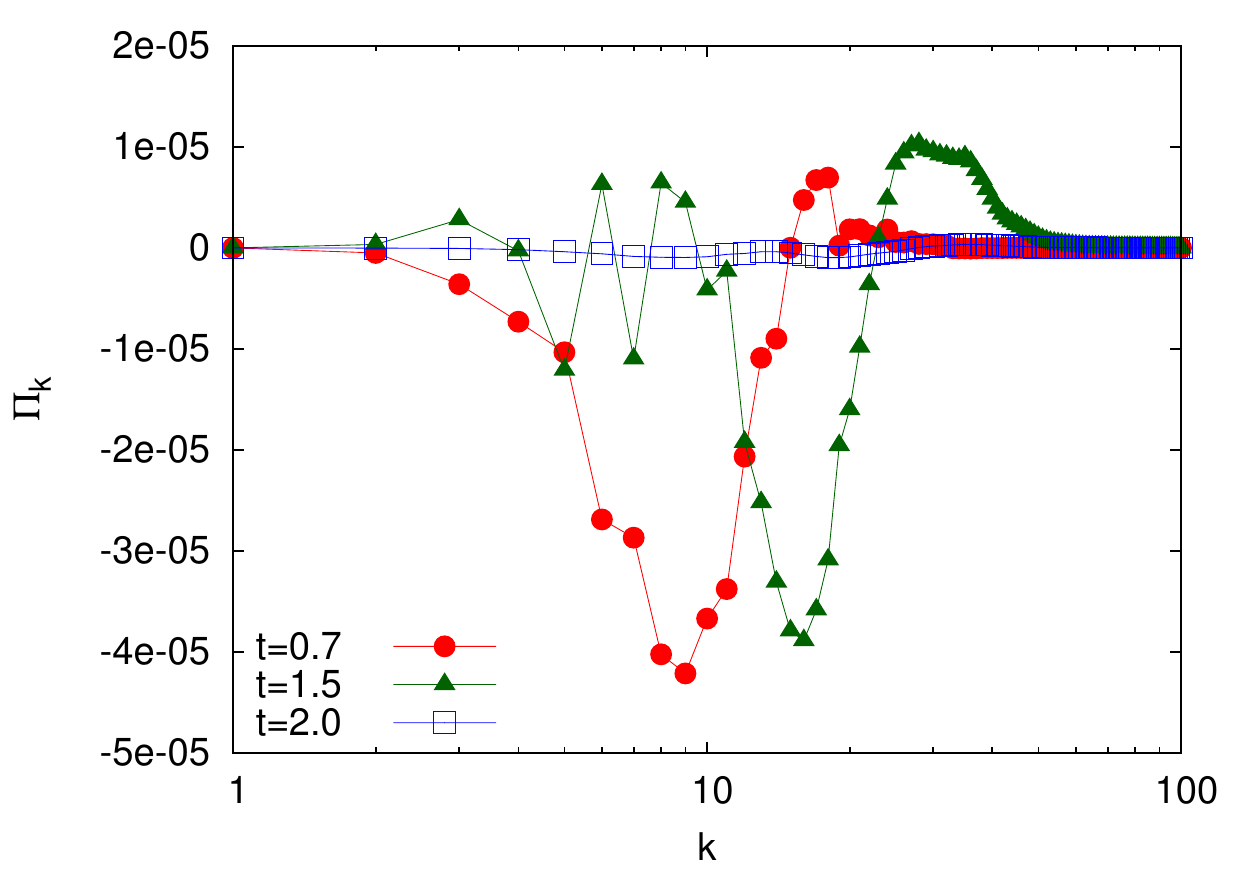}
\includegraphics[width=0.45\linewidth]{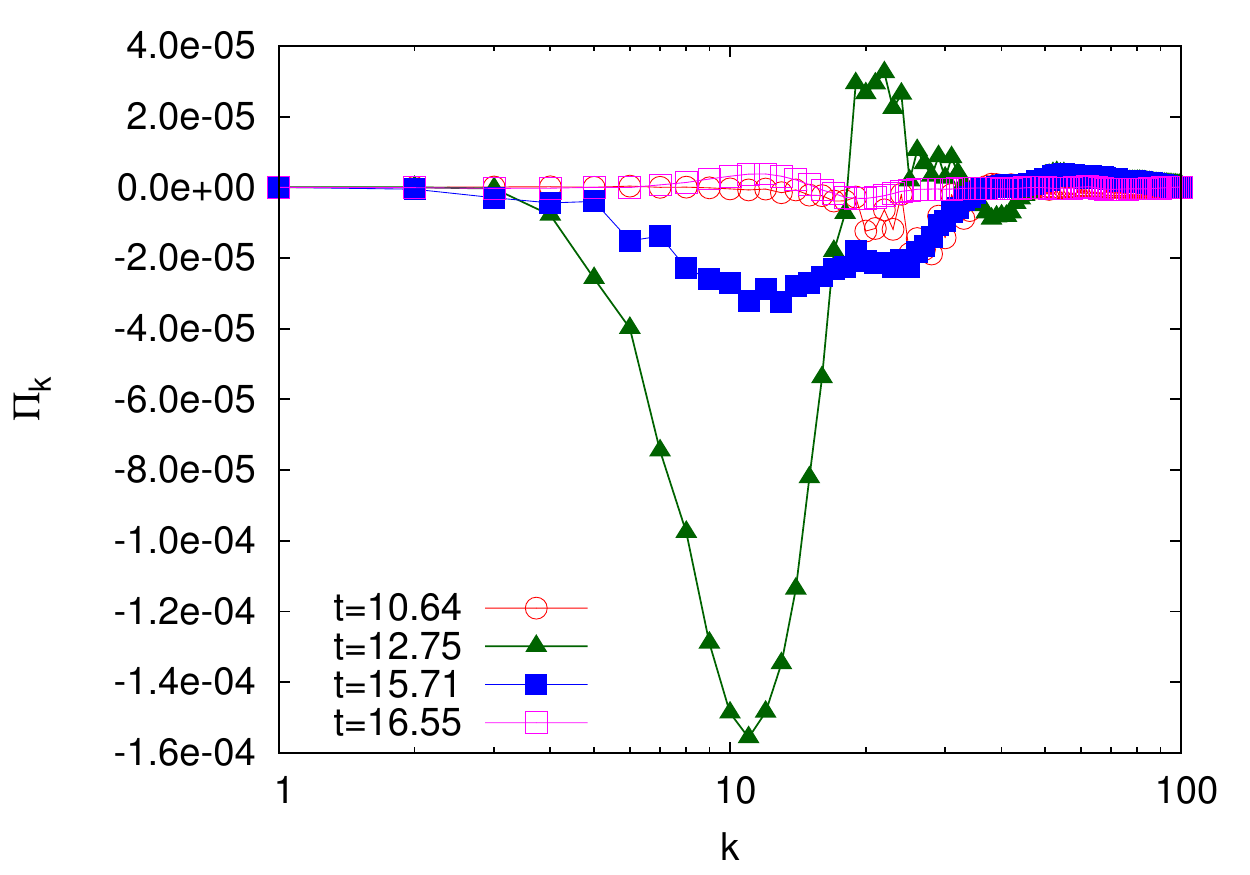}
\caption{Energy flux $\Pi(k)$ versus $k$ for the two vortex configuration  (left) and  the six vortex configuration (right).}
\label{fig:energyflux}
\end{center}
\end{figure}

\subsection{Energy flux in vortex merger}
\ADD{We now discuss the merger of vortices by looking at the energy flux. The plot in Figure~$4$ and $10$ shows the evolution of two ($\Rey_\Gamma=4\cdot 10^3$) and six ($\Rey_\Gamma=1.5\cdot 10^5$) vortex configurations.  The corresponding plot for the energy flux $\Pi_k= \sum_{k'=k}^{N/2} \hat{\bm v}_{-{\bm k'}} \widehat{[{\bm v} \cdot \nabla {\bm v}]}_{\bm k'}$ is shown in Fig. \ref{fig:energyflux}. We find that energy flux remains negative, indicating transfer of energy from small to large scales, for the duration of the merger events. At intermediate times, as the filamentary structures start to develop around the vortices, we also observe a positive enstrophy cascade at high wavenumber. }

\section{The generalized Lamb-Oseen vortex}

We have seen that at moderate Reynolds numbers an annular vortex is formed. This vortex can be written down analytically as an exact solution of the axisymmetric Navier-Stokes equations. The analytical solution for the annular vortical structure from the viscous evolution of a circular vortex sheet was derived by \cite{Kossin_Schubert2003}. We provide an alternative derivation in the appendix. We call this solution the generalized Lamb-Oseen (GLO) vortex with vorticity concentrated within an annulus. Such a vortex results from the evolution of a cylindrical vortex sheet at a radius $a$ with a circulation of $\Gamma_{o}$, given at $t=0$ by
\begin{align}
\label{eq:incondition}
\omega\left(r,0\right)& = \frac{\Gamma_{o}}{2\pi r}\delta\left(r-a\right),
\end{align}
by the axisymmetric vorticity equation
\begin{align}
\label{eq:vorticity}
\frac{\partial\omega}{\partial t}=\frac{\nu}{r}\frac{\partial}{\partial r}\left(r\frac{\partial \omega}{\partial r}\right).
\end{align}
The boundary conditions are symmetry at $r=0$ and a vorticity-free far-field. The generalized Lamb-Oseen vortex (see Appendix) is then obtained as
\begin{align}
\label{finalvort}
\omega=\frac{\Gamma_{o}}{4\pi \nu t}e^{-\left(\frac{a^2+r^2}{4\nu t}\right)}I_{0}\Big\{\frac{ar}{2\nu t}\Big\},
\end{align}
where $I_0$ is the modified Bessel function of the first kind and zeroth order. Setting $a$ to zero reduces the above to a standard Lamb-Oseen vortex. It is useful to note the non-dimensional parameter $\nu t/a^2=b^2 $, which controls the profile of the GLO. Further, it is interesting to note that differentiating Eq.~\ref{finalvort} with respect to time will yield additional solutions which satisfy the Navier-Stokes, including some that have net zero circulation.

\begin{figure}
\centering
\includegraphics[scale=0.35]{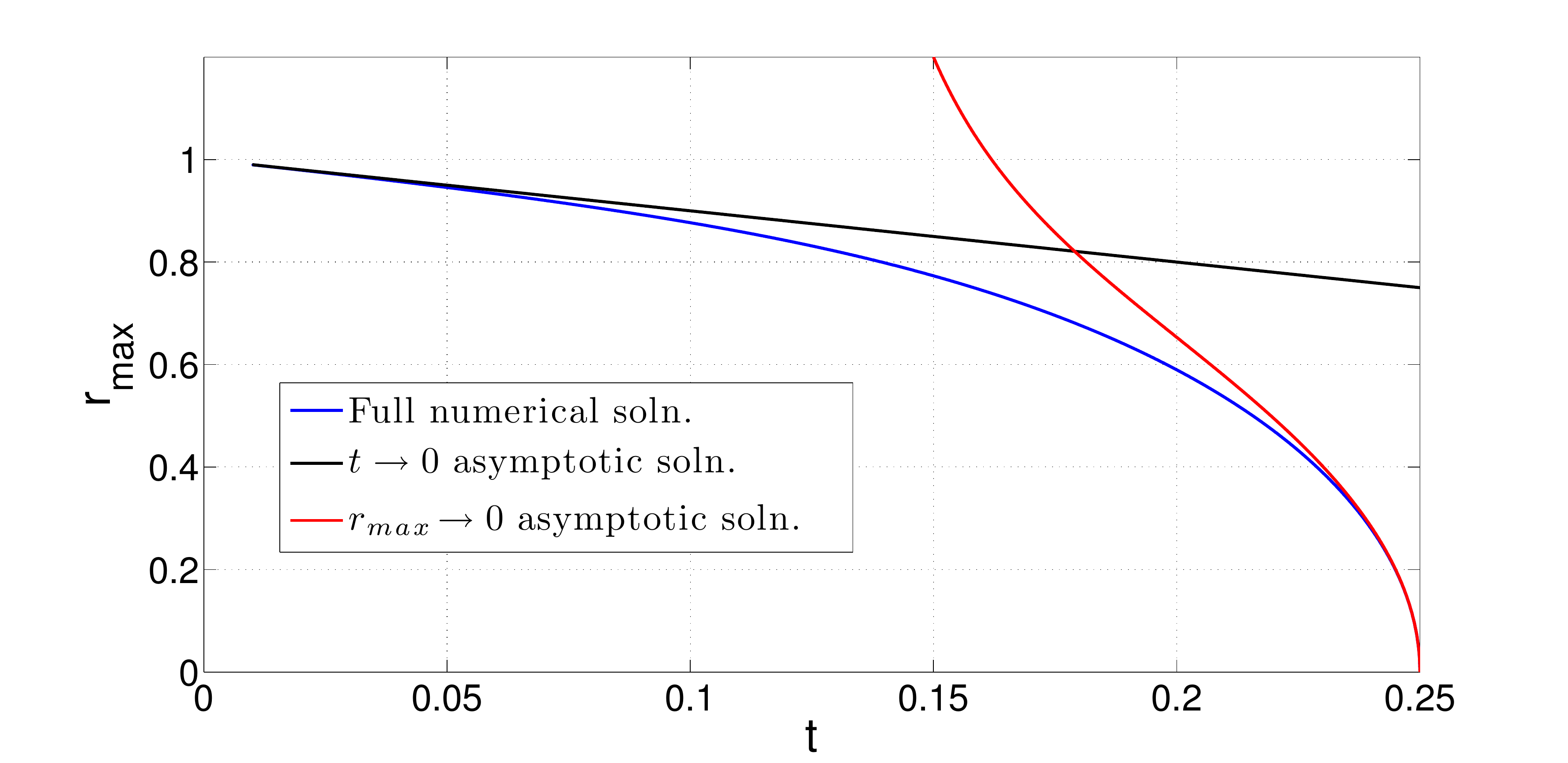}
\caption{Location of vorticity maximum $r_{max}$. Time is non-dimensionalised by $a^2/\nu$.}
\label{fig:vortmaxrad}
\end{figure}

At $t \to 0$, we may write the vorticity and, with some algebra, the location $r_{max}$ of the vorticity maximum, at leading order in $t$ as
\begin{equation}
\omega = \frac{\Gamma_{o}}{4 \pi^{3/2} a\sqrt{\nu t}}\exp\left[-\frac{(r-a)^2}{4 \nu t}\right], \quad
r_{max}=a-\frac{1}{a}\nu t, \hspace{0.1in} t\to0.
\label{eq:vortmaxtimezero}
\end{equation}
Thus the vorticity maximum will move inwards purely due to diffusion, initially in a ballistic manner. Next, we may estimate the behaviour as $r_{max} \to 0$ by writing, at leading order,
\begin{align}
r_{max}=\frac{4\sqrt{2} \nu t}{a}\sqrt{\frac{a^2-4\nu t}{8 \nu t- a^2}}.
\label{eq:vortmaxradzero}
\end{align}
The time taken for the vorticity maximum to reach the centre is thus $a^2/(4\nu)$. Thus, for $\nu t/a^2=b^2 < 0.25$, the GLO is an annular vortex sheet with the vorticity maxima away from the axis. For such a profile, a smaller $b^2$ corresponds to a thinner annular sheet. However, for $b^2 \ge 0.25$, we obtain a vortex with a vorticity maximum at the axis.
figure~\ref{fig:vortmaxrad} contains plots for the evolution of $r_{max}$ and its early and late behaviour, given by equations \ref{eq:vortmaxtimezero} and \ref{eq:vortmaxradzero}. 

In our simulations of merger above, in the cases when an annular vortex was arrived at, we found that the annular structure was, to a good approximation, the GLO vortex. In the following section we begin with a GLO vortex and perform numerical simulations with the method described above to check the analytical solution against the dynamics obtained numerically, and to study the stability of the annular structure. 

\subsection{Direct Numerical Simulations of the Generalised Lamb Oseen vortex}

The linear and nonlinear stability of annular vortices has been well studied under the inviscid framework. \citet{Dritschel1986} considered an annular Rankine vortex with a fixed outer radius and variable inner radius, and showed it to linearly unstable when the inner radius exceeds half the outer radius. The cases which are linearly stable go through a steepening and sharpening of disturbance waves when nonlinearity is included, but reach a stable state nevertheless. The nonlinear instability leads to a breaking up into multiple vortices, an example of a five vortex steady state emerging from an instability of the annular vortex is demonstrated. 

The existing literature contains studies that report annular vortices becoming unstable to several modes of imposed perturbations. \cite{Hendricks_Schubert2010} report that initially thin annular vortices breakup in an $m=5$ mode, forming mesovortices, and initially thick annular vortices break up in an $m=3$ mode. \cite{Menelaou2013a} study the stability of an annular vortex under imposed $m=2$ perturbations and find that wave-interactions may damp the growth of the perturbations. \cite{Menelaou2013b} find the existence of $m=3$ and $m=4$ modes in an actual Hurricane.

\begin{figure}
\begin{center}
\includegraphics[scale=0.3]{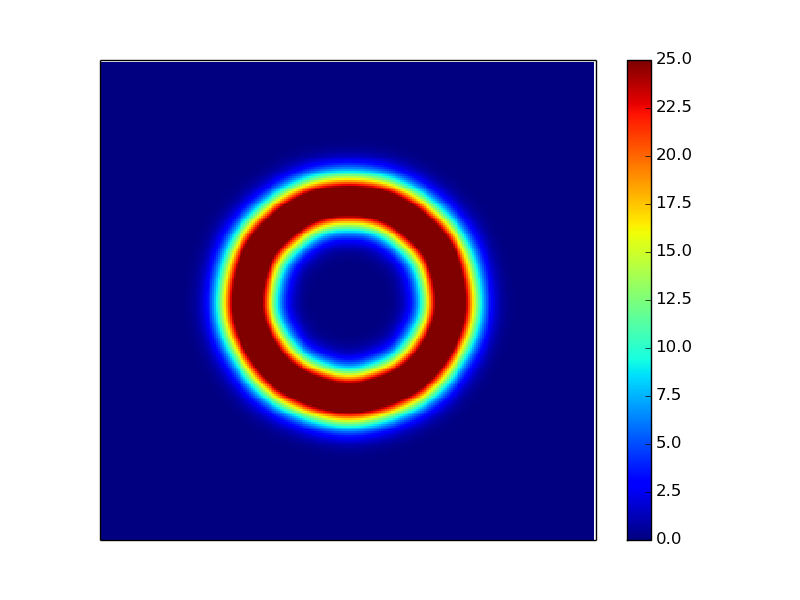}
\end{center}
\caption{\label{fig_init}  Initial GLO vortex (in colour online). The dark (red) regions indicate regions of intense vorticity whereas the light (blue) regions indicate regions of low vorticity. In all figures showing vorticity contours in the rest of the paper we have used the same colourbar.}
\end{figure}

We conduct a series of direct numerical simulations of the two-dimensional Navier-Stokes (NS) equations to study the evolution of the GLO vortex, and its stability to small perturbations of odd and even azimuthal wavenumber $m$. In what follows, the Reynolds number $\Rey_{\Gamma}\equiv \Gamma/\nu$ was varied by reducing the viscosity $\nu$ and keeping the circulation $\Gamma$ fixed at a value of $6$. To generate the  GLO, we set the parameter $\nu_t\equiv \nu t=0.001$, the distance from origin $a=0.25$, and fix $(\pi,\pi)$ as the centre. The vertical $y$ and horizontal $x$  distances are measured from this centre. We prescribe different values for the initial perturbation $\epsilon$ (see the remainder of section 4 and section 5).  Radial perturbations are added by modifying the radius of a given vorticity to $r\to r+\epsilon \sin[m\theta]$, where $m$ defines the azimuthal wavenumber of the perturbation, and $\theta$ is the azimuthal angle. The initial vortex structure for a GLO with a small perturbation is shown in figure~\ref{fig_init}. 
As it is unrealistic to scan the whole $(m,\Rey_\Gamma)$ phase-space with numerical simulations, we choose a few representative Reynolds numbers for four modes of disturbance, the even modes $m=4$ and $m=8$ and the odd modes $m=5$ and $m= 9$. 

\begin{figure*}
\includegraphics[scale=0.17]{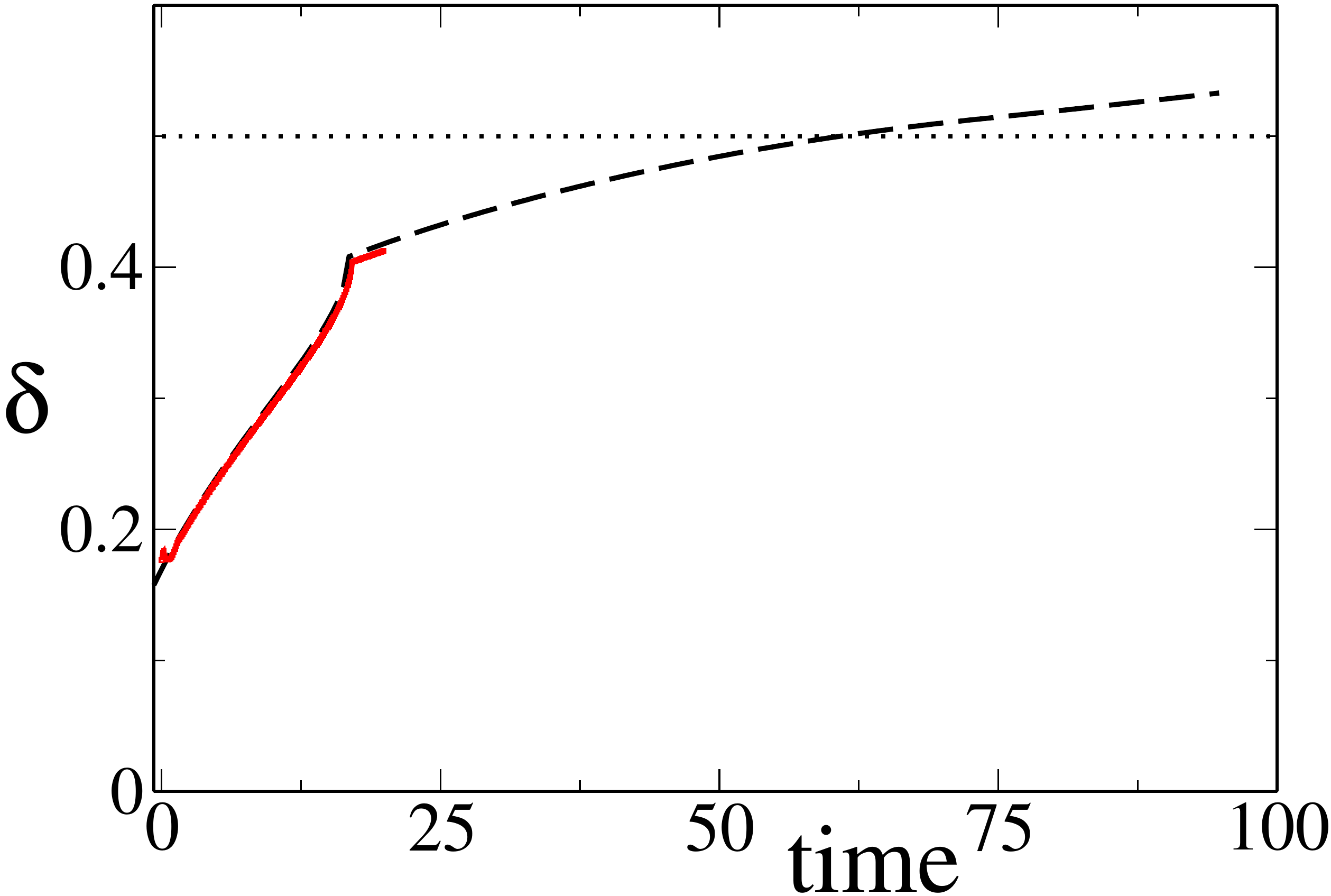}
\includegraphics[scale=0.17]{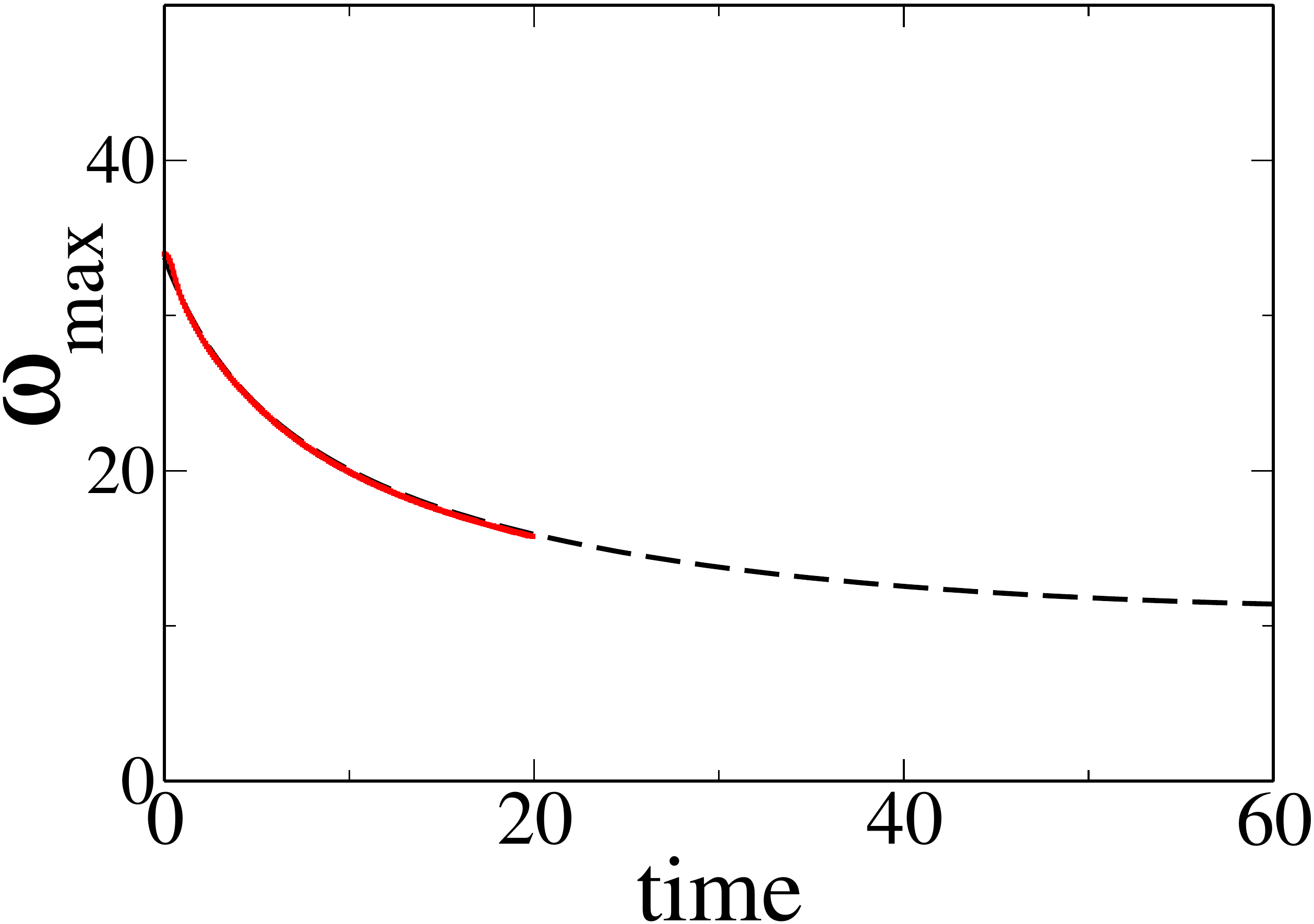}
\includegraphics[scale=0.17]{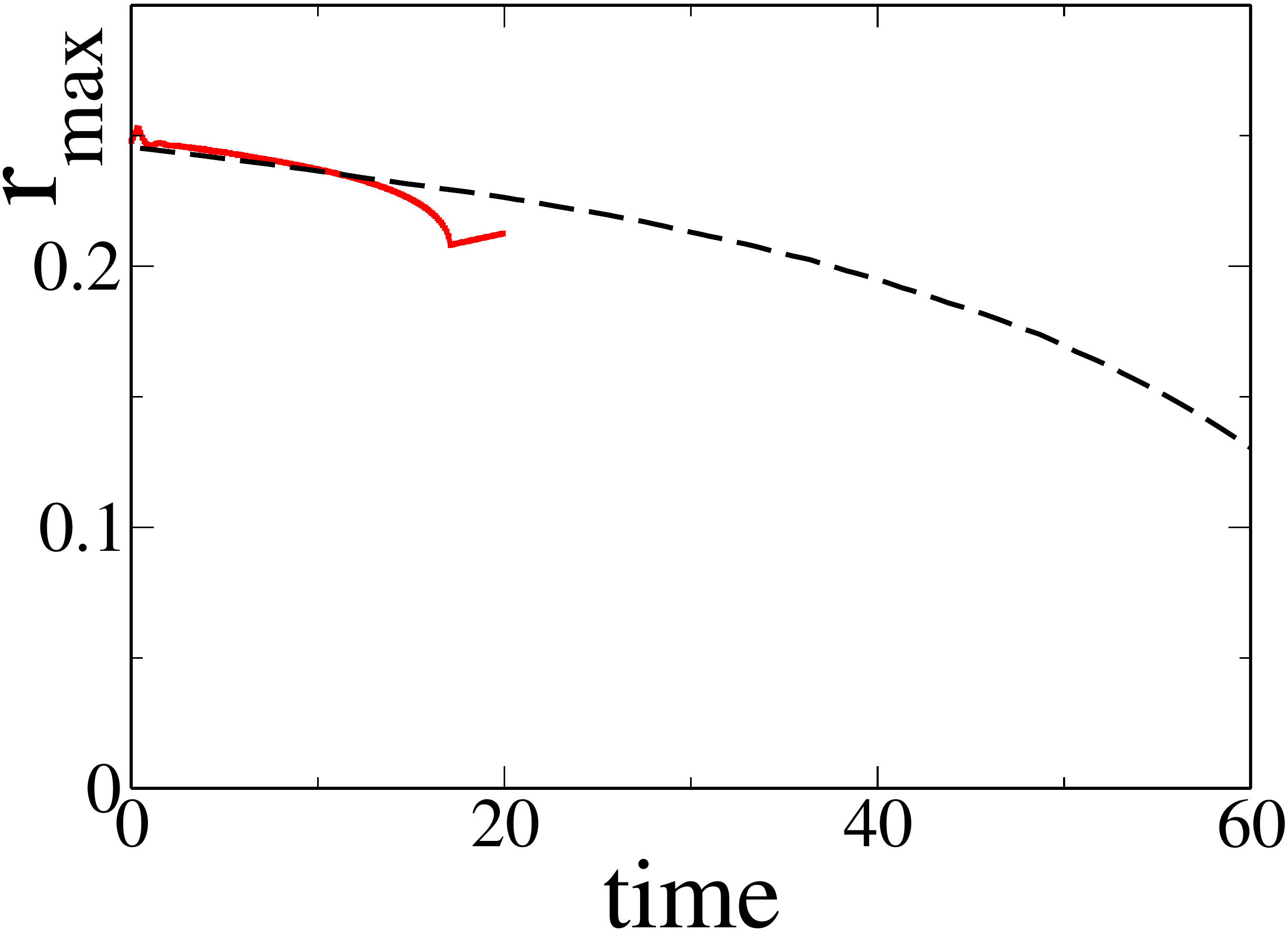}
\caption{\label{amax} Comparison of analytical evolution of the generalized Lamb-Oseen vortex with direct numerical simulations for $\Rey=30000$. (Left) Width of the annular region, (centre) Maximum vorticity, (right) Location of the vorticity maximum. The (black) dashed lines are from the analytical expression whereas the solid (red) lines are from the simulations.}
\end{figure*}

To test the consistency of our numerics, we compare, in figure \ref{amax}, numerical simulations of the dynamics of an annular vortex against the predicted evolution of the GLO from equation (\ref{finalvort}) both at a Reynolds number of $30000$. We expect the annulus to be stable at this Reynolds number. These simulations were done with $\epsilon=0.001$, and broadband noise. In this case, the perturbations do not grow visibly so we expect a good comparison with the theoretical GLO. The initial vortex in both is the same, and the width $\delta$ of the annular vortex in both cases is measured between the radial locations at which the vorticity drops to $1/e$ of its maximum value. Note that there are no fitting parameters. The widening with time of the annular region, as well as corresponding reduction in maximum vorticity of the numerical solution agrees extremely well with the theoretical GLO. In the radial location of the maximum vorticity, the agreement between theory and simulations is very good at early times, but a departure from GLO behaviour is seen at later times. The reason for this small departure could be small growths in the initial perturbations.

\begin{figure*}
\includegraphics[scale=0.35]{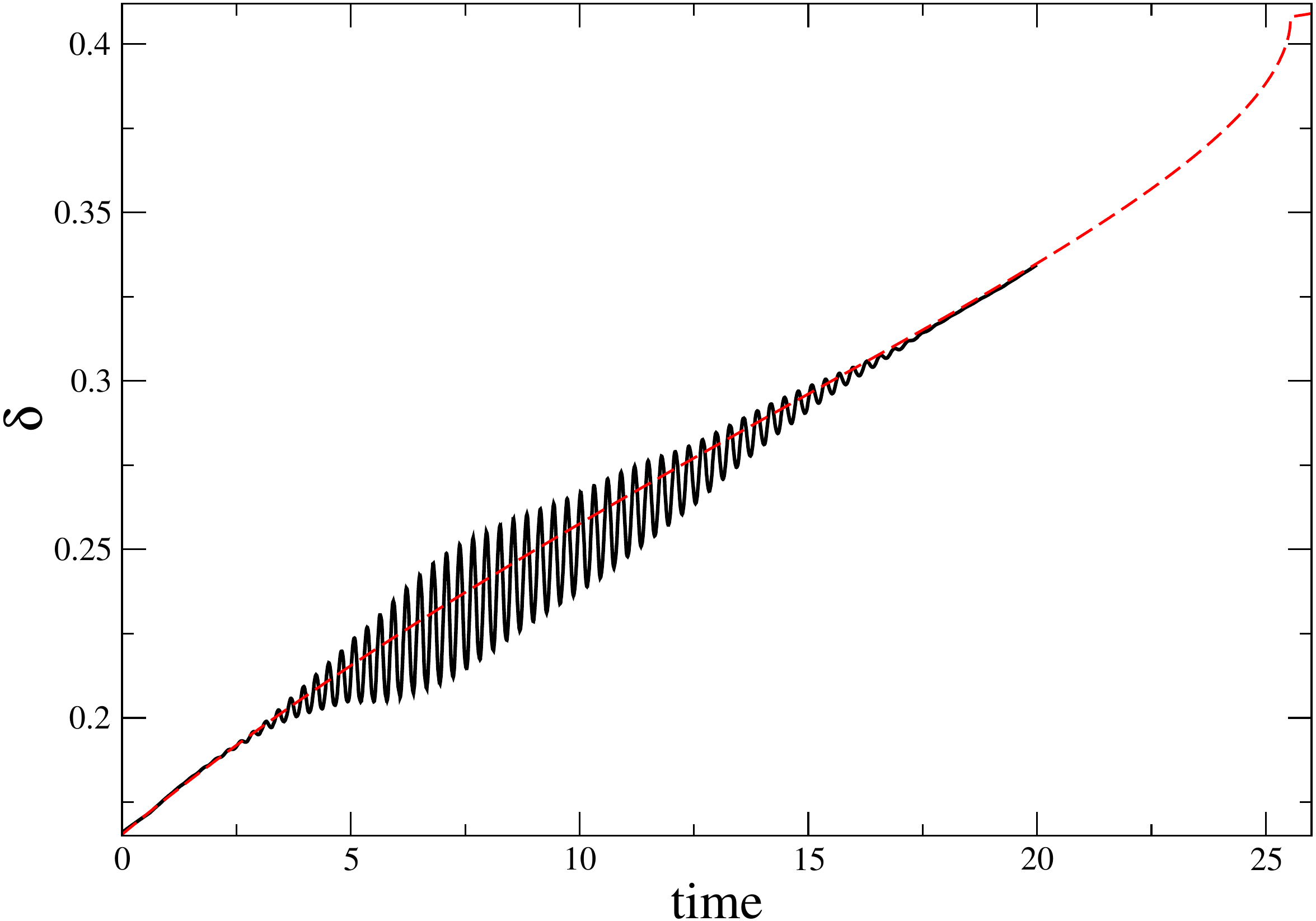}
\caption{\label{width45k} Variation of the width of an annular vortex of $\Rey=45000$ with time (solid black line). The initial condition contains an $m=4$ perturbation of amplitude $10^{-6}$ of the reference vorticity. The analytical GLO evolution is shown by the (red) dashed line.}
\end{figure*}

We next present, in figure \ref{width45k}, how the width of the annular vortex varies with time at a given azimuthal angle at a Reynolds number of $45000$, for an $m=4$ perturbation of amplitude $10^{-6}$ of the reference vorticity. At this Reynolds number, oscillations in the width are evident at moderate times. The axisymmetric analytical solution is shown for comparison and is seen to be a good indicator of the growing width of the annular region. These oscillations grow for some time, and then decay.

Like the multiple-vortex systems above that behave differently for odd and even numbers of vortices, the annular vortex responds differently to perturbations of odd and even azimuthal wavenumbers. We now discuss some cases to illustrate this.

\subsubsection{Even mode radial perturbations: $m=8$}

\begin{figure*}
\includegraphics[scale=0.3]{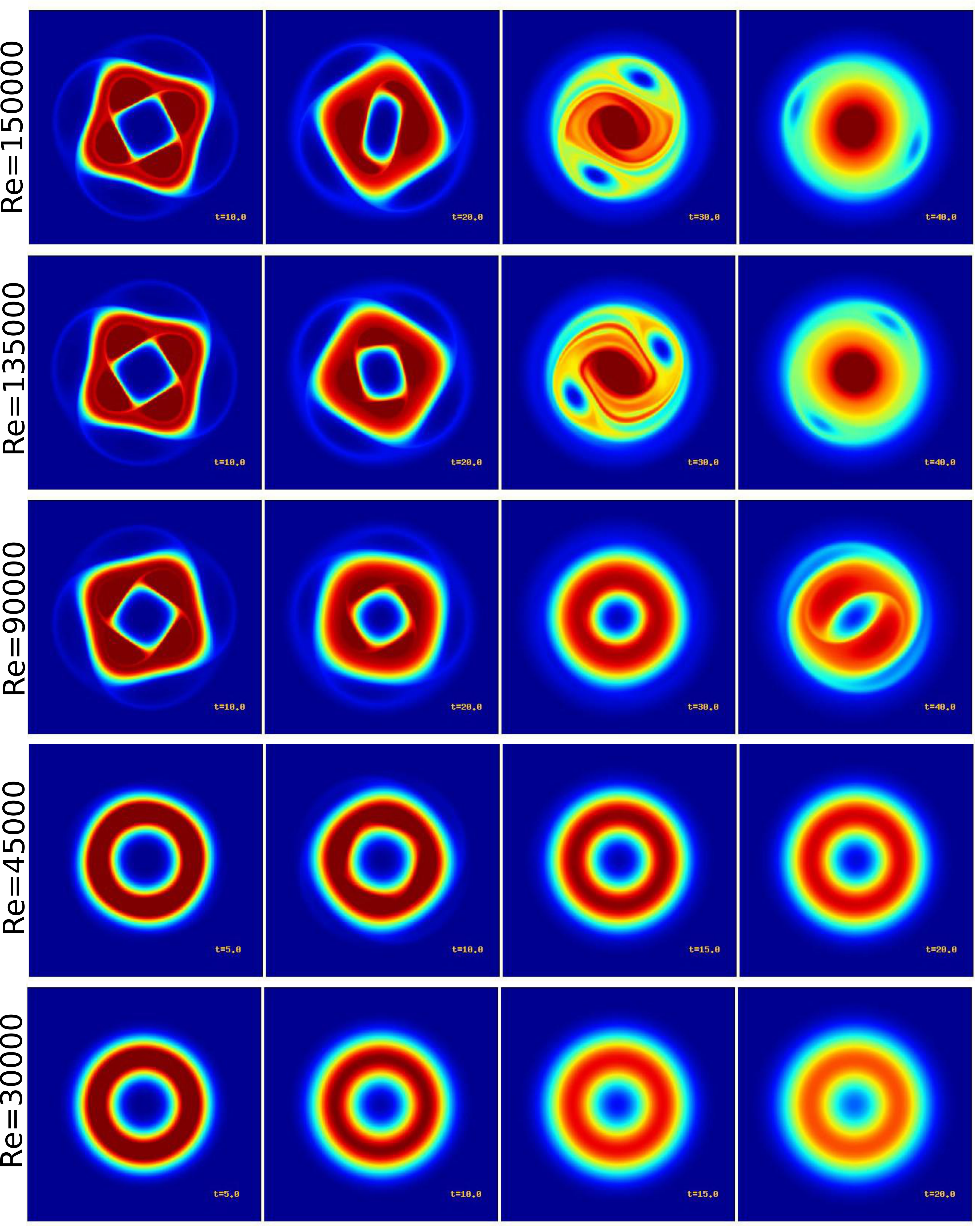}
\caption{\label{fig:fig1} Time evolution of the GLO annular vortex with an $m=8$ mode perturbation for various Reynolds number (shown in colour online; the same colourbar as in figure \ref{fig_init} is used). For small Reynolds number the GLO is stable and the radial perturbations decay. At $\Rey= 9\cdot 10^4$ instabilities start to appear. For even larger Reynolds numbers $\Rey = [1.35,1.5]\cdot 10^5$, the GLO breaks into a $4$  vortex structure that later evolves into a tripolar vortex structure.}
\end{figure*}

\begin{figure*}
\includegraphics[scale=0.35]{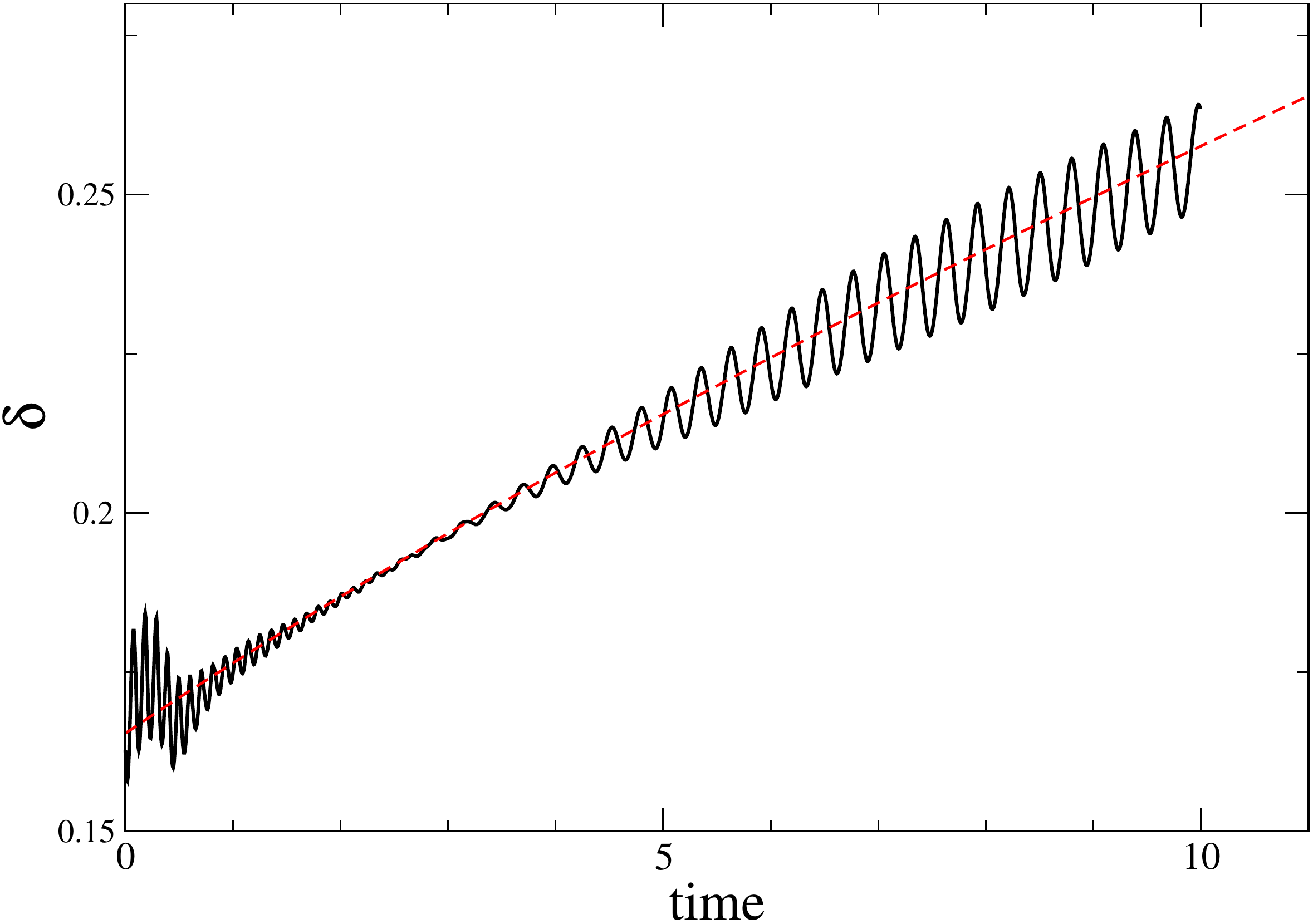}
\caption{\label{model_grow} Time variation of the width of an annular vortex at $\Rey=45000$ with an initial $m=8$ perturbation. Oscillations of this azimuthal wavenumber are seen at early times, which die down and give way to $m=4$ perturbations visible at later times. The dashed (red) line is the analytical GLO evolution.}
\end{figure*}

The plot in figure~\ref{fig:fig1} shows the evolution of a GLO vortex with an $m=8$ mode perturbation.  For Reynolds numbers  $\Rey \le 4.5\cdot 10^4$, the vortex ring is stable to initial perturbations. On increasing the Reynolds number beyond $\Rey_\Gamma=90000$ we observe that an initial instability sets in that makes a ``$4$ vortex necklace'' like structure. A similar structure had been seen in the early simulations of \cite{Schubert_etal1999} (who used $\Rey=10^5$ and an azimuthally broadband initial perturbation). For $\Rey=9\cdot10^4$ the vortex necklace disappears at intermediate times, but another instability seems to appear at late times. On the other hand, for $\Rey \geq 1.35\cdot10^5$ the vortex necklace breaks into a tripolar vortex.  Figure \ref{model_grow} shows the evolution of the width at a particular azimuthal angle of the annulus for $\Rey=45000$ and perturbations of azimuthal wavenumber $m=8$. It is seen that high frequency oscillations occur at early times, which decay and give rise to low frequency oscillations at later times. The initial oscillations correspond to an $m=8$ mode, as prescribed, whereas the low frequency oscillations are of $m=4$, as is evident in figure~\ref{fig:fig1} for this Reynolds number at a time of $10$. Our provision of an $m=8$ perturbation thus facilitates the growth of $m=4$ perturbations. These in turn die down at this Reynolds number, as was seen in figure~\ref{fig:fig1} at later times, and viscous processes take over. 

\subsubsection{Odd mode radial perturbation: $m=5$ and $9$}
\begin{figure*}
\includegraphics[scale=0.3]{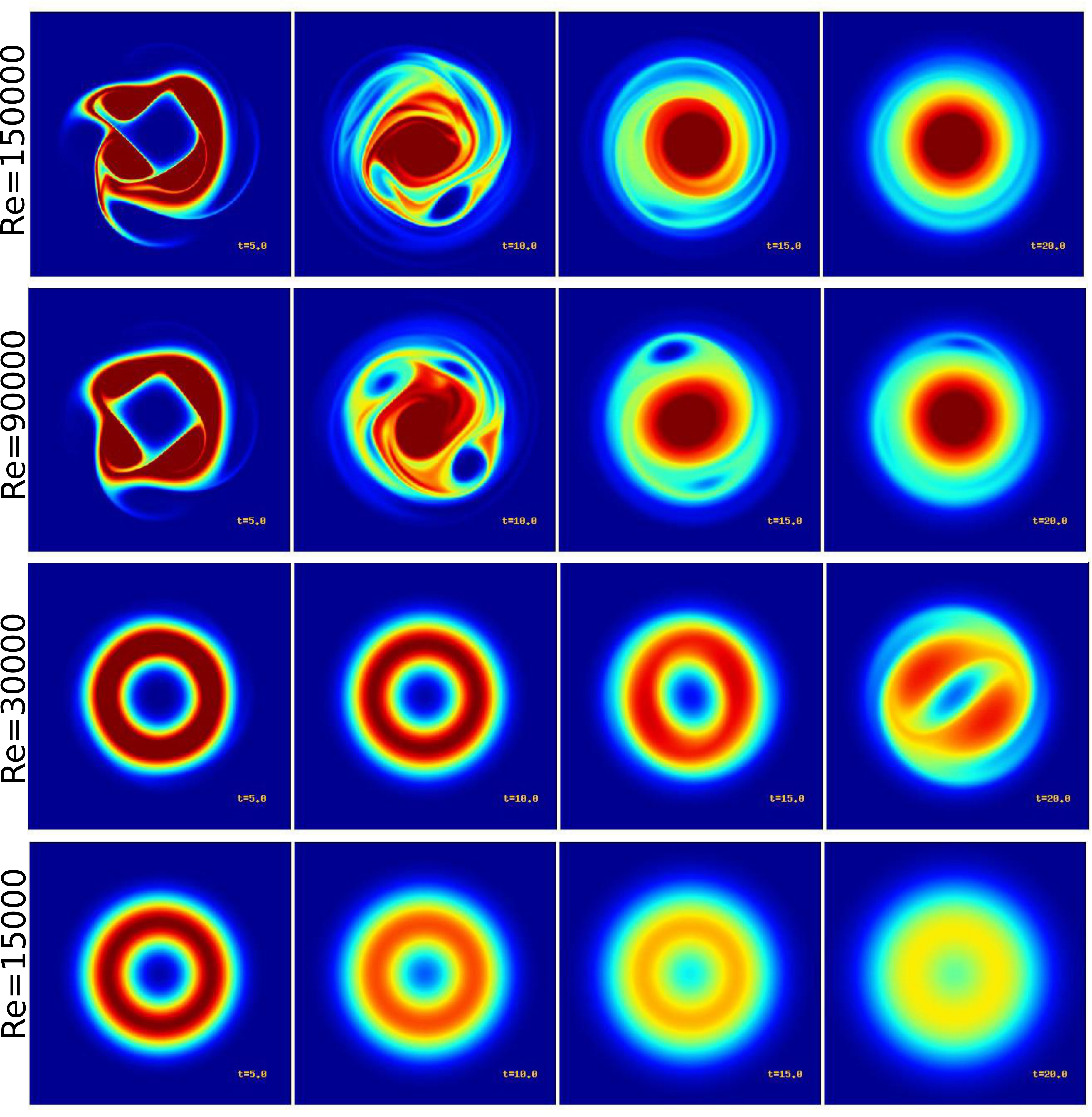}
\caption{\label{fig:fig2} Time evolution of the GLO annular vortex with an $m=5$ mode perturbation for various Reynolds number (shown in colour online; the same colourbar as in figure \ref{fig_init} is used). For small Reynolds number the GLO is stable and the radial perturbations decay. However, at larger Reynolds number the GLO breaks into a $4$ vortex necklace. On further increasing the Reynolds number, this structure breaks up into a turbulent patch.}
\end{figure*}

\begin{figure*}
\includegraphics[scale=0.3]{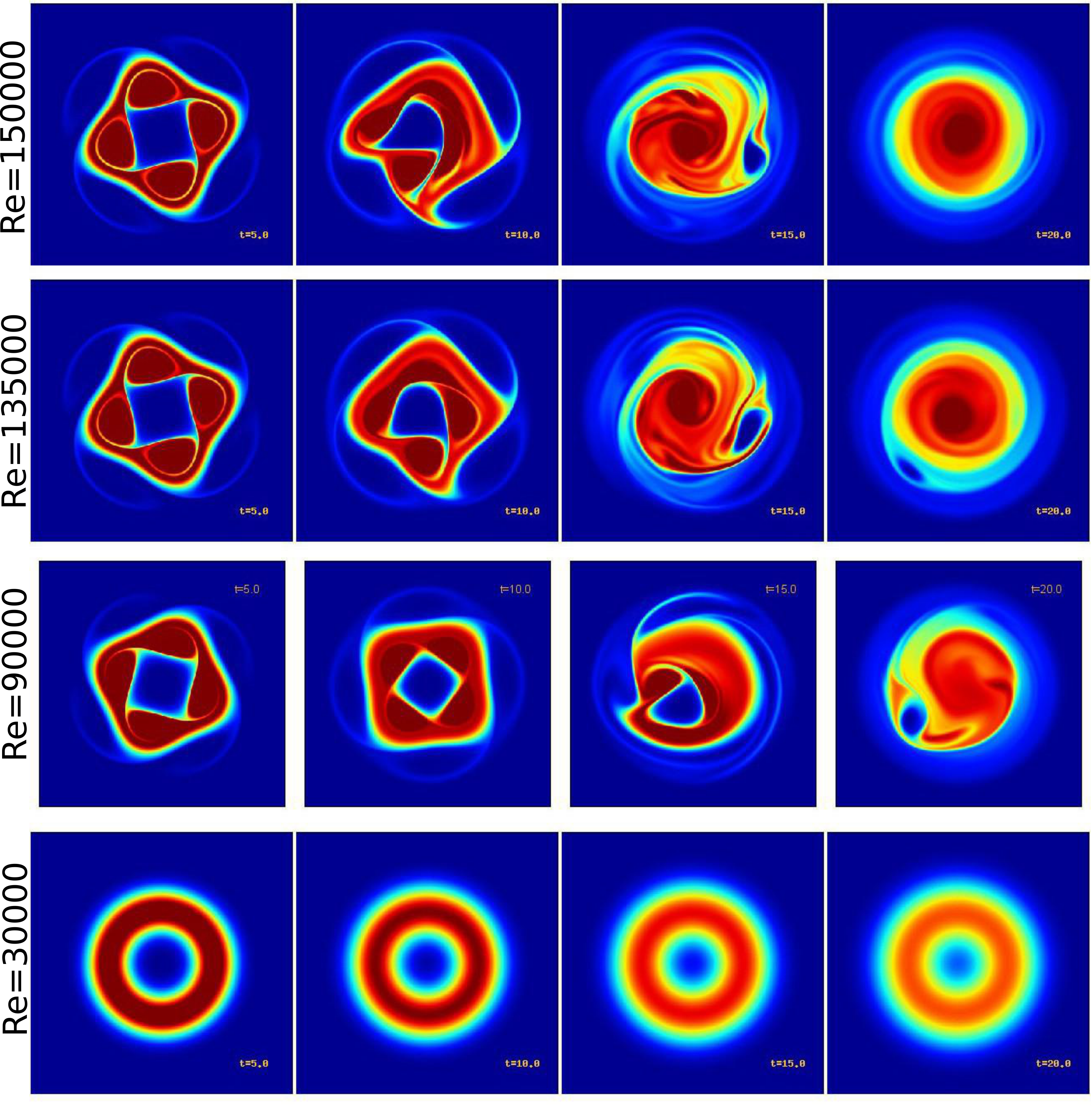}
\caption{\label{fig:fig3} Time evolution of the GLO annular vortex with an $m=9$ mode perturbation for various Reynolds number (shown in colour online; the same colourbar as in figure \ref{fig_init} is used). For small Reynolds number the GLO is stable and the radial perturbations decay. However, at larger Reynolds number the GLO breaks into a $4$ vortex necklace. On further increasing the Reynolds number, this structure breaks up into a turbulent patch.}
\end{figure*}

The plots in Figs.~\ref{fig:fig2} and \ref{fig:fig3} show the time evolution of a GLO vortex with $m=5$ and $m=9$ mode perturbations at different Reynolds number. At $\Rey=30000$ and above, the $m=9$ mode is linearly stable (see section \ref{linstab}), but triggers the $m=4$ mode which is unstable. We therefore see a $4$-vortex necklace. The $m=5$ mode is linearly unstable above $\Rey=30000$ and becomes turbulent without forming the $4$-vortex necklace.

A striking feature is that even the qualitative nature of evolution of instabilities is different for odd and even mode perturbations. For small values of Reynolds number the vortex ring is stable to odd-mode perturbations, as to even mode. On increasing the Reynolds number beyond $\Rey=90000$, we do observe, with odd mode perturbations too, that an initial instability sets in that makes a ``$4$ vortex necklace'' like structure. At a later stage however, in contrast to the $m=8$ mode perturbation, this structure breaks into filamentary structures and forms a turbulent patch. This feature stems from the fact that the vortex necklace is now less symmetric.

We next perform a linear stability analysis of the annular GLO vortex in order to better understand the results from our nonlinear simulations. We do not, however, expect the linear stability analysis to capture subtle differences between odd and even perturbation wavenumbers.

\section{Linear stability analysis of an annular vortex}
\label{linstab}

A thin cylindrical vortex sheet is linearly stable to axisymmetric perturbations in inviscid flow, as given by the condition for inviscid stability of \citet{Rayleigh1917}, that $\left(rV\right)^2$ ($V$ being the swirl velocity), should nowhere decrease as the radius $r$ increases. \citet{Michalke_Timme1967} found such a vortex sheet to be unstable to non-axisymmetric perturbations ($m\ge1$, where $m$ is the azimuthal wave number), so while Rayleigh's criterion is necessary and sufficient for axisymmetric disturbances, for non-axisymmetric disturbances it is necessary but not sufficient. \citet{Rotunno1978,Liebovich_Stewartson1983,Gent_McWilliams1986, Flierl1988,Terwey_Montgomery2002} have studied different aspects of inviscid stability in two and three dimensions, and it is evident that a cylindrical vortex is inviscidly unstable to non-axisymmetric perturbations. There are also several inviscid stability studies in the context of hurricanes and tornadoes. \cite{Schubert_etal1999} report that their inviscid stability calculations show a maximum growth rate for the $m=4$ mode. No studies, to our knowledge, have addressed the question of what happens in the viscous case, and so we conduct a viscous linear stability analysis. 

The present problem is a time-varying one, and a linear stability study can only be conducted by making the frozen flow approximation. We use as our base flow a vorticity profile frozen at a given instant of time. This quasi-steady approximation is the time-analogue of a parallel flow approximation made for spatially developing flows. \ADD{The assumption here is that changes in the basic flow, caused because the annular vortex is growing in thickness due to diffusion, are far slower than the perturbation time scale. In other words, during one oscillation of the perturbation wave, the change in thickness is so small that the GLO vortex can be taken to be constant during this time. The validity of the assumption for the present flow is estimated below by comparing actual perturbation time scales to the rate at which the GLO vortex thickness grows, and it will be seen that these time scales are separated by two orders of magnitude. } The GLO vortex produces an entirely axisymmetric flow in which the nonlinear terms vanish. Perturbations imposed on the axisymmetric initial state bring the nonlinear terms into play. As the GLO vortex gets smoothed out by viscosity, we expect that the flow becomes less unstable to perturbations.

Following the standard procedure, we decompose the total motion into a background state and express the perturbations in their normal mode form, characterized by an azimuthal wavenumber $m$ and a complex frequency $f$, as, e.g. for the azimuthal component of velocity
\begin{align}
u_{\theta_{tot}}&=U_{\theta}\left(r\right) + \hat{u}_{\theta}\left(r\right) e^{i\left(m\theta-ft\right)}.
\end{align}

Substituting such expressions into the two-dimensional incompressible continuity and Navier-Stokes equations in plane polar co-ordinates, neglecting nonlinear terms in the perturbations, and eliminating pressure and $\hat u_{\theta}$, we are left with an eigenvalue problem in the radial velocity component $u_r$:
\begin{align}
A\hat{u}_{r}=fB\hat{u}_{r}.
\label{eq:linearised}
\end{align}
The system is stable if the imaginary part of $f$, i.e., $f_{i}<0$ and unstable if $f_{i}>0$.  $A$ and $B$ are linear differential operators given by

\begin{align}
\label{eq:linear_lhs}
A&= i\nu\left\{r^2D^4+6rD^3+\left(5-2m^2\right) D^2-\frac{\left(1+2m^2\right)D}{r}+\frac{\left(m^2-1\right)^2}{r^2}\right\}
\nonumber\\
&+m\Omega \left\{r^2D^2+3rD+\left(1-m^2\right)\right\}-mrZ^{\prime},
\end{align}
and
\begin{align}
\label{eq:linear_rhs}
B&=\left\{r^2D^2+3rD+\left(1-m^2\right)\right\},
\end{align}
where the operator $\displaystyle D\equiv {d/dr}$, angular velocity $\displaystyle \Omega=U_{\theta}/r$, and base flow vorticity $\displaystyle Z=U_{\theta}^{\prime}+U_{\theta}/r$. Primes denote derivative with respect to $r$. The linear stability equations (\ref{eq:linearised}, \ref{eq:linear_lhs}, \ref{eq:linear_rhs}) are the same as equation (3.1) in \cite{Harish2010}, written for constant density. The boundary conditions depend on the mean profiles, which obey $U_\theta\sim r$ as $r\to0$, $\displaystyle U_\theta\sim 1/r$ as $r\to\infty$. The boundary conditions are then deduced from the leading order terms of the Taylor series about $r=0$ and $r \to \infty$, satisfying, for large Reynolds numbers,
\begin{align}
\hat{u}_{r}\sim r^{|m|-1}, {\rm for}~ r\to0, \quad {\rm and} ~ \hat{u}_{r} \sim r^{-|m|-1} ~{\rm as} ~ r\to \infty.
\end{align}
A similar exercise for three-dimensional perturbations yields exponential, rather than algebraic, decay at large $r$. These boundary conditions were originally derived by \citet{Batchelor_Gill1962} in the form of compatibility relations, and were formalized later on by \citet{Lessen_Paillet1974} and \citet{Khomalash1989}.

We study the linear stability of the GLO vortex of equation (\ref{finalvort}). 
We have made detailed comparisons with a smoothed annular Rankine profile and get no qualitative difference in the answers. We solve the eigenvalue problem above using a Chebyshev collocation technique. To specify a grid in the domain $\left[0,R_{max}\right]$, and to cluster grid points into the vicinity of the annular vortex, we use a modified version of the algebraic stretching used by \citet{Khomalash1989}, given by
\begin{equation}
y=\frac{x\left(1-\xi^3\right)}{q_1+q_2\xi+(1-q_2)\xi^3},\quad q_1=1+\frac{2x}{R_{max}},\quad x=\frac{pR_{max}}{R_{max}-2p},
\label{eq:algb}
\end{equation}
where $\xi$ is the Chebyshev coordinate lying in [-1,1], $p$ is the radius at which clustering is required, and $q_2$ is chosen to lie between $0.5$ and $0.8$. We have also used the exponential stretching of \cite{Rama2004} and obtained the same answers. Our computational domain extends up to $R_{max}=1000$, and with the number of Chebyshev collocation points $N$ of up to $1000$.

\begin{figure*}
\includegraphics[scale=0.35]{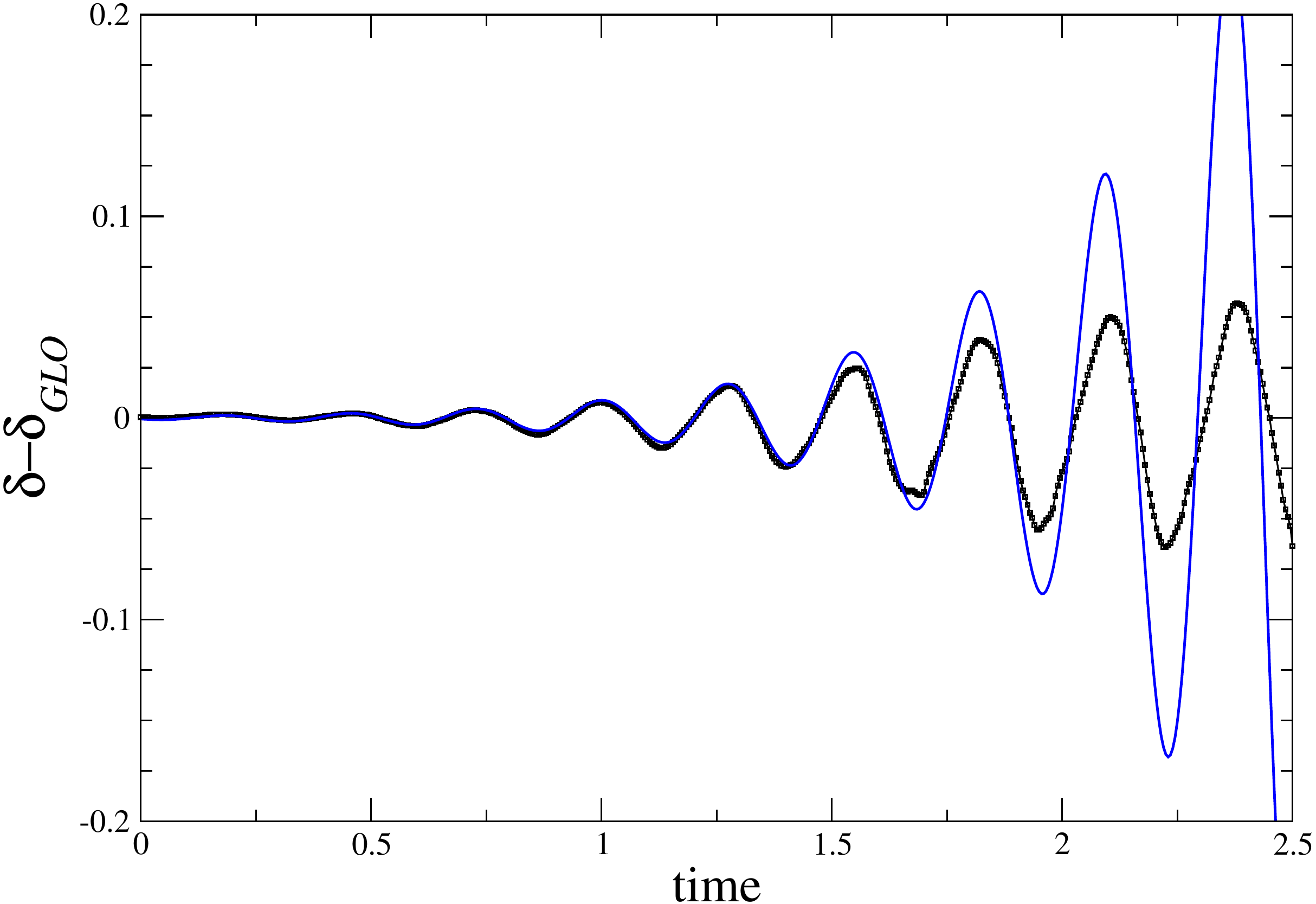}
\caption{\label{fig:growth_best_fit} A best fit of the oscillations seen at moderate times (shown in colour online). The (black) filled squares are numerical simulations at $\Rey=45000$, with an $m=4$ perturbations of amplitude $10^{-3}$ in the vorticity. The (blue) line is a fit, given by equation (\ref{fit001}).}
\end{figure*}

\begin{figure*}
\includegraphics[scale=0.25]{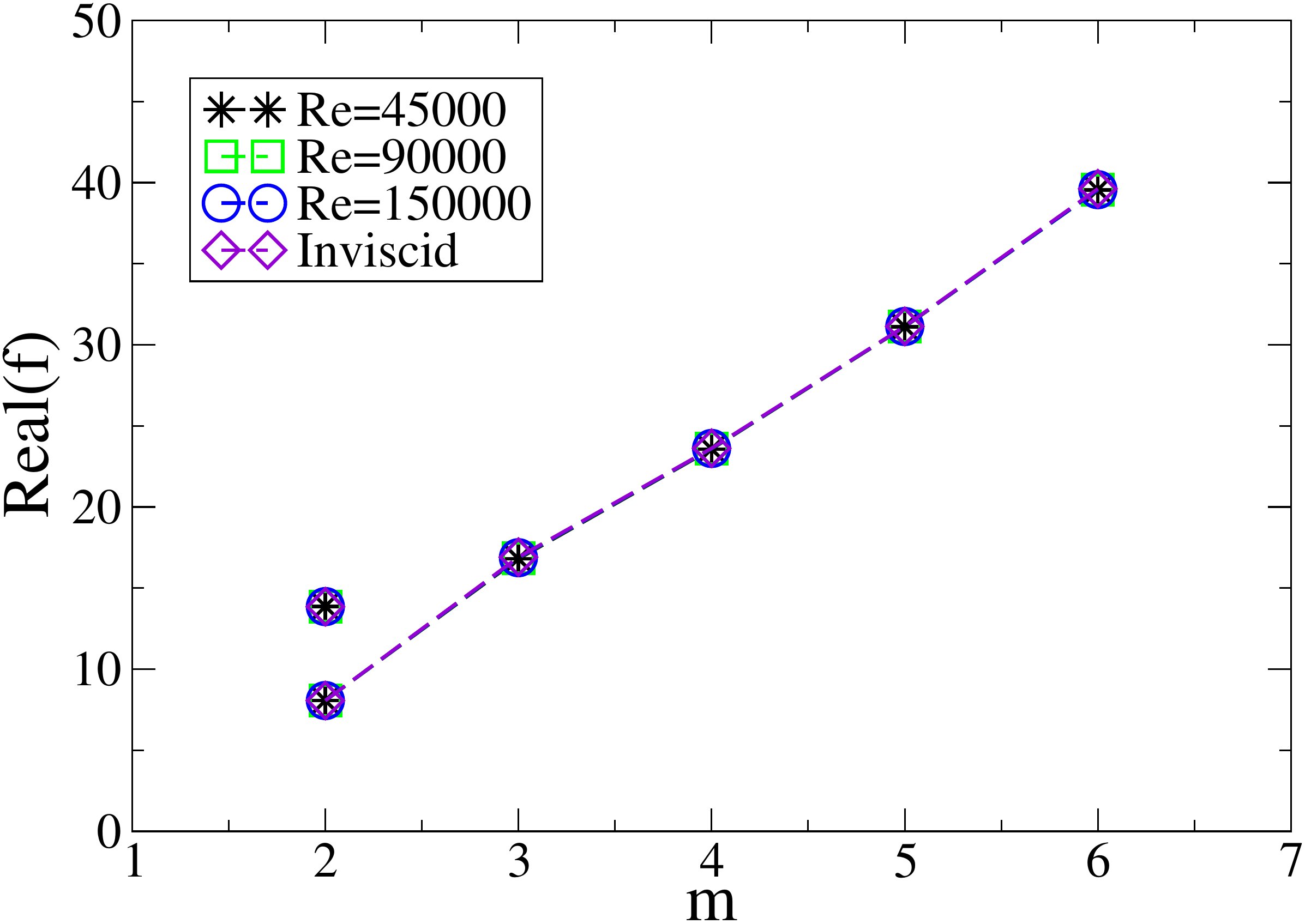}
\includegraphics[scale=0.25]{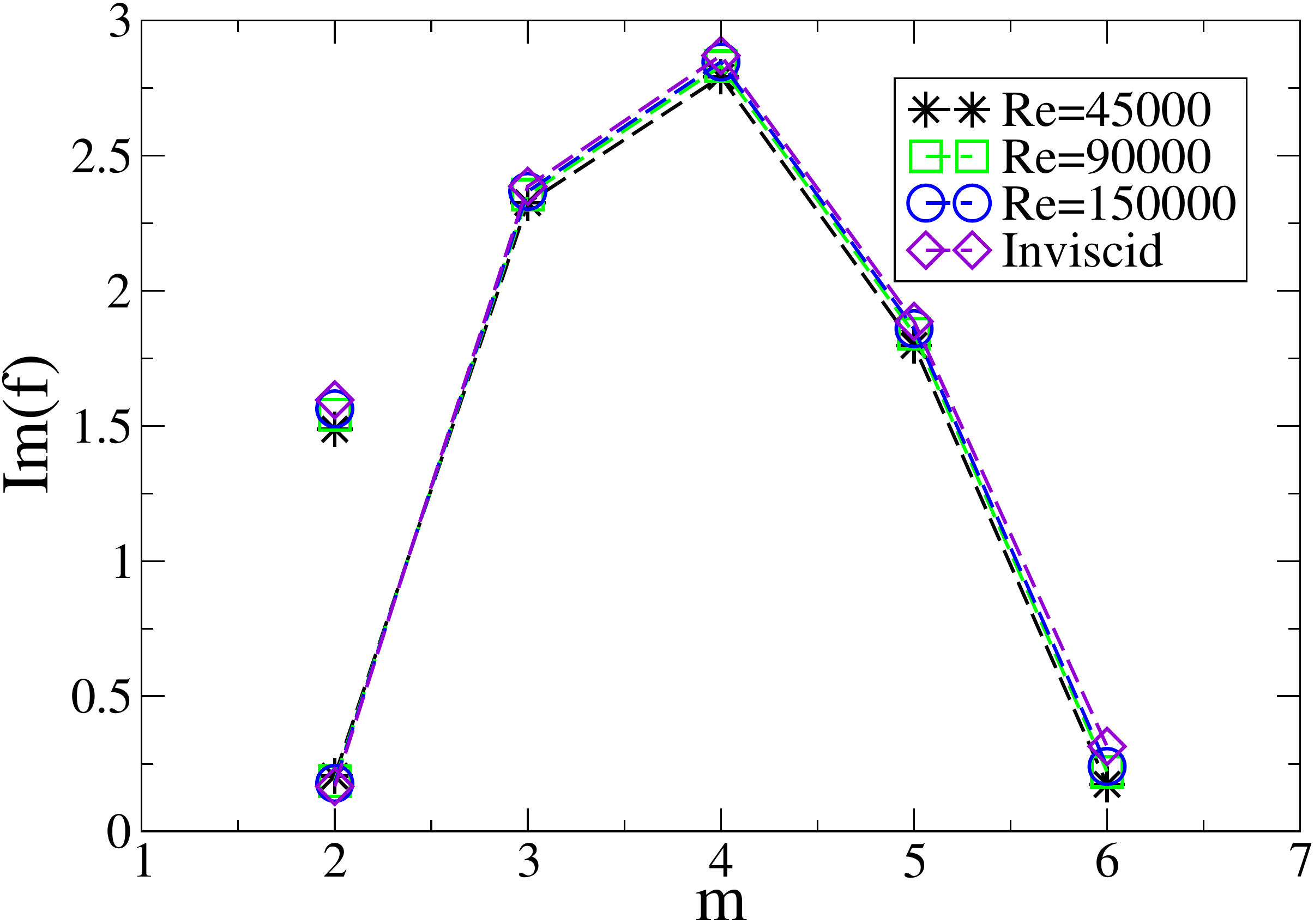}
\caption{\label{fig:linear_growth_frequency}  Circular frequency (left) and growth rate (right) as a function of azimuthal wavenumber from the linear stability analysis. The $m=4$ mode has the highest growth rate, and there are two unstable $m=2$ modes (see text).}
\end{figure*}

Instability growth rates, as well as the azimuthal wavenumbers, increase with increasing Reynolds number and, at high Reynolds numbers, are very similar to inviscid predictions. We discuss linear stability results for a Reynolds number of $45000$, which is high enough that disturbance growth is visible to the naked eye in the simulations, and small enough that at long times, viscous effects take over and the annulus returns to a GLO vortex. The circular frequency and growth rate of the most unstable perturbations from linear stability predictions, in the non dimensional units of the simulation, are given in figure \ref{fig:linear_growth_frequency}.
\begin{figure*}
\includegraphics[scale=0.25]{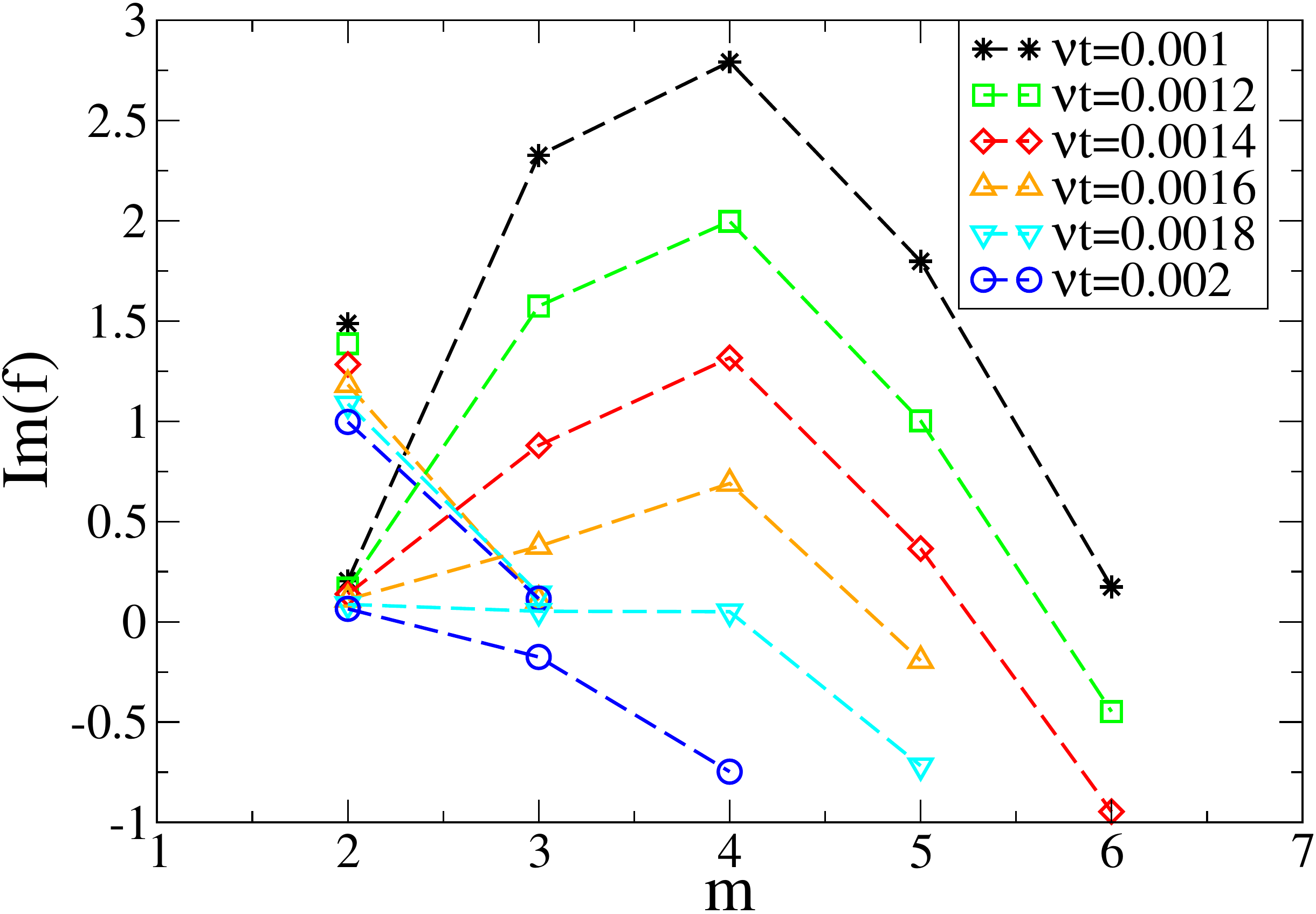}
\caption{\label{fig:vary_nut_or_delta} Growth rate as a function of azimuthal wavenumber for varying initial $\delta$ or $b^2=\nu t/a^2$ from the linear stability analysis for a Reynolds number of $45000$. $\nu t=0.001$ corresponds to our initial profile or a simulation time of zero. $\nu t=0.002$ corresponds to a simulation time of $7.5$.}
\end{figure*}

Figure \ref{fig:linear_growth_frequency} also shows the existence of two $m=2$ modes that are unstable. The  slower-growing of the two is the same as that of \cite{Michalke_Timme1967}. A more detailed analysis of the faster growing $m=2$ mode is done elsewhere, and will not be discussed further in the present study.

To compare the linear stability results with nonlinear dynamics, we return to the oscillations seen in figure \ref{width45k}. We subtract the width of the GLO vortex at every time, and plot the oscillations at early times in figure \ref{fig:growth_best_fit}. Shown in the same figure is a best-fit curve, given by
\begin{equation}
\delta-\delta_{GLO} = 0.0008 e^{2.5t}\sin\left[23t-2.5\right].
\label{fit001}
\end{equation}
It is seen that the frequency of the oscillation is close to the value of $23.6$ predicted by linear theory for $m=4$. This frequency is maintained constant over a long time. However, we do not see any reasonable stretch of exponential growth, so nonlinearities are important throughout the evolution. We have performed $m=4$ simulations with three amplitudes of initial perturbation, $10^{-6}$, $10^{-3}$ and $10^{-2}$. All three show oscillations whose frequency is extremely well fitted by a value of $23$. Exponential growth only occurs over short stretches of time in each case, and these best-fit growth rates are different, at $1.75$, $2.5$ and $3.5$ respectively. These growth rates compare well with the linear stability growth rate of $2.8$. The simulation of \cite{Schubert_etal1999} was for a Reynolds number of $10^5$, and although they used only $512^2$ grid points at this high Reynolds number, the growth rate they report is also comparable. In the figure \ref{model_grow}, we find initial oscillations of azimuthal wavenumber $8$ whose frequency is very close to the value of $58.17$ obtained by linear stability, but no evidence of exponential growth. The $m=4$ oscillations at later time are of frequency close to $24$. Such behaviour is not uncommon in shear flows, with the frequency predicted by linear stability studies being quite prominent in the dynamics even after nonlinearities have become dominant.

\red{We return to re-examine the frozen-flow assumption by the typical case of $m=4$ seen above. Equation (\ref{fit001}) gives the perturbation time scale as $1.7^{-1}$, and it may be seen from figure \ref{width45k} that during the time period of one oscillation of the perturbation, the analytical base state of the annular GLO vortex, shown by the dashed line, grows in thickness by around $1$ percent. Thus we see that the time-scales of the base state and the perturbation are vastly separated and the frozen-flow approximation is valid.}

\ADD{\section{Discussion and Conclusion}}

To summarise, \ADD{ we first study the viscous dynamics of $n \ge 3$ vortices placed at the vertices of a regular polygon. In inviscid flow, arrangements of patches of uniform vorticity are known to behave similarly to point vortices as long as their radii are smaller than a critical threshold. Above this threshold, the vortices deform significantly and tend toward merger. The role of viscosity in this initial stage of the dynamics is to grow the radii of the vortices past the critical radii. } At moderate Reynolds number, the vortices align themselves into a long-lived annular structure. The annular vortex may be described analytically as a generalised Lamb Oseen vortex, which diffuses slowly inwards to form a single Gaussian vortex at long times. The annular stage dominates the merger process more as the number of vortices increases, effectively delaying the merger. At high Reynolds numbers, a single vortex at the centre is attained at large times, but there is no annular structure formation. Remarkably, and the behaviour for odd and even number of vortices is qualitatively different. When $n$ is even, the merger process at high Reynolds numbers is characterised by vortex pairing events rather than annulus formation. The option of complete vortex pairing is not available when $n$ is odd, and a breakdown into chaotic motion is seen. In a much shorter time than when $n$ is even, a single central vortex is formed. \ADD{As expected, the merger events cause energy to first cascade to larger length scales, whereas the formation of filamentary structures causes a cascade of energy to smaller scales.}

We then study the annular vortex in some detail, performing numerical simulations and (viscous) stability analysis. The time evolution of our numerical simulations of the annular vortex agrees extremely well with our generalised Lamb Oseen solution for low Reynolds numbers. Our numerical simulations show that the annular vortex is stable to imposed perturbations at $\Rey=45000$ and below, while instabilities clearly influence dynamics at $\Rey=90000$ and above. The nonlinear evolution of even and odd mode perturbations is strikingly different. Again the odd mode perturbations result in a more chaotic flow, and collapse to a single vortex at much shorter times, whereas even modes proceed via reduction of wavenumber to effect a large slow down of the inverse cascade process.

This work, we hope, will motivate experimenters and simulators of two-dimensional turbulence to evaluate the contributions to the inverse cascade of multiple vortex interactions.

\bigskip

{\bf Appendix}

Taking the Laplace transform in time of Eqs.~\ref{eq:vorticity} and using Eq.~\ref{eq:incondition}, we get
\begin{align}
\frac{d^2\hat{\omega}}{dr^2}+\frac{1}{r}\frac{d\hat{\omega}}{dr}-\frac{s\hat{\omega}}{\nu}&=-\frac{\omega\left(r,0\right)}{\nu}=f\left(r\right),
\label{symmetrybc}
\end{align}
where $\hat{\omega}$ is the Laplace transform of $\omega$. The homogeneous part is a modified Bessel equation in the variable $r\sqrt{\frac{s}{\nu}}$, whose solution is
\begin{align}
\hat{\omega}_{h}=c_{1}I_{0}\left(r \sqrt{\frac{s}{\nu}}\ \right)+c_{2}K_{0} \left(r \sqrt{\frac{s}{\nu}}\ \right),
\end{align}
where $I_{0}$ and $K_{0}$ are modified Bessels functions of the first and second kind respectively \citep{Abramowitz&Stegun1965}. The particular solution is obtained by variation of parameters, as
\begin{align}
\hat{\omega}_{p}=I_{0}\left(r\sqrt{\frac{s}{\nu}}\right)\int_0^r r^{\prime}f\left(r^{\prime}\right)K_{0}\left(r^{\prime}\sqrt{\frac{s}{\nu}}\right)dr^{\prime}-K_{0}\left(r\sqrt{\frac{s}{\nu}}\right)\int_0^r r^{\prime}f\left(r^{\prime}\right)I_{0}\left(r^{\prime}\sqrt{\frac{s}{\nu}}\right)\,dr^{\prime}~,
\end{align}
The complete solution $\hat{\omega}=\hat{\omega}_{h}+\hat{\omega}_{p}$. Applying the boundary conditions, which require that $\hat \omega$ be finite at $r=0$, and that $\hat{\omega}\to0$ as $r\to\infty$, and remembering that $K_0$ and $I_0$ diverge respectively as $r \to 0$ and $\infty$, we get
\begin{align}
\label{lapfinalvort}
\hat{\omega}=\begin{cases}
\displaystyle \frac{\Gamma_{o}}{2\pi \nu}I_{0}\left(r\sqrt{\frac{s}{\nu}}\right)K_{0}\left(a\sqrt{\frac{s}{\nu}}\right)& \textnormal{for } r \leq a~, \\[1em]
\displaystyle \frac{\Gamma_{o}}{2\pi \nu}I_{0}\left(a\sqrt{\frac{s}{\nu}}\right)K_{0}\left(r\sqrt{\frac{s}{\nu}}\right)& \textnormal{for } r > a~.
\end{cases}
\end{align}
It can be seen that the symmetry condition (Eq. \ref{symmetrybc}) is automatically satisfied. Taking the inverse Laplace transform, and using the following standard result from \citet{Erdelyi1954}, 
\begin{align}
L^{-1}\bigg\{K_{n}\left[\sqrt{s}\left(\sqrt{\alpha}+\sqrt{\beta}\right)\right]I_{n}\left[\sqrt{s}\left(\sqrt{\alpha}-\sqrt{\beta}\right)\right]\bigg\}=\frac{1}{2t}e^{-\left(\frac{\alpha+\beta}{2t}\right)}I_{n}\bigg\{\frac{\alpha+\beta}{2t}\bigg\}~,
\end{align}
we get equation (\ref{finalvort}).

\bibliographystyle{revtex4-1}
%{aiprev4-1}
\bibliography{annular_pre}
\end{document}